\documentclass[]{elsarticle}
\usepackage{geometry}
\usepackage[titletoc,page]{appendix}
\usepackage{amsmath,amssymb}
\usepackage{graphicx,xcolor}
\usepackage{subcaption}
\usepackage{hyperref}
\usepackage[capitalise]{cleveref}
\usepackage{natbib}
\usepackage{braket}
\usepackage{wrapfig}
\usepackage{mathabx}

\usepackage{lineno}
\modulolinenumbers[5]

\newcommand{\fdir}{.}
\newcommand{\tops}[2]{\texorpdfstring{#1}{#2}}

\DeclareMathOperator{\sgn}{sgn}

\newcommand{\iu}{{\mathrm i}}

\newcommand{\inv}[1]{\ensuremath{\frac{1}{#1}}}
\newcommand{\intd}{\ensuremath{\mathrm{d}}}
\newcommand{\xip}{\ensuremath{f^+}}
\newcommand{\xim}{\ensuremath{f^-}}
\newcommand{\xipm}{\ensuremath{f^{\pm}}}

\newcommand{\xic}{\ensuremath{f^c}}
\newcommand{\xitp}{\ensuremath{f_t^+}}
\newcommand{\xitpm}{\ensuremath{f_t^{\pm}}}

\renewcommand{\vec}{\ensuremath{\mathbf}}

\graphicspath{{graphics/classstat/}}

\biboptions{sort&compress}
\journal{Nuclear Physics B}

\begin{document}
\title{Spectral functions and dynamic critical behavior of relativistic \tops{$Z_{2}$}{Z\_2} theories}

\address[gie]{Institut f\"ur Theoretische Physik, Justus-Liebig-Universit\"at, Heinrich-Buff-Ring 16, 35392 Gie{\ss}en, Germany}
\address[hfhf]{Helmholtz Research Academy Hessen for FAIR (HFHF), Campus Gie{\ss}en, 35392 Gie{\ss}en, Germany}
\address[bie]{Fakult\"at f\"ur Physik, Universit\"at Bielefeld, D-33615 Bielefeld, Germany}

\author[gie]{Dominik Schweitzer}
\ead{dominik.schweitzer@theo.physik.uni-giessen.de}
\author[bie]{S\"{o}ren Schlichting}
\ead{sschlichting@physik.uni-bielefeld.de}
\author[gie,hfhf]{Lorenz von Smekal}
\ead{lorenz.smekal@physik.uni-giessen.de}

\begin{abstract}
    We investigate the dynamic critical behaviour of a relativistic scalar field theory with $ Z_2$ symmetry by calculating spectral functions of the order parameter at zero and non-vanishing momenta from first-principles classical-statistical lattice simulations in real-time.
    We find that at temperatures above the critical point ($T>T_c$), the spectral functions are well described by relativistic quasi-particle peaks.
    Close to the transition temperature ($T\sim T_c$), we observe strong infrared contributions building up.
    In the ordered phase at low temperatures ($T<T_c$), in addition to the quasi-particle peak, we observe a soft mode with a dispersion relation indicative of collective excitations.
    Investigating the spectral functions close to $T_c$, we demonstrate that the behavior in the vicinity of the critical point is controlled by dynamic scaling functions and the dynamic critical exponent $z$, which we determine from our simulations.
    By considering the equations of motion for a closed system and a system coupled to a heat bath, we extract the dynamic critical behavior for two different dynamic universality classes (Models A \& C) in two and three spatial dimensions.
\end{abstract}


\maketitle    
\tableofcontents
\clearpage

\section{Introduction}\label{sec:intro}
When a thermodynamic system comes close to a critical point, one finds anomalous behaviour in a multitude of its properties.
Large fluctuations lead to scale-invariant physics, not only in static quantities, but also in dynamic observables such as multi-time correlation functions and relaxation rates \cite{hohenberg_theory_1977}.
These observables can be measured by a range of experiments, e.g.~in condensed-matter applications like ultra-thin films \cite{dunlavy_critical_2005}, and will likely become relevant in future heavy-ion collision experiments closing in on the QCD critical endpoint (CEP) \cite{akiba_hot_2015}.

Over the last few years, the search for the QCD critical endpoint has gained a considerable amount of theoretical and experimental attention \cite{odyniec_rhic_2013,bzdak_mapping_2020}. 
Dedicated heavy-ion collision experiments at the Relativistic Heavy-Ion Collider (RHIC), and the future GSI-FAIR and NICA facilities are designed to probe the relevant region of the QCD phase diagram, so the task is set for theorists to predict how the QCD critical point can be located with the obtained data.
Since first principle calculations based on lattice QCD are currently unable to address the interesting high-density region at large baryon chemical potential, one has to employ effective theories, making use of prior knowledge about its critical behaviour \cite{bluhm_dynamics_2020}.

The scale-invariant nature of dynamic critical phenomena leads to universal behaviour of microscopically different systems, and to an identical value of the dynamic critical exponent $z$.
But in addition to the well-known universality classes governing static critical phenomena, one must take into account properties of the equations of motion of the system.
By factoring in conservation laws and mode couplings, Hohenberg and Halperin \cite{hohenberg_theory_1977} developed a scheme of dynamic universality classes, called ``Models.''
In some of these models, the dynamic critical exponent $z$ follows directly from the static critical exponents via simple scaling relations. 
This is for example the case for Model C, a relaxational model with a conserved density. 
In general, however, the precise determination of $z$ turns out to be significantly more challenging. 
Even for the rather simple 2D Ising model with Glauber dynamics (Model A), rather sophisticated methods \cite{nightingale_dynamic_1996} were needed to obtain a reasonably precise result for $z$.

In this study we investigate a relativistic scalar field theory in the static universality class of the QCD CEP ($ Z_2$ Ising), yet in a different dynamic one.
While the dynamic universality class at the QCD critical endpoint is believed to be that of Model H \cite{son_dynamic_2004}, we will focus here on the conceptually simpler relaxational Models A and C which do not require the explicit treatment of an additional conserved vector field and its dynamic coupling to a conserved order parameter.
Nevertheless, we believe that investigating the simpler models provides a first important step towards characterizing the real-time dynamics in the vicinity of a critical point in a systematic way, ready to be extended to more complicated models such as Model H in the future, which will then be of direct relevance to QCD.

Our analysis of dynamic critical phenomena is based on the real-time correlations of the order-parameter field, specifically, its spectral function.
Spectral functions are of great interest in many areas of condensed matter, nuclear and particle physics.
They play an important role for our understanding of the physics of heavy-ion collisions, the Early Universe and more.
Containing the spectrum of quasi-particle, multi-particle and collective excitations of the system in the given channel, they can be used to identify the relevant degrees of freedom e.g.~for transport or hydrodynamic descriptions.
In addition, transport coefficients are typically also obtained 
from particular low-energy limits of spectral functions \cite{moore_bulk_2008}.
Naturally, close to a critical point, the emergent dynamic critical phenomena have a strong effect on spectral functions.
Since slow modes dominate the dynamics, infrared-divergent power laws arise, whose spectral index is controlled by the dynamic critical exponents $z$ and the static critical exponents $\alpha,\beta,\gamma,\delta$, and $\nu$.
Due to scale invariance in the vicinity critical point, one further expects that, similar to static quantities, the low-energy behaviour of the spectral function is fully described by universal scaling functions.

Of course, to capture such highly non-trivial infrared phenomena, one needs to calculate spectral functions non-perturbatively.
Quite generally, however, this is very challenging for all equilibrium field theory methods in Euclidean space-time, as for example first-principles lattice QCD simulations, because it requires an analytic continuation of Euclidean correlations back to the real-time domain which is a numerically ill-posed inverse problem.
While different reconstruction schemes exist for different purposes, ranging from Maximum Entropy methods \cite{jarrell_bayesian_1996,asakawa_maximum_2001,burnier_bayesian_2013}, the Bachus-Gilbert method \cite{brandt_pion_2015}, Tikhonov regularization \cite{dudal_kallen-lehmann_2014} or the Schlessinger-Point (or Resonances-via-Padé) method \cite{schlessinger_use_1968,tripolt_threshold_2017}, all of these come with 
limited ranges of applicability (see, e.g.~Ref.~\cite{tripolt_numerical_2019} for a recent comparison).
To calculate spectral functions from non-perturbative functional methods, e.g.~based on Dyson-Schwinger or Functional Renormalization Group equations \cite{roeder_selfconsistent_2006,mueller_quark_2010,floerchinger_analytic_2012,kamikado_real-time_2014,tripolt_spectral_2014,tripolt_flow_2014,mesterhazy_quantum_2015,pawlowski_real_2015,strodthoff_self-consistent_2017,pawlowski_finite_2018,fischer_bayesian_2018} can provide a powerful alternative, especially when the functional equations are analytically continued or even formulated directly on the closed-time path. 
However, in one way or another, they all require prescriptions to obtain a closed set of equations form an  originally infinite hierarchy.
Controlling truncation errors then becomes important for the systematics, and any a priori knowledge of the structure of correlations in the theory is obviously beneficial as additional input or benchmark.

In this work, we calculate the spectral function of a single-component scalar field theory employing classical-statistical lattice simulations in real-time \cite{aarts_spectral_2001,berges_dynamic_2010,schlichting_spectral_2019}.
We use the fluctuation-dissipation relation or Kubo-Martin-Schwinger (KMS, \cite{kubo_statistical-mechanical_1957,martin_theory_1959}) condition to obtain the spectral function from the statistical function which in the classical limit can be calculated from an elementary unequal-time correlation function of classical fields.
While the method was successfully employed in a pre-cursor study \cite{berges_dynamic_2010}, where the dynamic critical exponent of a scalar field theory in two spatial dimensions with conservative dynamics (Model C) was calculated, we extend this study in several regards and systematically compare the results in 2+1 and 3+1 dimensions. 
By extending the analysis of spectral functions to finite spatial momenta, we extract the universal scaling functions that govern the dynamics in the vicinity of the critical point. 
We also quantify the divergence of the characteristic time scale $\xi_t$ at the critical point and extract the critical exponent $z$ for the different dynamic universality classes.

This paper is organized as follows:
After giving a short overview of our model and simulation setup in \cref{sec:setup}, the classical-statistical approach to calculate spectral functions is explained in \cref{sec:specfunc}.
We continue with a deeper look into the static critical behaviour of the system in \cref{sec:scale}, where we also extract the non-universal amplitudes of critical power laws, which allow us to present our subsequent results in terms of dimensionless scaling variables.
Starting with \cref{sec:noncrit}, we present results for spectral functions at different points in the phase diagram, and analyze their behaviour away from criticality.
We proceed with a detailed investigation of the dynamic critical behavior in \cref{sec:critdyn}, where we extract the dynamic scaling functions and scaling exponents. 
Our conclusions as well as an outlook for further applications are provided in \cref{sec:conclusion}.
Some details on our analysis of the diverging time scales as well as the general structure of the non-critical spectral functions are given in the appendices.

\section{Simulation setup}\label{sec:setup}
We simulate a relativistic $Z_2$ scalar field theory defined by the following lattice Hamiltonian:
\begin{equation}
    H = \sum\limits_x a^d \left\{ \frac{1}{2} \pi_x^2 - \frac{1}{2 a^2} \sum\limits_{y\sim x} \phi_x \phi_{y} + \left( \frac{m^2}{2} + \frac{d}{a^2} \right)\phi_x^2 + \frac{\lambda}{4!}\phi_x^4 + J\phi_x \right\},
    \label{eq:z2_hamiltonian}
\end{equation}
where $\phi_x$ are single-component scalar field variables at lattice site $x$, and $\pi_x$ their respective conjugate momenta.
By $\sum_{y\sim x}$ we denote the sum over all lattice sites $y$ adjacent to lattice site $x$, i.e.~the nearest neighbours of $x$, while the sum $\sum_x$ runs over all lattice sites $x$ in the volume $V$.
The spatial lattice spacing is denoted by $a$, which we set to $a=1$ and drop from equations and units from now on.
The parameter $d \in \left\{ 2,3 \right\}$ counts the spatial dimensions.
In the absence of an explicit symmetry breaking ($J=0$), the Hamiltonian in \cref{eq:z2_hamiltonian} is invariant under the $Z_2$ transformation ($\phi \to -\phi$).
However, for negative values of $m^2<0$, the symmetry is spontaneously broken for temperatures $T<T_c$, and restored above $T_c$ via a second order phase-transition in the $Z_2$ (Ising) universality class.
Here, we investigate the critical behaviour of this Ising-like model in both 2+1 and 3+1 dimensions. 
The numerical values of the mass and coupling constant can in principle be tuned to enhance the critical signals and minimize corrections to critical scaling \cite{hasenbusch_monte_1999}.
In order to assess the impact of the parameters on the relevant critical amplitudes, we have studied, in 2+1 dimensions, as set of couplings $\lambda \in \left\{ 1, 3.2, 10 \right\}$ for $m^2 = -1$ as well as $\lambda = -m^2 = 0.5$.
Because we did not observe any significant changes in the relevant amplitudes, however, we have fixed our model parameters simply to $m^2 = -1$ and $\lambda = 1$, together with $J=0$ if not explicitly stated otherwise.

In contrast to quantum field theory, one can compute time-dependent observables in classical field theory directly.
In particular, there is no fluctuating complex action inducing a sign problem, so one can obtain the real-time evolution by integrating the classical equations of motion of the system and averaging an ensemble over initial conditions.
Clearly, there are many ways to obtain a thermal set of initial conditions; in practice, we use a Hybrid Monte-Carlo method \cite{duane_hybrid_1987}.

We will consider both isolated systems and systems coupled to a heat bath.
Generally, to compute  a generic observable $O(\phi(t), \pi(t))$ of the phase space variables $\phi$ and $\pi$, we solve Langevin-type equations
\begin{equation}
    \partial_t \phi_x = \frac{\partial H}{\partial \pi_x}, \quad \partial_t \pi_x = -\frac{\partial H}{\partial \phi_x} - \gamma \, \pi_x + \sqrt{2\gamma T}\, \eta_x(t),
    \label{eq:Langevin_eom}
\end{equation}
where
\begin{equation}
    \frac{\partial H}{\partial \pi_x} = \pi_x, \quad \frac{\partial H}{\partial \phi_x} = - \sum_{y\sim x} \left( \phi_y - \phi_x \right) + \left( m^2 + \frac{\lambda}{6} \phi_x^2 \right)\phi_x + J,
    \label{eq:H_derivatives}
\end{equation}
and $\eta$ is a Gaussian white noise with zero mean and $\Braket{\eta_x(t)\eta_y(t')} = \delta_{xy}\delta(t-t')$. We subsequently evaluate the observable $O(\phi(t), \pi(t))$ on the classical trajectory, and finally perform an averaging over the classical-statistical ensemble.

We note that in \cref{eq:Langevin_eom} we have introduced an additional parameter $\gamma$ which characterizes the strength of the coupling to the heat bath.
Due to the fluctuation dissipation relation, the coefficient of the noise term is then fixed by $\gamma T$ and ensures a thermal Gaussian distribution of the conjugate momentum field variables $\pi_x$ at temperature $T$.
Evidently, for $\gamma \to 0$ one recovers the Hamiltonian equations of motion, such that the total energy of the system $H$ in \cref{eq:z2_hamiltonian} is explicitly conserved.
In this case, individual realizations of the system evolve micro-canonically and the temperature of the system only enters in the characterization of the initial ensemble.

Based on the structure of the equations of motion in \cref{eq:Langevin_eom} and \cref{eq:H_derivatives}, one concludes that our system belongs to the class of relaxational models (Models A-D) in the classification scheme of Hohenberg and Halperin \cite{hohenberg_theory_1977}.
Since our equations of motion do not conserve the order parameter, this is further limited to Model A if there are no conservation laws ($\gamma > 0$) or Model~C with conserved energy ($\gamma=0$).

\section{Spectral function}\label{sec:specfunc}
Spectral functions of bosonic operators $\hat{O}\left( t, \vec x \right)$ in a quantum field theory are defined by a decomposition of the two-point function (with translational invariance in space and time):
\begin{align}
    G_O(t-t', \vec x - \vec x') &= \Braket{\mathcal T \hat O(t, \vec x) \hat O^{\dagger}(t', \vec x')} \nonumber\\
    &= F(t-t', \vec x - \vec x') - \frac{\iu}{2} \rho(t-t', \vec x - \vec x') \sgn\left( t-t' \right),
\end{align}
where $\mathcal T$ denotes the time-ordering operator, and we have introduced the spectral ($\rho$) and statistical ($F$) correlation functions
\begin{align}
    \rho \left(t-t', \vec x - \vec x' \right) &= \iu \Braket{\left[ \hat O(t, \vec x), \hat O^{\dagger}(t', \vec x')\right]}, \label{eq:def_rho} \\
    F \left(t-t', \vec x - \vec x' \right) &= \frac{1}{2} \Braket{\left[ \hat O(t, \vec x), \hat O^{\dagger}(t', \vec x')\right]_+ } - \braket{\hat O(t, \vec x)}\braket{\hat O^{\dagger}(t', \vec x')}, \label{eq:def_F}
\end{align}
with $[\,,\,]_+$ representing the anti-commutator.
While the spectral function $\rho$ is relevant to describe the (linear) response of the system to an external perturbation, and thus provides the spectrum of possible excitations, the statistical correlation function $F$ in general describes the quantum and thermal statistical fluctuations present in the system.

Based on the Kubo-Martin-Schwinger (KMS) \cite{kubo_statistical-mechanical_1957,martin_theory_1959,parisi_statistical_1989} condition in thermal equilibrium, $F$ and $\rho$ are related via the fluctuation-dissipation relation.
\begin{align}
    F(\omega, \vec p, T) &= \left( n_T(\omega) + \inv 2 \right) \rho(\omega, \vec p, T) \label{eq:fluct-diss-appr}
\end{align}
Note that, for practical reasons, we defined the Fourier transformations of $F$ and $\rho$ as
\begin{align}
    F(\omega, \vec p, T) &= \int \intd t \intd^dx \, e^{\iu(\omega t - \vec p \vec x)} F(t, \vec x, T), \label{eq:FT_F}\\
    \rho(\omega, \vec p, T) &= -\iu \int \intd t \intd^dx \, e^{i(\omega t - \vec p \vec x)} \rho(t, \vec x, T), \label{eq:FT_rho}
\end{align}
where the additional factor of $-i$ in \eqref{eq:FT_rho} ensures that the spectral function $\rho(t)$, which is an odd function under space-time reflections $\rho(-t,-x) = -\rho(t, x)$, is real in both the time ($\rho(t)$) and frequency domain ($\rho(\omega)$).
In the classical limit, the thermal distribution becomes 
\begin{equation}
    n_T(\omega) + 1/2 = \frac{1}{e^{\beta\omega} + 1} + \frac{1}{2}\approx T/\omega, 
\end{equation}
and the spectral can be computed directly as
\begin{equation}
    \rho(t, \vec p, T) = -\frac{1}{T}\partial_t F(t, \vec p, T)
    \label{eq:rho_from_class_fdt}
\end{equation}

Close to the critical point, the dynamics of the system is dominated by its slow infrared modes with frequencies $\omega \ll T$ very small compared to the temperature scale where quantum effects are relevant.
Therefore, the classical description suffices to fully capture the dynamic critical behaviour of the model.

We are interested in the spectral function of the order parameter field $\phi$, which in the classical-statistical field theory is formally defined as
\begin{equation}
    \rho(t-t_0, \vec x- \vec x_0) = -\Braket{ \left\{ \phi(t,  \vec x), \phi(t_0,  \vec x_0) \right\} }_{\text{cl}}.
    \label{eq:def_rho_class}
\end{equation}
While it is in principle possible to evaluate the Poisson brackets directly (see e.g.~\cite{boguslavski_unraveling_2019,pineiro_orioli_breaking_2019}), we follow previous works and instead exploit the classical KMS condition to calculate the spectral function in thermal equilibrium \cite{aarts_spectral_2001,berges_dynamic_2010,schlichting_spectral_2019}.
By virtue of \cref{eq:rho_from_class_fdt}, we can compute this spectral function $\rho(t, \vec x, T)$ via the statistical two-point function $F(t, \vec x, T)$ directly from the classical lattice fields,
\begin{equation}
    \rho(t, \vec x, T) = -\frac{1}{2T}\Braket{\pi(t, \vec x) \phi(0, \vec 0) - \phi(t, \vec x) \pi(0, \vec 0)}.
    \label{eq:rho_from_F}
\end{equation}

\section{Static universality and scale setting}\label{sec:scale}
Before we address the real-time dynamics of the system, we briefly investigate the static critical behaviour of our model in 2 and 3 spatial dimensions.%
\footnote{Since we investigate a classical system, there is no additional Euclidean time direction.
Stated differently, from the perspective of Euclidean thermal field theory, the classical theory only contains the contribution from the zeroth Matsubara mode.}
Measuring the non-universal amplitudes will enable us to express our later results in terms of dimensionless quantities, for a more direct comparison as well as a plausibility check for our numerics setup.

\begin{table}
    \centering
    \caption{Ising critical exponents.
        In 2D, they are analytically known from Onsager's solution \cite{onsager_crystal_1944}.
    The high-precision calculations for the 3D Ising exponents were obtained from the conformal bootstrap approach \cite{kos_precision_2016,komargodski_random-bond_2017}.}
    \label{tab:exponents}
    \begin{tabular}{l r r}
        & \multicolumn{1}{c}{2D} & \multicolumn{1}{c}{3D} \\
        \hline\hline
        $\beta$ & 0.125 & 0.326419 \\
$\gamma$ & 1.75 & 1.237075 \\
$\delta$ & 15.0 & 4.78984 \\
$\nu$ & 1.0 & 0.629971 \\
$\eta$ & 0.25 & 0.036298 \\
$\omega$ & 2.0 & 0.82966 \\
     \end{tabular}
\end{table}

The basic observables we want to look at are the order parameter in form of the magnetization $M$ and the corresponding susceptibility $\chi$, which are defined as
\begin{align}
    M &= \frac{1}{V}\sum\limits_x \phi_x \label{eq:mag}, \\
    \chi &= \frac{\partial \Braket{M}}{\partial J} = \frac{V}{T} \left( \Braket{ M^2 } - \Braket{M}^2 \right).\label{eq:susc}
\end{align}
Since in the absence of explicit symmetry breaking ($J=0$) the magnetization defined in \cref{eq:mag} vanishes identically $\Braket{M} = 0$ in any finite volume, we instead consider
\begin{equation}
    M(J=0) = \frac{1}{V} \left|\sum\limits_x\phi_x\right| \label{eq:absmag}.
\end{equation}

When extracting the spatial correlation length $\xi$, we consider the plane-correlation function $\bar G(n)$ between the field-average $S(n)$ over lines for $d=2$ (respectively planes for $d=3$):
\begin{align}
    S(n) &\equiv L^{1-d} \sum\limits_{j(,k)}^N \phi_{n\vec e_x + j\vec e_y (+k\vec e_z)},  \\
    \bar G(n) &= L^{d-1} \left( \Braket{S(n)S(0)} - M^2 \right),
    \label{eq:slice_G}
\end{align}
which for sufficiently large separations $n$ is expected to follow and exponential behaviour of the form
\begin{equation}
    \bar G(n) = A \left( \exp(-n/\xi) + \exp\left( (n-L)/\xi  \right) \right).
    \label{eq:exp_ansatz_G}
\end{equation}
Based on this behaviour, we deduce an effective correlation length $\xi_{\text{eff.}}(n)$ from the data by considering the logarithmic derivative
\begin{equation}
    \xi_{\text{eff.}}(n) = \frac{-1}{\ln\left( \bar G(n+1)/\bar G(n) \right)}.
    \label{eq:xieff}
\end{equation}
and subsequently look for a plateau in a range of separations $n$, which is then used to determine the spatial correlation length $\xi$.\footnote{%
Note that we do not actually measure $S(n)$, but modes of the order parameter field with finite spatial momentum in one spatial direction $\vec p \equiv |p| \vec e_x$:
\begin{equation}
    \phi(\vec p) = \frac{1}{V}\sum\limits_x e^{-i \vec p \cdot \vec x} \phi_x.
    \label{eq:phi_p}
\end{equation}
However, one can trivially obtain $\bar G(n)$ by a Fourier transformation:
\begin{equation}
    \bar G(n) = \sum\limits_p \Braket{\phi(p\,\vec e_x)\phi(-p\,\vec e_x)} \cos(2\pi p n/L).
    \label{eq:Gn_from_Gp}
\end{equation}
}

\begin{figure*}
    \graphicspath{{\fdir}}
    \begin{minipage}[t]{.50\linewidth}
        \centering{$d=2$}
    \end{minipage}
    \begin{minipage}[t]{.50\linewidth}
        \centering{$d=3$}
    \end{minipage}
    \begin{minipage}[t]{.50\linewidth}
        \includegraphics{\fdir/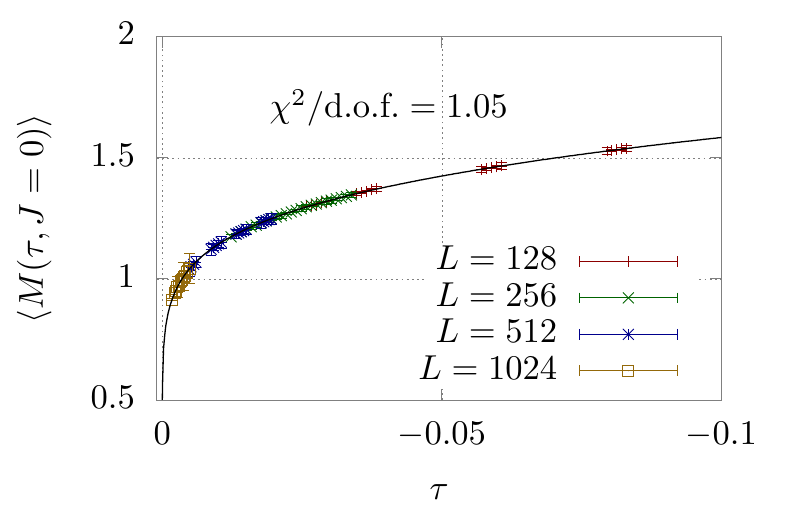}
    \end{minipage}
    \begin{minipage}[t]{.50\linewidth}
        \includegraphics{\fdir/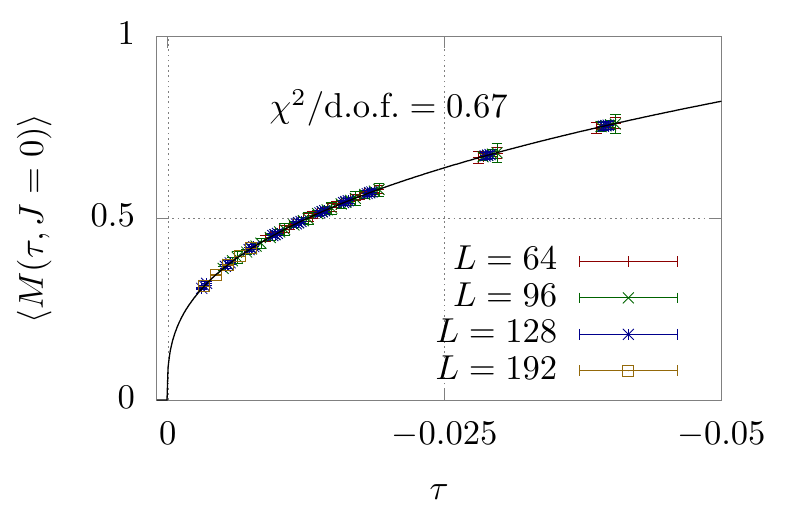}
    \end{minipage}
    \begin{minipage}[t]{.50\linewidth}
        \includegraphics{\fdir/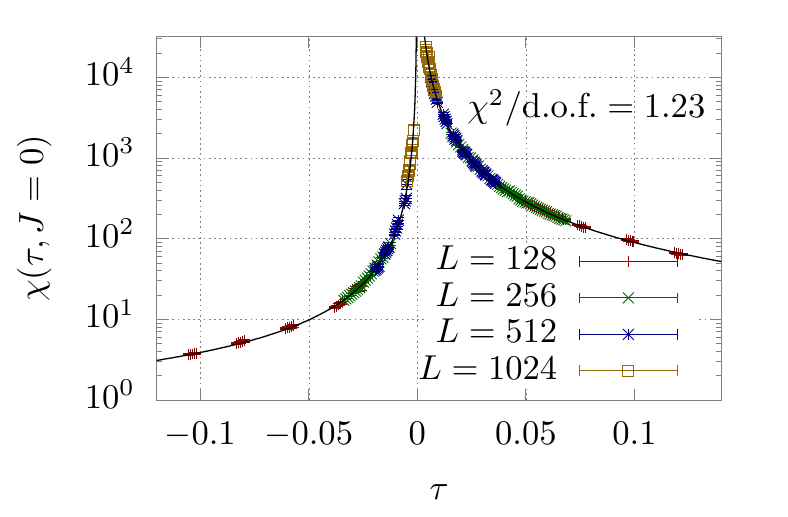}
    \end{minipage}
    \begin{minipage}[t]{.50\linewidth}
        \includegraphics{\fdir/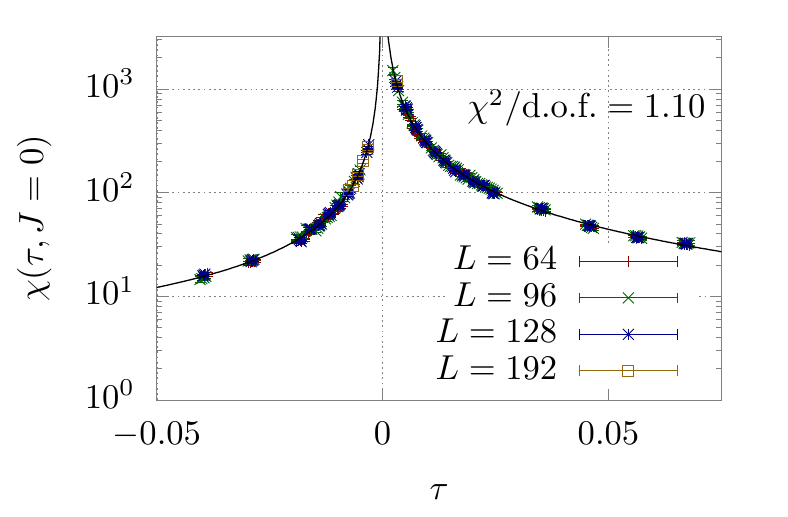}
    \end{minipage}
    \begin{minipage}[t]{.50\linewidth}
        \includegraphics{\fdir/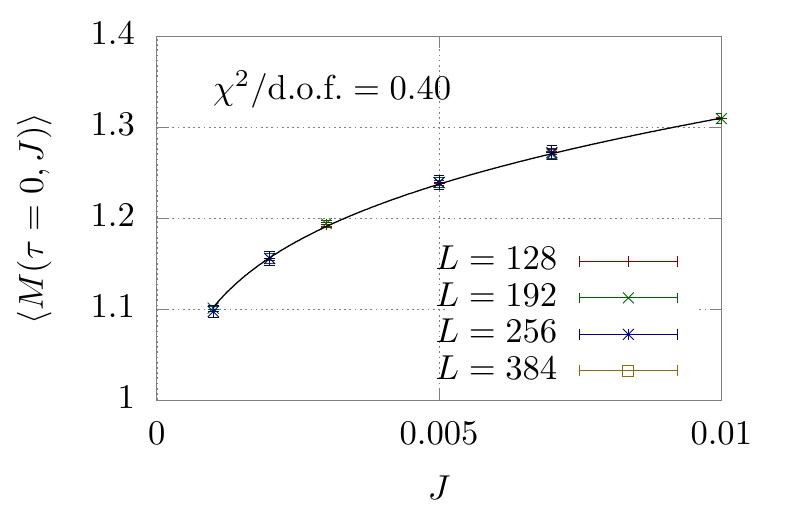}
    \end{minipage}
    \begin{minipage}[t]{.50\linewidth}
        \includegraphics{\fdir/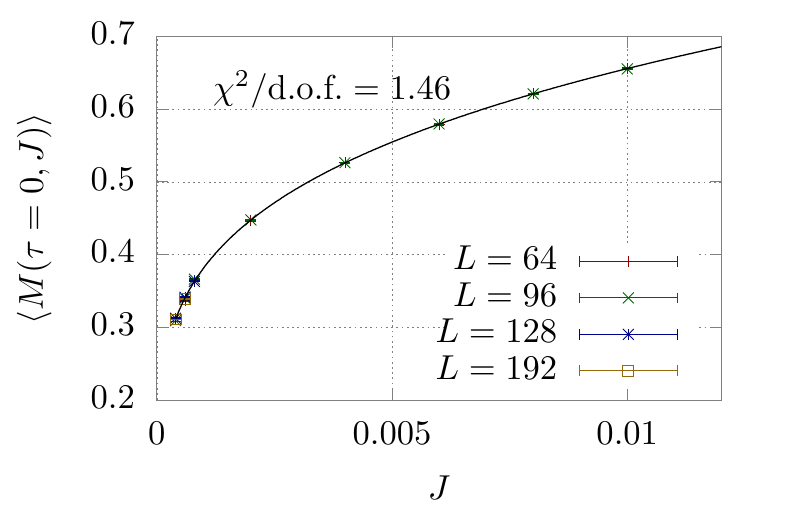}
    \end{minipage}
    \begin{minipage}[t]{.50\linewidth}
        \includegraphics{\fdir/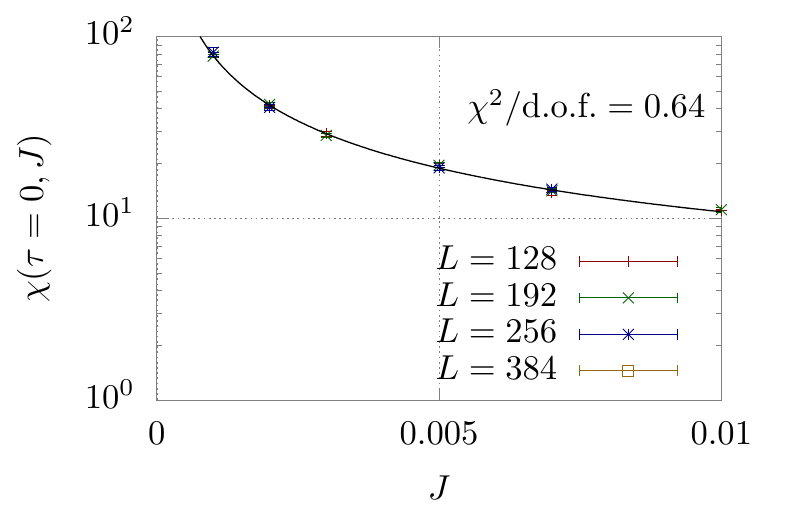}
    \end{minipage}
    \begin{minipage}[t]{.50\linewidth}
        \includegraphics{\fdir/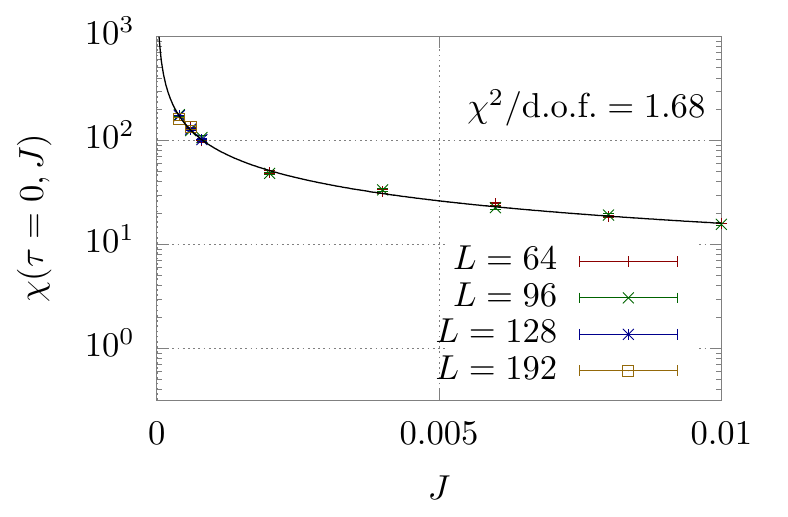}
    \end{minipage}
    \caption{Near-critical behaviour of the order parameter $M$ and susceptibilities $\chi$ in $d=2$ (left panels) and $d=3$ (right panels) spatial dimensions.
        Solid lines show fits to \cref{eqal:fit_ansatz} ff., with the corresponding $\chi^2$-values given in each figure.
    By taking into account the first two sub-leading corrections to scaling, we obtain excellent agreement between data and fits.}
    \label{fig:amplitudes}
\end{figure*}

\begin{figure*}
    \graphicspath{{\fdir}}
    \begin{minipage}[t]{.50\linewidth}
        \centering{$d=2$}
    \end{minipage}
    \begin{minipage}[t]{.50\linewidth}
        \centering{$d=3$}
    \end{minipage}
    \begin{minipage}[t]{.5\linewidth}
        \includegraphics{\fdir/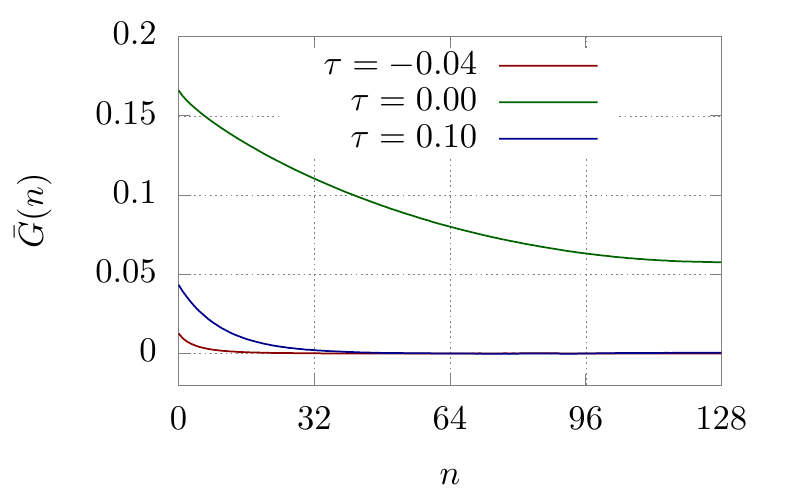}
    \end{minipage}
    \begin{minipage}[t]{.5\linewidth}
        \includegraphics{\fdir/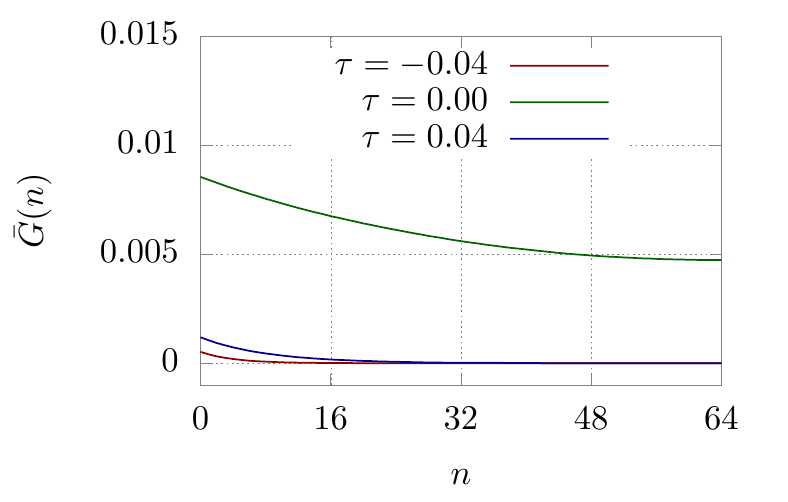}
    \end{minipage}
    \begin{minipage}[t]{.5\linewidth}
        \includegraphics{\fdir/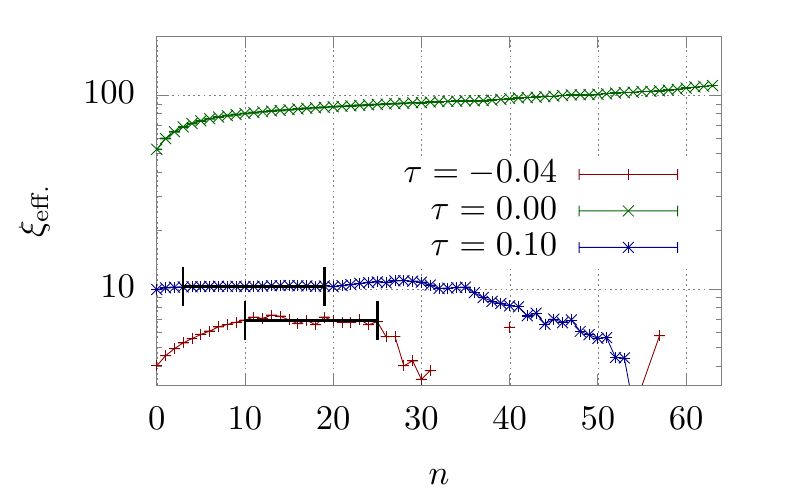}
    \end{minipage}
    \begin{minipage}[t]{.5\linewidth}
        \includegraphics{\fdir/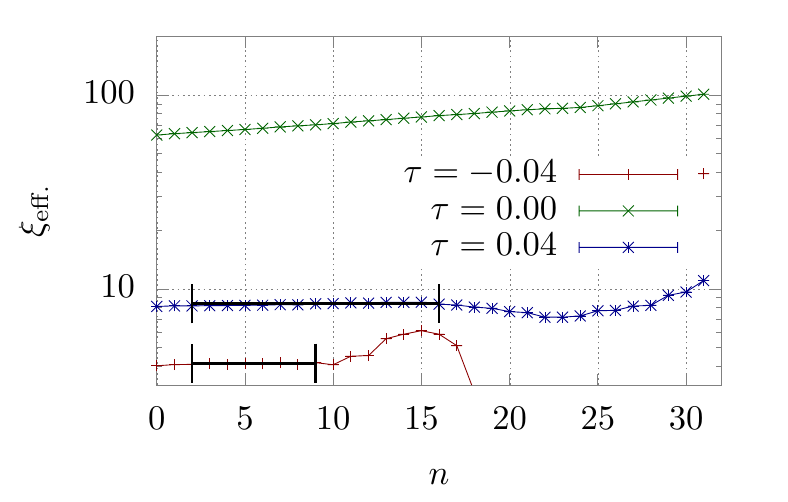}
    \end{minipage}
    \caption{Extraction of the correlation length $\xi(\tau, J)$ from the two-point function $\bar G(n)$.
        Upper panels show exemplary data for the slice correlator $\bar G(n)$ at different temperatures $\tau$; lower panels show the effective correlation length $\xi_\text{eff.}(x)$ defined by \cref{eq:xieff}.
        We extract the correlation length $\xi(\tau, J)$ by looking for a constant plateau in $\xi_{\text{eff.}}(x)$ and fitting a constant, if possible.
        Black lines with vertical marks in the lower plots indicate both range and result of the fit.}
\label{fig:ppfit}
\end{figure*}

\begin{figure*}
    \graphicspath{{\fdir}}
    \begin{minipage}[t]{.50\linewidth}
        \centering{$d=2$}
    \end{minipage}
    \begin{minipage}[t]{.50\linewidth}
        \centering{$d=3$}
    \end{minipage}
    \begin{minipage}[t]{.50\linewidth}
        \includegraphics{\fdir/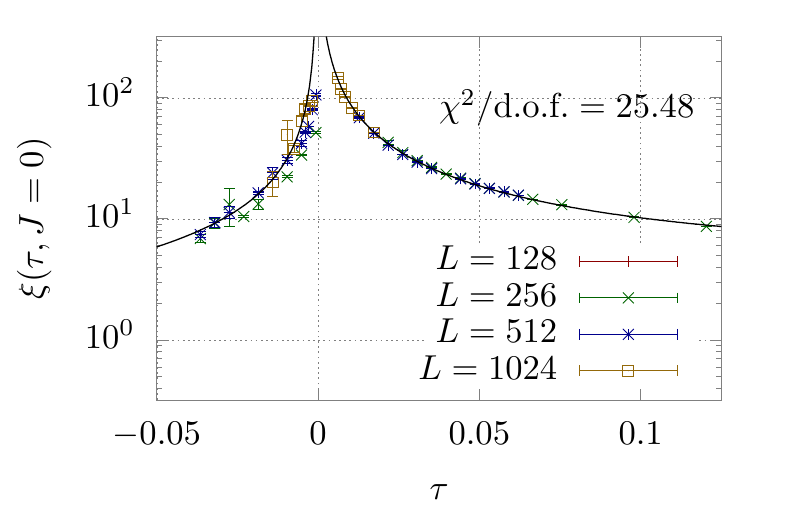}
    \end{minipage}
    \begin{minipage}[t]{.50\linewidth}
        \includegraphics{\fdir/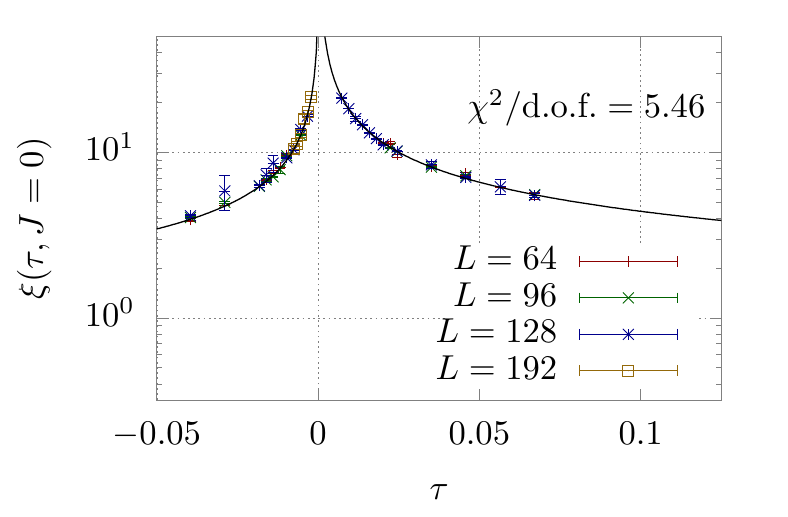}
    \end{minipage}
    \begin{minipage}[t]{.50\linewidth}
        \includegraphics{\fdir/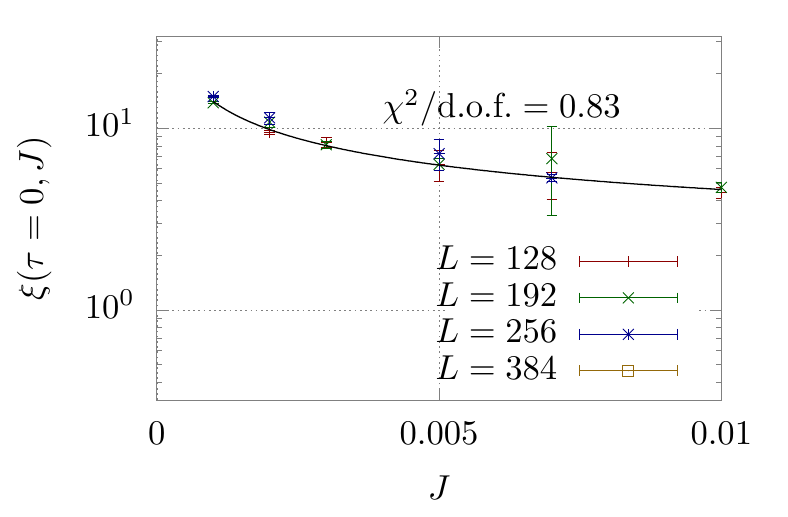}
    \end{minipage}
    \begin{minipage}[t]{.50\linewidth}
        \includegraphics{\fdir/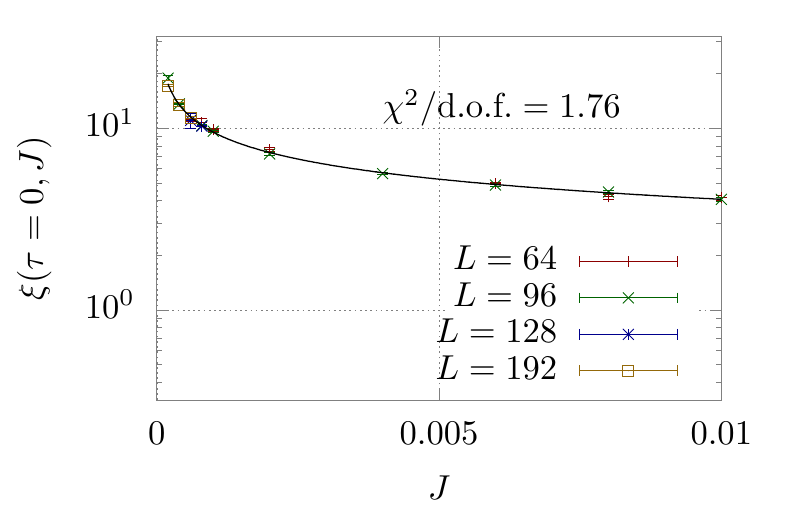}
    \end{minipage}
    \caption{Near-critical behaviour of the spatial correlation length $\xi(\tau, J)$ as function of $\tau$ at $J=0$ (top panels), respectively as function of $J$ at $\tau=0$ (lower panels).
        Solid lines show fits to \cref{eq:xi_ansatz} with the corresponding $\chi^2$-values indicated in each figure.
}
    \label{fig:clengths}
\end{figure*}

We provide a compact summary of our static results in \cref{fig:amplitudes,fig:ppfit,fig:clengths}, where we show the behaviour of the order parameter $M$ and susceptibilities $\chi$ (\cref{fig:amplitudes}), visualize the extraction of the spatial correlation length (\cref{fig:ppfit}), and show the behaviour of the correlation length $\xi$ as a function of temperature $T$ and external field $J$ (\cref{fig:clengths}).
For all of the aforementioned observables, we approach the thermodynamic limit:
By comparing data points of at least two different lattice volumes for any given point in the phase diagram, we can select a range of data where finite volume effects are negligible.
We follow common procedure and express the temperature dependence of our results in terms of the reduced temperature $\tau$
\begin{equation}
    \tau = \frac{T - T_c}{T_c},
    \label{eq:def_tau}
\end{equation}
where $T_c$ is the critical temperature given in \cref{tab:amp_corrections}.

\begin{table}
    \caption{Asymptotic amplitudes and corrections to scaling, as obtained from fits to \cref{eqal:fit_ansatz} ff.
    The given uncertainties only include statistical errors.
    If a correction amplitude does not improve the $\chi^2/\text{d.o.f.}$ of a fit, it is set to zero and denoted by a dash.}
    \label{tab:amp_corrections}
    \begin{minipage}[t]{.5\linewidth}
        \centering
        \begin{tabular}{l r r}
            & \multicolumn{1}{c}{2D} & \multicolumn{1}{c}{3D} \\
            \hline\hline
            $T_c$ & 4.4629(10) & 9.37074(28) \\
            $B$ & 2.0203(16) & 1.937(17) \\
            $C^+$ & 1.4059(63) & 0.830(42) \\
            $\xip$ & 0.9176(76) & 0.918(27) \\
            $B^c$ & 1.7425(13) & 1.5291(22) \\
            $C^c$ & 0.1222(16) & 0.3173(97) \\
            $\xic$ & 0.3489(95) & 0.537(16) \\
         \end{tabular}
    \end{minipage}
    \begin{minipage}[t]{.5\linewidth}
        \centering
        \begin{tabular}{l r r}
            & \multicolumn{1}{c}{2D} & \multicolumn{1}{c}{3D} \\
            \hline\hline
            $B_1$ & -1.1(2) & 0.9(1) \\
            $B_2$ & 0.56(3) & -1.0(3) \\
            $B_1^c$ & - & 0.83(1) \\
            $B_2^c$ & 20.4(8) & - \\
            $C_1^+$ & - & 2.8(8) \\
            $C_2^+$ & 1.33(8) & -5.2(1.9) \\
            $C_1^-$ & - & 2.6(2) \\
            $C_2^-$ & -9.7(3) & - \\
            $C_1^c$ & 302(48) & 2.2(4) \\
            $\xip_1$ & 11.6(1.9) & 0.4(2) \\
            $\xim_1$ & - & 0.5(3) \\
            $\xic_1$ & - & 1.5(6) \\
         \end{tabular}
    \end{minipage}
\end{table}

Based on our results in \cref{fig:amplitudes,fig:clengths}, we extract the non-universal amplitudes of the critical power laws either as functions of $\tau$ at vanishing external field, $J=0$, or as functions of a dimensionless $\bar J\equiv J/J_0\ge 0$ at $\tau = 0$.
We follow the notation of \cite{pelissetto_critical_2002} and distinguish the various non-universal amplitudes by superscripts, marking the region of the respective phase diagram that they describe.
Specifically, $X^{\pm}$ indicates the sign of $\tau$ and marks the amplitude on the $J=0$ axis, whereas $X^c$ denotes the corresponding amplitude on the $\tau=0$ axis.
Motivated by \cite{pelissetto_critical_2002} and \cite{engels_numerical_2003}, we extract the critical temperature and non-universal amplitudes, including the leading scaling corrections, by fitting the following ans\"atze to correspondingly selected points.
\begin{align}
    \Braket{M(\tau)} &= B (-\tau)^{\beta} \left( 1 + B_1 (-\tau)^{\omega \nu} - B_2 \tau \right), \;\; \tau < 0\, , \label{eqal:fit_ansatz} \\
    \Braket{M(\bar J)} &= B^c J_0^{1/\delta} \,  \bar J^{1/\delta} \left( 1 + B_1^c \bar J^{\omega \nu_c} + B_2^c \bar J \right), \\
    \chi(\tau) &= C^{\pm} |\tau|^{-\gamma} \left( 1 + C^{\pm}_1 |\tau|^{\omega \nu} + C_2^{\pm} |\tau|\right), \\
    \chi(\bar J) &= C^c J_0^{-\gamma_c} \, \bar J^{-\gamma_c} \left( 1 + C^c_1 \bar J^{\omega \nu_c}\right), \\
    \xi(\tau) &= \xipm |\tau|^{-\nu} \left( 1 + \xipm_1 |\tau|^{\omega \nu}\right) \label{eq:xi_ansatz},\\
    \xi(\bar J) &= f^cJ_0^{-\nu_c} \,  \bar J^{-\nu_c} \left( 1 + f_1^c \bar J^{\omega \nu_c}\right),
\end{align}
where $\gamma_c \equiv \gamma/\beta\delta = 1 - 1/\delta $ and $\nu_c \equiv \nu/\beta\delta = (1+ 1/\delta)/d$ from scaling and hyperscaling relations.
We note that $J_0$ is chosen such that $ B^c J_0^{1/\delta} = B$, and $C^c J_0^{-\gamma_c}= B/(\delta J_0)$ (as well as $J_0^{-\nu_c} = (B^c/B)^{\nu/\beta} $); and in the absence of scaling corrections, the magnetic equation of state can be written in the Widom-Griffiths form, $y = f(x)$, with dimensionless magnetization $\widebar M = M/B$,
\begin{equation}
  y = \bar J /{\widebar M}^{\delta},  
  \;\; x = \tau  /{\widebar M}^{1/\beta} \, \;\;\; \mbox{and normalizations} \;\; f(0) = 1\,, \;\; f(-1) = 0 \,.
\end{equation}

By virtue of the $ Z_2$ symmetry of our model, we know that its critical behaviour is described within the Ising universality class.
Our model and method are not optimized for computing static critical exponents; we therefore reduce the degrees of freedom in our calculations by taking the critical exponents from the literature.
In $d=2$ spatial dimensions, the critical exponents are exactly known from the analytic solution by Onsager \cite{onsager_crystal_1944}.
For the numeric values of the $d=3$ exponents, we take those obtained by the conformal bootstrapping method \cite{kos_precision_2016,komargodski_random-bond_2017}, which are listed in \cref{tab:exponents} alongside their $d=2$ counterparts.

The asymptotic scaling amplitudes along with the coefficients of the scaling corrections extracted from the fits to the data are listed in \cref{tab:amp_corrections}. The comparison between ansatz and data is shown in \cref{fig:amplitudes,fig:clengths}.
For most quantities, we obtain a $\chi^2/\text{d.o.f.}$ close to unity, indicating sufficient agreement between fit and data.
For $d=2$, our procedure for the correlation length at $J=0$ does not work too well, especially for $\tau < 0$, leading to a relatively large resulting value of $\chi^2$. 
This is primarily due to heavily fluctuating data in the low-temperature phase, as well as probably an underestimation of the error of the correlation length $\xi$ in our extraction.

It is a useful cross-check to validate the measured amplitudes by comparing the known universal amplitude ratios.
We use a subset that relates all of the amplitudes we measure:
\begin{align}
    U_2 &= C^+/C^-, \\ 
    U_{\xi} &= f^+/f^-, \\ 
    R_{\chi} &= C^+ B^{\delta-1}/\left( B^c \right)^{\delta}, \\
    Q_2 &= \left( f^c/f^+ \right)^{2-\eta} C^+/C^-,\\
    \delta &= B^c/C^c.
\end{align}
The ratios we measured are shown alongside the amplitudes in \cref{tab:amplitude_ratios}.
For most of the ratios, we find results in the right ballpark, with exception of $U_{\xi}$ and $Q_2$ in $d=2$ dimensions:
Those ratios are the ones including amplitudes of $\xi$.

Different definitions of $\xi$ lead to slightly different $U_{\xi}$, see e.g.~the difference between second-moment correlation length and inverse mass gap in \cite{pelissetto_critical_2002}.
Since our definition again differs from the aforementioned ones, we are bound to get (slightly) different amplitudes.
Nevertheless, it is strongly related to the definition by the inverse mass gap, so we would expect $U_{\xi} \approx 2$.
Since we do not rely on exact estimates of the correlation length for the purpose of this study, we accept these residual ambiguities for now.

When studying dynamic critical behaviour, we will normalize our results using the non-universal amplitudes.
In this process, we will also obtain a timescale by fitting the amplitude $\xitp$ of the correlation time $\xi_t$, which will be determined in a later section  (see \cref{eq:xi_power_law} in Sec.~\ref{sec:xi_t}).
If not stated otherwise, we give all results in terms of dimensionless scaling variables, which are indicated by a bar and constructed as follows:
\begin{align}
    &\bar J = J/J_0 = J (B^c/B)^{\delta}, \\ 
    &\bar p = p \xip, \\
    &\bar t = t/\xitp, \\
    &\bar \omega = \omega \xitp.
    \label{eqal:scalingvars}
\end{align}

\begin{table*}
    \centering
    \caption{Universal amplitude ratios.
        The given uncertainties only include statistical errors.
    The ratios are shown alongside their literature values,  which were taken from \cite{pelissetto_critical_2002}.}
    \label{tab:amplitude_ratios}
    \begin{tabular}{l r r r r}
        & \multicolumn{2}{c}{2D} & \multicolumn{2}{c}{3D} \\
        & meas. & lit. & meas. & lit. \\
        \hline\hline
        $U_2$ & 40.05(35) & 37.69 & 4.29(24) & 4.76(2) \\
$U_{\xi}$ & 3.14(38) & 2 & 1.939(90) & 1.896(10) \\
$R_{\chi}$ & 6.40(10) & 6.778 & 1.331(80) & 1.660(4) \\
$Q_2$ & 2.12(11) & 2.836 & 0.913(93) & 1.195(10) \\
$\delta$ & 14.26(19) & 15 & 4.82(15) & 4.78984(1) \\
     \end{tabular}
\end{table*}

\section{Non-critical spectral functions}\label{sec:noncrit}
\begin{figure}
    \graphicspath{{\fdir}}
    \begin{minipage}[t]{.50\linewidth}
        \centering{$d=2$}
    \end{minipage}
    \begin{minipage}[t]{.50\linewidth}
        \centering{$d=3$}
    \end{minipage}
    \begin{minipage}[t]{.5\linewidth}
        \includegraphics{\fdir/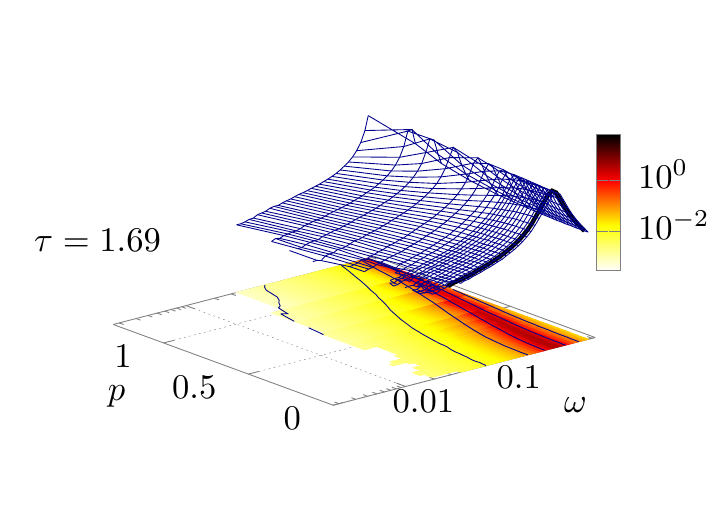}
    \end{minipage}
    \begin{minipage}[t]{.5\linewidth}
        \includegraphics{\fdir/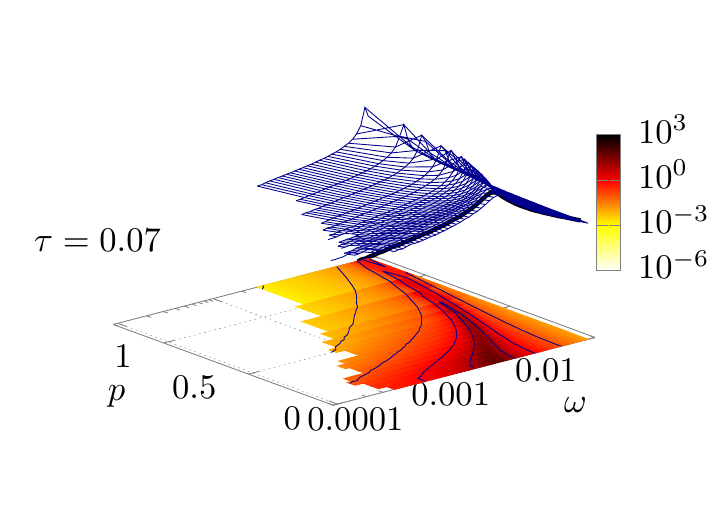}
    \end{minipage}
    \begin{minipage}[t]{.5\linewidth}
        \includegraphics{\fdir/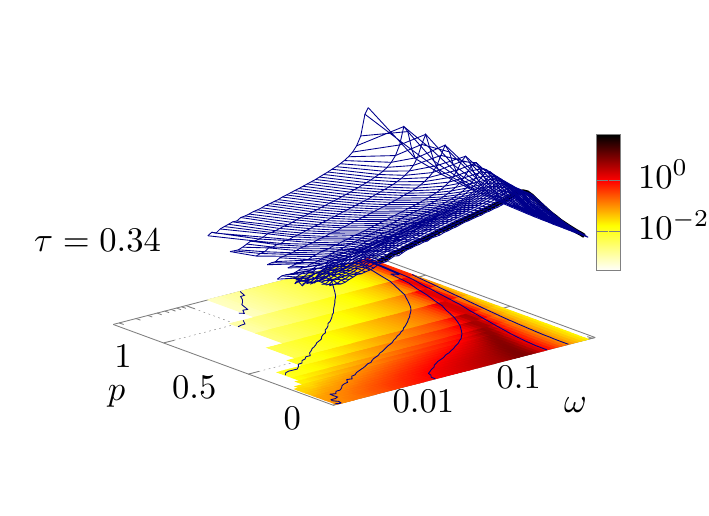}
    \end{minipage}
    \begin{minipage}[t]{.5\linewidth}
        \includegraphics{\fdir/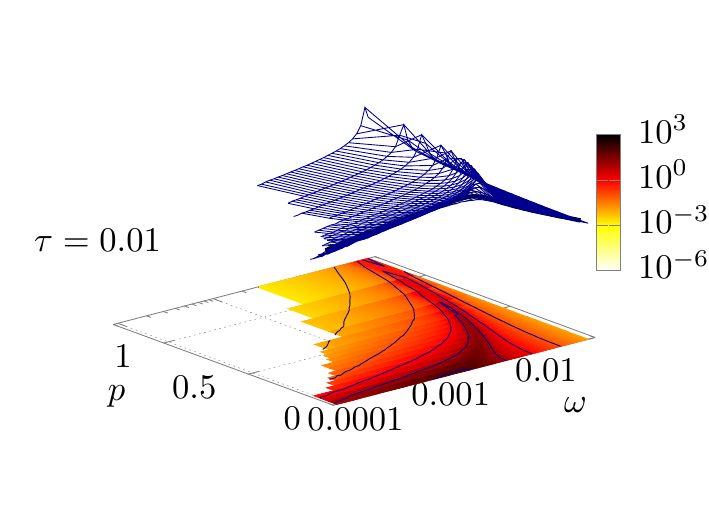}
    \end{minipage}
    \begin{minipage}[t]{.5\linewidth}
        \includegraphics{\fdir/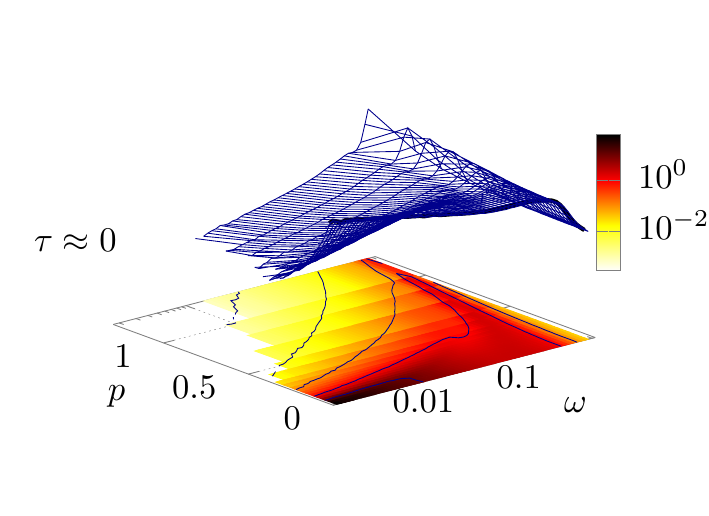}
    \end{minipage}
    \begin{minipage}[t]{.5\linewidth}
        \includegraphics{\fdir/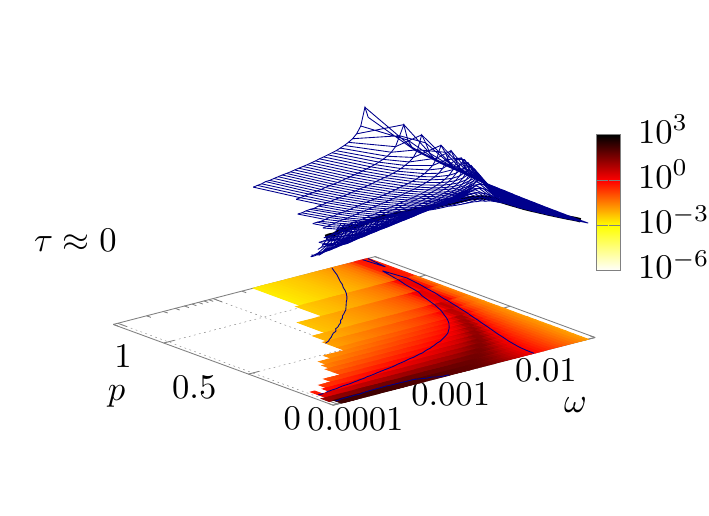}
    \end{minipage}
    \begin{minipage}[t]{.5\linewidth}
        \includegraphics{\fdir/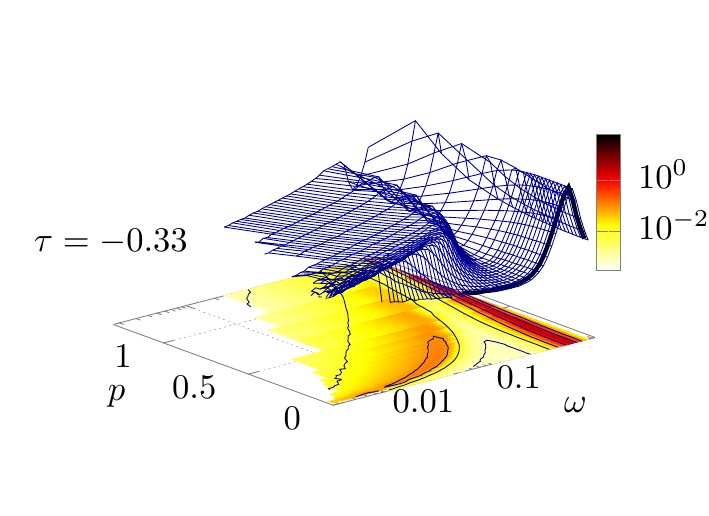}
    \end{minipage}
    \begin{minipage}[t]{.5\linewidth}
        \includegraphics{\fdir/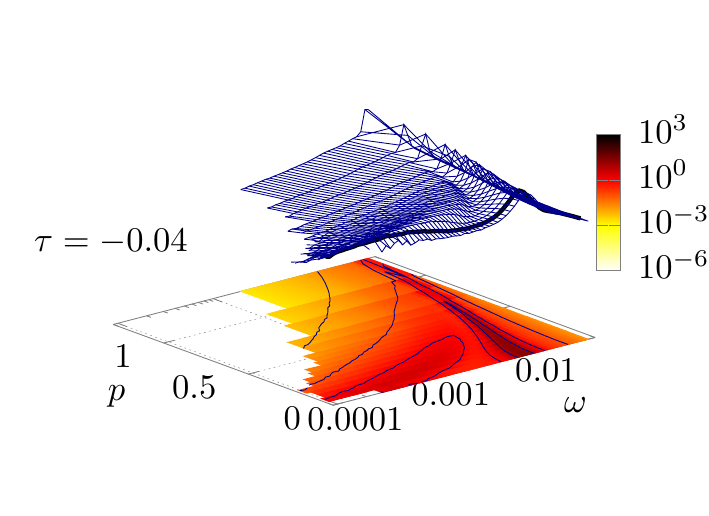}
    \end{minipage}
    \caption{Overview of the behaviour of the spectral function $\rho(\omega,p)$ for Hamiltonian dynamics (Model C, $\gamma = 0$) at different points in the phase diagram, above $T_c$ (top panels), near $T_c$ (central panels), and below $T_c$ (bottom panels) in 2+1 (left panels) and 3+1 (right panels) dimensions.
        Heat maps at the bottom of each panel visualize support and spectral strength in the $(p,\omega)$ plane.
        The frequency ($\omega$) and spectral function ($\rho$) axes are scaled logarithmically, the momentum ($p$) axis linearly.
        The spectral functions $\rho(\omega,p)$ for the smallest non-zero momentum mode are highlighted by a black solid line on the 3D surface.
        Spectral functions in the symmetric phase $T\gg T_c$ are dominated by a quasi-particle peak with relativistic dispersion relation.
        Near the critical point $T \approx T_c$, a strongly infrared-enhanced mode emerges in addition to the quasi-particle peak that persists at higher momenta.
        Below the critical temperature $T \ll T_c$, a soft second mode at low frequencies emerges for finite momenta.}
    \label{fig:sf-overview}
\end{figure}

Based on our analysis of static critical phenomena, we will now investigate the behaviour of the spectral function at different points in the phase diagram.
We proceed as outlined in \cref{sec:intro} and calculate the classical-statistical spectral function of the order parameter field using \cref{eq:rho_from_class_fdt} resp.~\cref{eq:rho_from_F}.
We prepare $\sim 30$ independent thermal configurations as initial conditions, which we then evolve for $\sim 10^4 - 10^5a$, using and Euler-Maruyama scheme.
Notably, we find that in the vicinity of the critical point, the time step $\Delta t$ in the integrator has to be chosen sufficiently small to avoid discretization errors.
If not stated otherwise, we employ $\Delta t  = .00625\, a$, and we have checked that discretization errors are negligible for the results presented here.

By recording the evolution of the Fourier modes of the order parameter field, we subsequently compute the spectral function $\rho(t,p)$ in \cref{eq:rho_from_class_fdt}.
Statistical errors are estimates form point-wise averages of $\rho(t,p)$ and respectively $\rho(\omega, p)$ over different configurations.
If not stated otherwise, the spectral functions shown in this section are obtained for vanishing external field $J=0$ on lattice volumes of $256^2$ respectively $128^3$, where we have the smallest remaining finite volume effects.

In \cref{fig:sf-overview}, we give an overview over the behaviour of spectral functions at different temperatures.
At high temperatures ($T \gg T_c$), the spectral function is well approximated by a relativistic Breit-Wigner peak shape, with the peak position shifting with $p$ according to the relativistic dispersion relation $\omega^2 - p^2 = m^2(T)$.
In \cite{aarts_spectral_2001}, the zero-momentum ($p=0$) mode of the spectral function was investigated in a scalar theory without phase transition. 
The study was done at high temperatures, so that $T \gg \omega$.
It was found that both, masses and widths of the measured spectral function agree well with the analytical predictions from resummed two-loop perturbation theory. We find here a very similar Breit-Wigner shape, and since we also calculate the spectral function at finite spatial momenta, we can check for the correct dispersion relation of the quasi-particle mass.

When the system approaches criticality from above, the effective mass $m^2(T)$ decreases, and the quasi-particle peak becomes less and less pronounced.
Close to the critical point, an infrared power law behaviour builds up at small frequencies and momenta which, as we will discuss shortly in \cref{sec:critdyn}, encodes the dynamic critical behaviour of the spectral function.\footnote{The slope of this power law is related to the dynamic critical exponent $z$, which is the main subject of the next section.}
With increasing spatial momentum $p$, the cut-off imposed by $p$ suppresses the critical contribution.
Specifically, at high momenta the spectral function retains its Breit-Wigner shape even in this near-critical regime, however with significantly smaller quasi-particle mass.
By comparing the different columns of \cref{fig:sf-overview} one further notices that the window of reduced temperatures $\tau$, where a critical enhancement of the spectral function can be observed, is much smaller in 3+1 than in 2+1 dimensions, as can be expected from the larger influence of infrared fluctuations in lower dimensions.

Below the critical temperature, in the ordered phase, the effective quasi-particle mass $m^2(T)$ increases again, and gradually approaches its mean-field value of $\sqrt{-2m^2}$ in the limit $T\to0$, where all thermal fluctuations are suppressed.
However, in addition to the quasi-particle peak, a second low-frequency excitation arises for finite spatial momenta $p$, with a different spectral shape and a dispersion relation indicative of soft collective excitations such as thermally driven capillary waves, see Appendix \ref{sec:capwav}.
While in 3+1 dimensions the contributions from this soft mode only carry a small fraction of the spectral weight, it is more pronounced in 2+1 dimensions, where by looking at the lower left panel of \cref{fig:sf-overview}, one can easily discern the valley in the spectral function that separates it from the quasi-particle peak.

By comparing the spectral functions in the symmetry-broken phase at $\tau <0$, and the vicinity of the critical point $\tau \approx 0$ in \cref{fig:sf-overview}, one is led to speculate that it is this second excitation at $\tau<0$ that might turn into the critical IR divergence close to the critical point.
Indeed, as we will demonstrate shortly, by tracing the maxima of both excitations, one observes an avoided-crossing behaviour near $T_c$, before the low-frequency mode eventually disappears at $\tau>0$. A similar avoided crossing was already observed in the classical-statistical lattice simulations of O(4)-model spectral functions in 3+1 dimesions in \cite{schlichting_spectral_2019}. Although this study was restricted to vanishing spatial momenta where in agreement with the results presented here, there is no comparable soft mode in the longitudinal $\sigma$-spectral function of the order parameter fluctuations, an avoided crossing does show up at zero momentum in the transverse $\pi$-spectral function of the O(4)-model, see Ref.~\cite{schlichting_spectral_2019} for details.  

\subsection{Dispersion relation of quasi-particle peak}
Based on the results in \cref{fig:sf-overview}, we will now investigate some properties of the spectral functions extracted from lattice data.
We start by confirming the relativistic dispersion relation of the quasi-particle peak, study the temperature dependence of the peak parameters and analyse the effect of the Langevin damping $\gamma$ on the spectral function.

In order to analyze the dispersion relation, we fit the peaks in the spectral function with a relativistic Breit-Wigner ansatz
\begin{equation}
    \rho_{\text{BW}}(\omega, p) = \frac{ \Gamma(p) \omega}{(\omega^2 - \omega_c^2(p))^2 + \omega^2 \Gamma^2(p)},
    \label{eq:breit-wigner}
\end{equation}
where the central frequency of the effective mass resonance is expected to follow the relativistic energy-momentum relation
\begin{equation}
    \omega_c^2(p) = m_\mathrm{eff}^2 + p^2.
    \label{eq:dispersion}
\end{equation}

\begin{figure}
    \graphicspath{{\fdir}}
    \begin{minipage}[t]{.5\linewidth}
        \includegraphics{\fdir/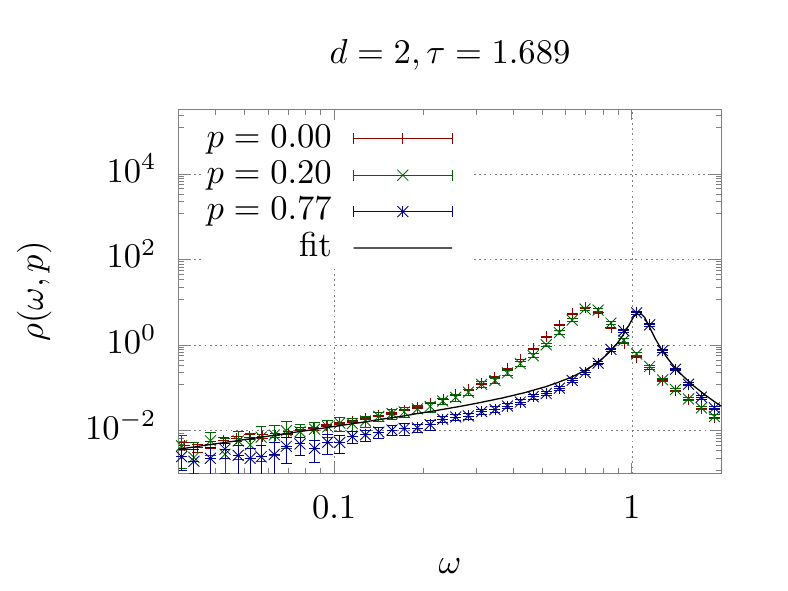}
    \end{minipage}
    \begin{minipage}[t]{.5\linewidth}
        \includegraphics{\fdir/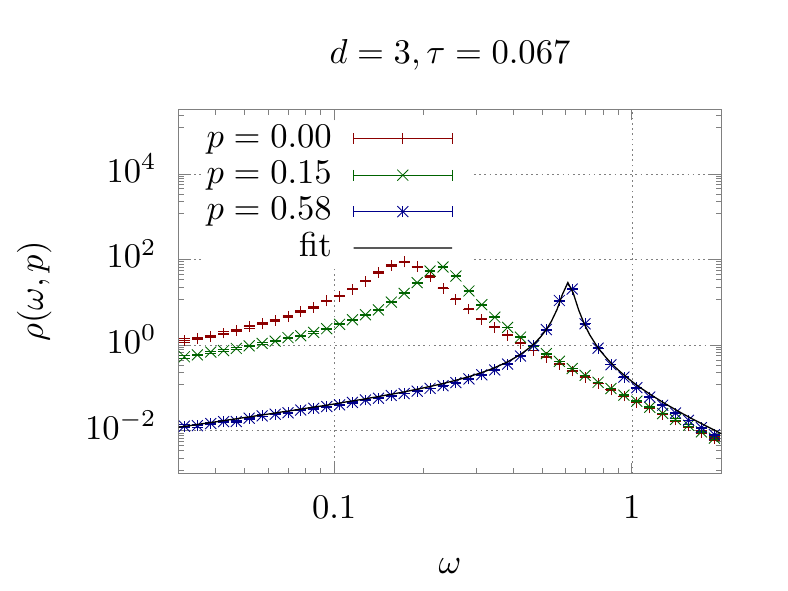}
    \end{minipage}
    \begin{minipage}[t]{.5\linewidth}
        \includegraphics{\fdir/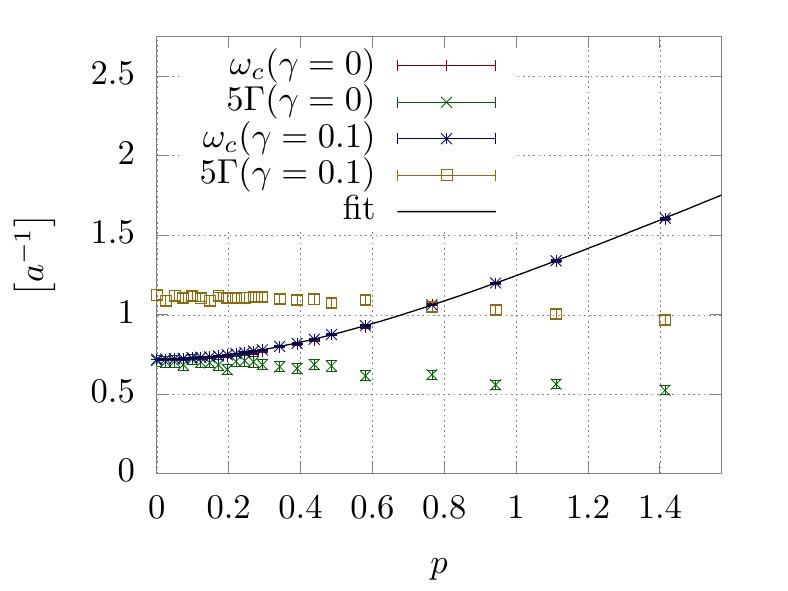}
    \end{minipage}
    \begin{minipage}[t]{.5\linewidth}
        \includegraphics{\fdir/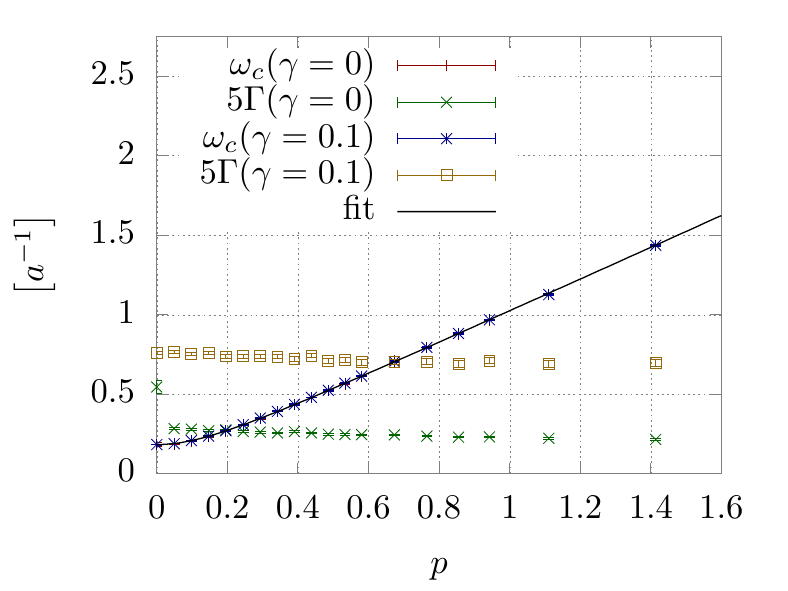}
    \end{minipage}
    \caption{Dispersion relation and damping rate in the symmetric phase.
        Upper panels show the spectral functions in lattice units for vanishing Langevin coupling $\gamma=0$ with an exemplary Breit-Wigner fit as solid line; lower panels show the resulting fit parameters in lattice units over the different spatial momenta, with a fit to a relativistic dispersion relation \cref{eq:dispersion}.
        Note that the width $\Gamma$ is scaled by a factor of 5 to improve legibility.}
    \label{fig:res-ht}
\end{figure}

\begin{figure}
    \graphicspath{{\fdir}}
    \begin{minipage}[t]{.5\linewidth}
        \includegraphics{\fdir/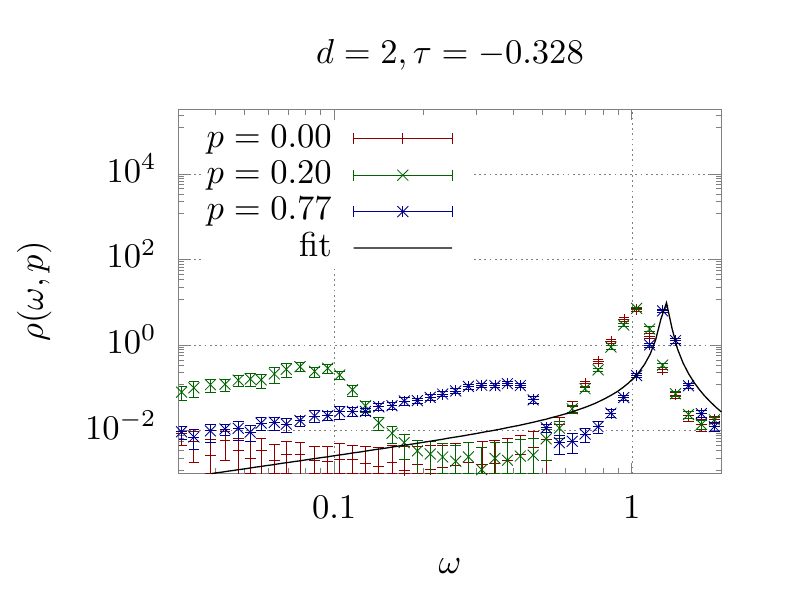}
    \end{minipage}
    \begin{minipage}[t]{.5\linewidth}
        \includegraphics{\fdir/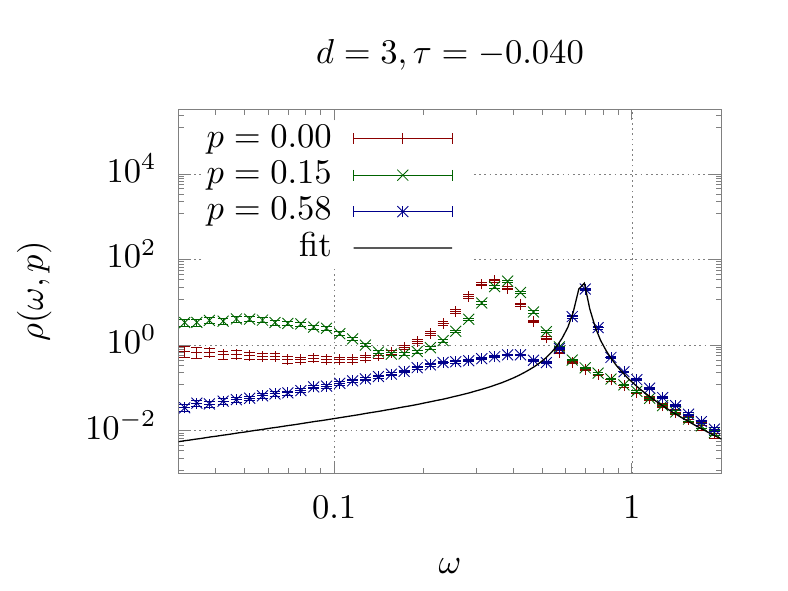}
    \end{minipage}
    \begin{minipage}[t]{.5\linewidth}
        \includegraphics{\fdir/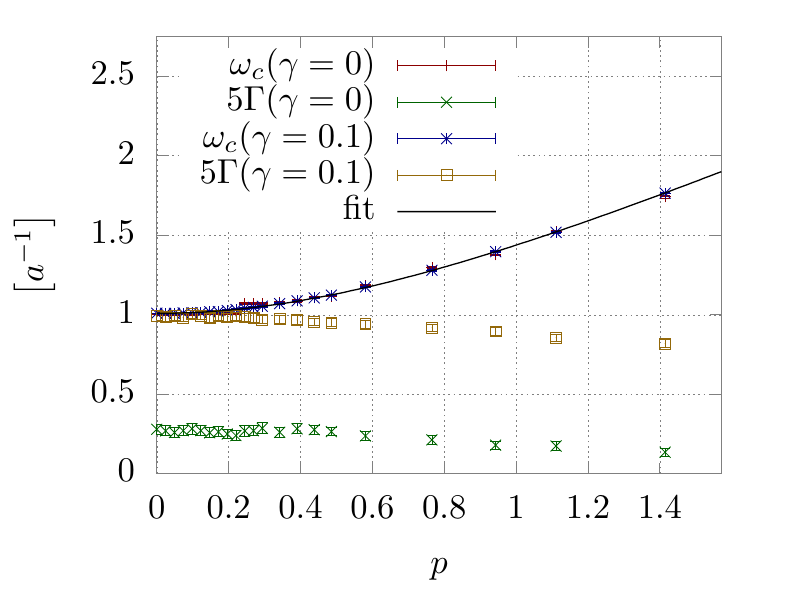}
    \end{minipage}
    \begin{minipage}[t]{.5\linewidth}
        \includegraphics{\fdir/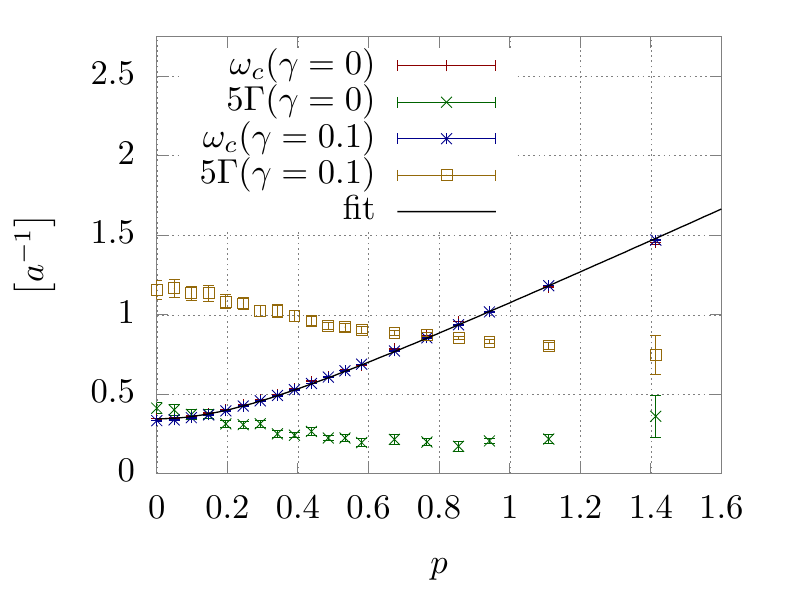}
    \end{minipage}
    \caption{Dispersion relation and damping rate in the ordered phase, illustrated analogously to \cref{fig:res-ht}.
        Due to the presence of the second excitation at lower $\omega$, the Breit-Wigner fit does no longer describe the spectral function data as well as in the symmetric phase.
    Hence, the fit window for the Breit-Wigner fit was narrowed to the high-frequency regime.}
    \label{fig:res-lt}
\end{figure}

Visualizations of the fits to the spectral function are given in the top panels of \cref{fig:res-ht} for the symmetric phase ($\tau>0$) and in \cref{fig:res-lt} for the symmetry broken phase ($\tau<0$), while the bottom panels of \cref{fig:res-ht,fig:res-lt} show the extracted values of the peak-frequency $\omega_c(p)$ and damping rates $\Gamma(p)$.
Clearly, the fit to the Breit-Wigner function describes the spectral function very well at high temperatures, where the quasi-particle excitation is the only discernible structure in spectral functions in \cref{fig:res-ht}.
While in the low temperature phase, shown in \cref{fig:res-lt}, the Breit-Wigner fit still describes the quasi-particle peak very well, one also clearly observes the additional low-energy excitation for finite spatial momenta ($p>0$).%

So far we have focused on the general behaviour of the spectral function for Hamiltonian dynamics (Model C), which we will now compare to the spectral functions for Langevin dynamics (Model A).
Since the key difference between Hamiltonian and Langevin dynamics lies in the introduction of an additional frequency-independent damping and noise coupling to the heat bath, one naturally expects the additional damping to contribute to the resonance decay width $\Gamma(p)$ of the (non-critical) spectral functions.
Explicit comparisons of the results for Hamiltonian ($\gamma=0$) and Langevin ($\gamma=0.1$) dynamics, shown in the bottom panels of \cref{fig:res-ht,fig:res-lt}, confirm this expectation, indicating further that differences in the numerically extracted damping rates $\Gamma_{\gamma=0.1}(p) - \Gamma_{\gamma=0}(p) \simeq \gamma$ are in fact close to the Langevin coupling $\gamma$.
While away from criticality, the qualitative and quantitative features of the spectral functions are very similar between conservative and weakly dissipative systems, the change in conservation laws does of course affect the dynamic critical behaviour, as we will discuss in more detail in \cref{sec:critdyn}.

\begin{figure}
    \graphicspath{{\fdir}}
    \begin{minipage}[t]{.5\linewidth}
        \includegraphics{\fdir/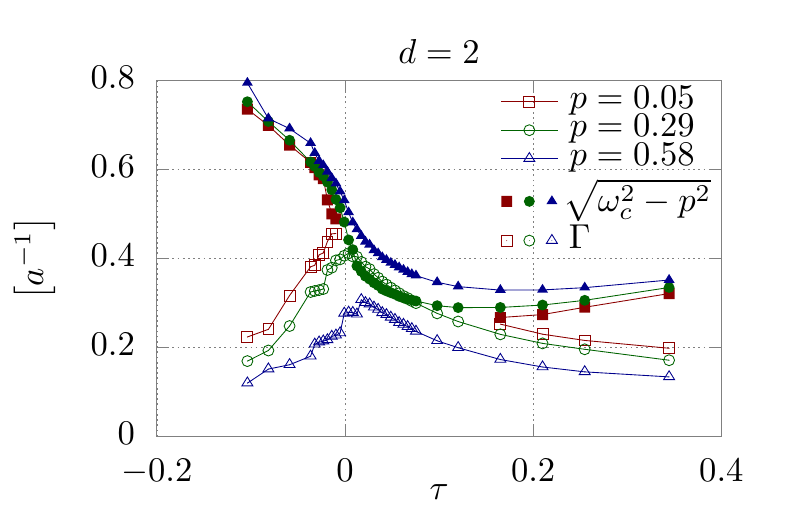}
    \end{minipage}
    \begin{minipage}[t]{.5\linewidth}
    \includegraphics{\fdir/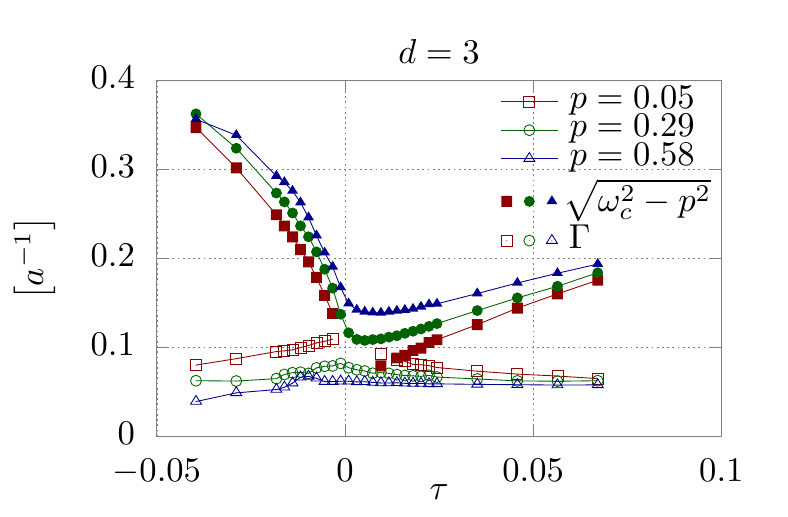} \end{minipage}
    \caption{Effective quasi-particle masses and damping rates as a function of reduced temperature.
        Near the critical point ($\tau \approx 0$) the effective mass drops and the damping rate increases; while at low momentum $p$ the spectral function is no longer described by a simple quasi-particle peak, the behaviour of high momentum modes is remarkably smooth across the phase transition.
    }
    \label{fig:phplots}
\end{figure}

We conclude our discussion of the non-critical spectral functions by investigating the temperature dependence of the central frequencies and decay widths of the quasi-particle peak, which are summarized in \cref{fig:phplots}.
By defining $m_{\text{eff}} = \sqrt{\omega_c(p)-p^2}$, curves for different momenta $p$ should coincide whenever the quasi-particle peaks satisfy the dispersion relation in \cref{eq:dispersion}.
Starting from the low temperature phase, the resonances turn into sharp $\delta$-peaks in the limit $T\to0$, where $m_{\text{eff}}(T=0) = \sqrt{-2m^2}$ assumes its mean-field value and the decay width $\Gamma(T=0) = 0$ vanishes.

Even at the lowest temperatures measured, the effective masses are still rather large but then decrease as the system approaches the critical temperature, while the decay width $\Gamma(T)$ increases simultaneously.
Near the critical temperature, the behavior of the spectral function at low momentum can no longer be described by a simple quasi-particle structure, as already seen in \cref{fig:sf-overview}.
However, when increasing the temperature above the critical point, the quasi particle structure at low momentum is restored, with the mass increasing monotonously as a function of temperature for $T>T_c$.
Conversely, for the modes with large spatial momenta, this process is continuous.
While the effective mass reaches a finite minimum at $\tau=0$, the spectral function retains its Breit-Wigner form across the transition.

In summary we find that, in the non-critical regime, the measured spectral functions behave as expected.
The relativistic dispersion relation is fulfilled, and we find the expected temperature dependence on both sides of the phase transition.
Is is interesting to see how, for low spatial momenta $p \ll 1$, the spectral shape changes at different temperatures and shows critical scaling, while at larger spatial momenta $p > 1$, the spectral function retains its Breit-Wigner shape and moves across the transition in a completely continuous fashion.

In the non-critical regime, introducing a finite Langevin damping $\gamma$ leads to broadening of resonances, as one expects from the structure of the equations of motion.
Close to the critical point, we see the same effect of the Langevin damping $\gamma$ on the remaining quasi-particle contribution to the spectral function, whereas the infrared-divergent critical part appears qualitatively unchanged.
However, as we will show in detail in \cref{sec:critdyn}, the slope of the infrared divergent power law is modified slightly.

\section{Critical dynamics and scaling functions}\label{sec:critdyn}

In the following section we investigate the critical behaviour of spectral functions.
We focus on the determination of a universal scaling function for the spectral function, as well as the extraction of the dynamic critical exponent $z$.

Dynamic critical phenomena have been studied since the late 60s, with the first numerical results for Ising models in the 90s \cite{dammann_dynamical_1993,lacasse_new_1993,matz_dynamic_1994,wang_study_1995,li_finite_1996,nightingale_dynamic_1996,silverio_soares_numerical_1997,nightingale_universal_1998}.
These studies were mostly concerned with finding the dynamic critical exponent $z$ of the Ising model with Glauber dynamics (2D Model A), where a multitude of results ranging from $z=1.7$ to $z=2.7$ have been published, however slowly converging towards $z \approx 2.2$.
Nightingale and Bl\"ote \cite{nightingale_dynamic_1996} were the first to calculate $z$ in 2D Model A in an Ising model with high precision using a variance-reducing Monte Carlo algorithm, albeit on rather small lattices up to $15 \times 15$.
They found $z=2.1665(12)$, quoting a two-sigma error, in accordance with the former trend.

In 2004, Dunlavy and Venus measured critical slowing down in ferromagnetic ultra-thin films \cite{dunlavy_critical_2005}, governed by 2D Model A dynamics as well.
The resulting critical exponent of $2.09(6)$, giving the 95\% confidence interval, is close to the Monte Carlo result, but the difference of nearly $2.5\sigma$ is somewhat large.

The most recent result by Zhong et al.~\cite{zhong_critical_2018} for a two-dimensional scalar $\phi^4$ model with local Metropolis updates seems to confirm the results for 2D Model A by Nightingale and Bl\"ote  from \cite{nightingale_dynamic_1996}, as they find $z_A = 2.17(3)$ and $z_A = 2.19(3)$ for two different values of the coupling constant.
Additionally, they demonstrated that quantities derived from the two-point function follow a scaling behaviour.

In 2010, a precursor study \cite{berges_dynamic_2010} of the present one first tried to confirm the dynamic critical exponent in 2D Model C, $z = 2 + \alpha/\nu = 2$ for an Ising-like scalar field theory with conservative dynamics, and found $z=2.0(1)$.

Naturally, numerical simulations in $d=3$ are more expensive than in $d=2$, and thus there are fewer and less precise results for the dynamic critical exponents.
The study by Matz et al.~\cite{matz_dynamic_1994} hints at $z=2.05(5)$ for 3D Model A, with logarithmic corrections; Wang et al.~\cite{wang_study_1995} reach a compatible result of $z = 2.09(4)$.

We are not aware of any previous Monte-Carlo studies on the dynamic critical exponent for Model C in $d=3$; however, based on the classification scheme by Halperin and Hohenberg the dynamic critical exponent there is known by virtue of the same scaling relation ($z=2+\frac{\alpha}{\nu}$) as in $d=2$, which amounts to $z \approx 2.17$ for a model in the 3D Ising universality class.

\begin{figure}
    \graphicspath{{\fdir}}
    \begin{minipage}[t]{.5\linewidth}
        \includegraphics{\fdir/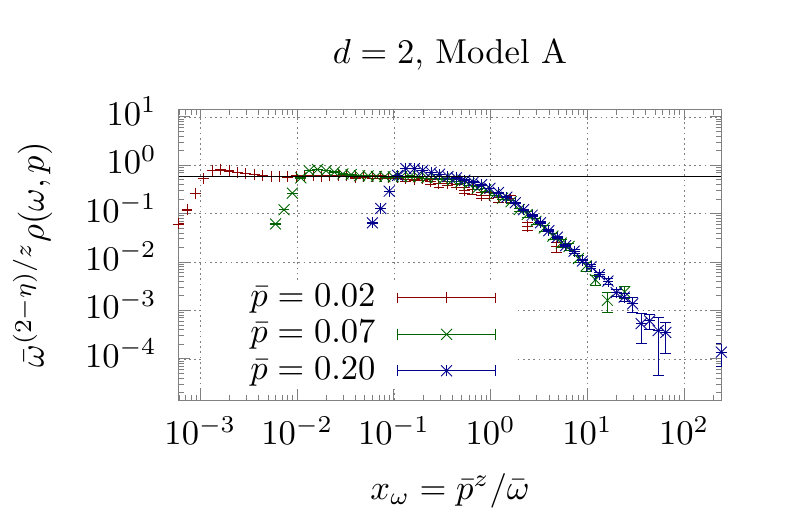}
    \end{minipage}
    \begin{minipage}[t]{.5\linewidth}
        \includegraphics{\fdir/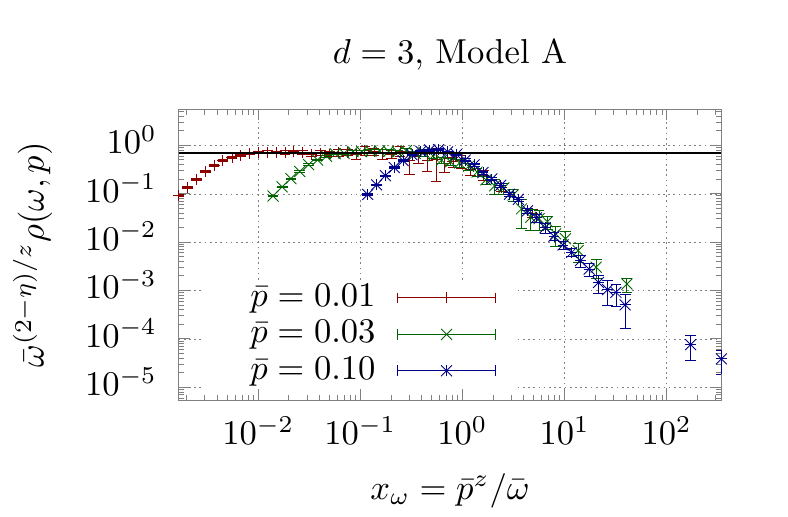}
    \end{minipage}
    \begin{minipage}[t]{.5\linewidth}
        \includegraphics{\fdir/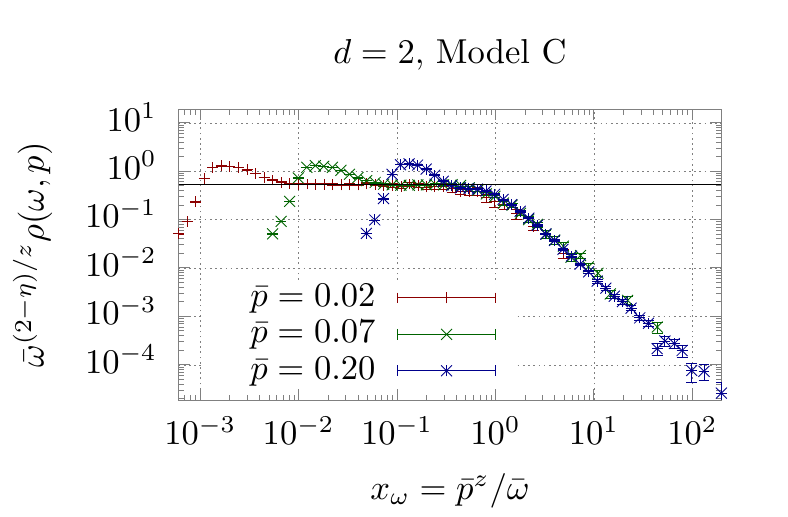}
    \end{minipage}
    \begin{minipage}[t]{.5\linewidth}
        \includegraphics{\fdir/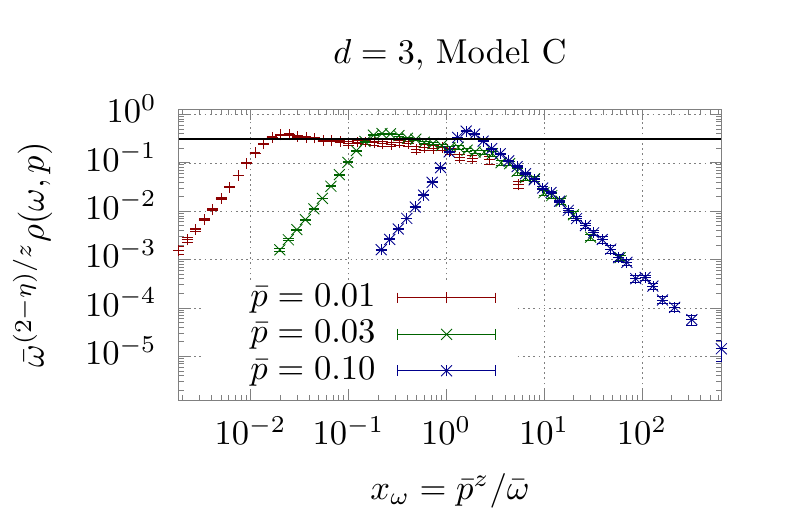}
    \end{minipage}
    \caption{Scaling function $f_{\omega}(x_{\omega}, 0)$ of the critical spectral function at non-zero spatial momentum for the four different critical scenarios.
        Solid black lines represent the amplitude $f_{\omega}$ obtained from the fit of the universal scaling function in \cref{fig:p_scaling_functions} at large $x_{p}$.
        Due to availability of data, we use $\tau = 0.0009(2)$ ($d=2$) resp.~$\tau = -0.00008(5)$ (d=3) as a proxy for the critical temperature.
        Note that, despite $\tau \approx 0$, finite size effects are not relevant here due to finite spatial momentum.
        However, to achieve the comparatively very low $p$ in $d=3$, a single set of data at $N=512$ very close to $T_c$ was generated.}
    \label{fig:f_omega_p}
\end{figure}

\begin{figure}
    \graphicspath{{\fdir}}
    \begin{minipage}[t]{.5\linewidth}
        \includegraphics{\fdir/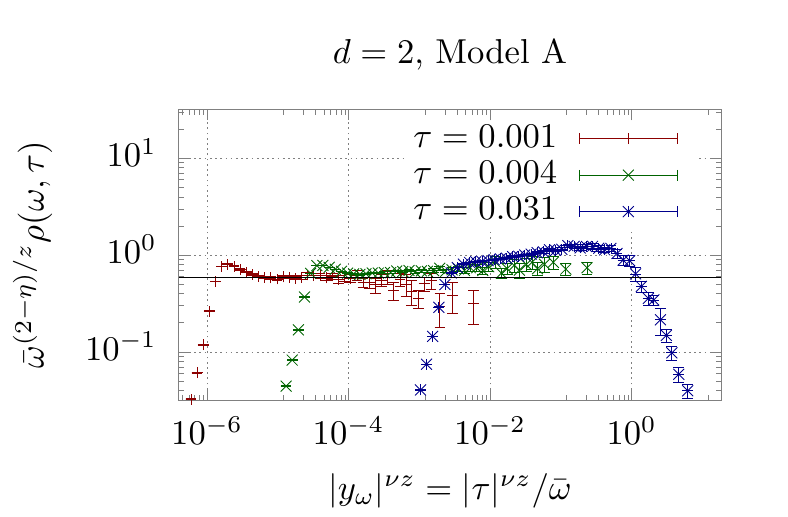}
    \end{minipage}
    \begin{minipage}[t]{.5\linewidth}
        \includegraphics{\fdir/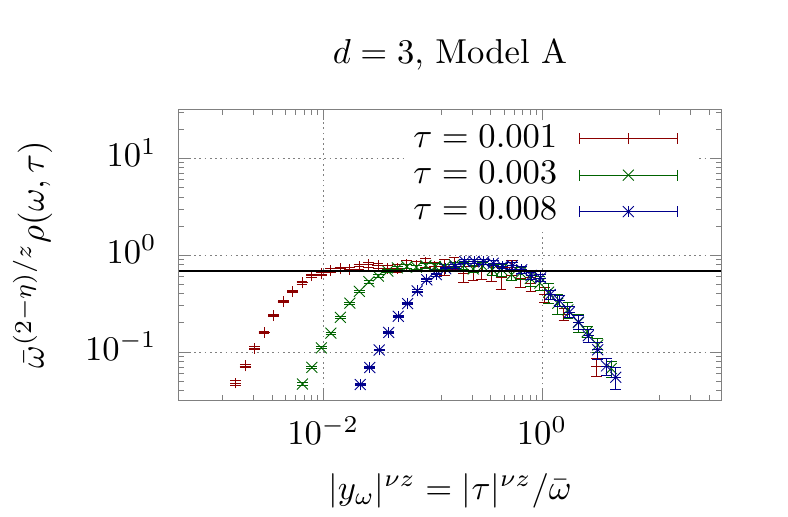}
    \end{minipage}
    \begin{minipage}[t]{.5\linewidth}
        \includegraphics{\fdir/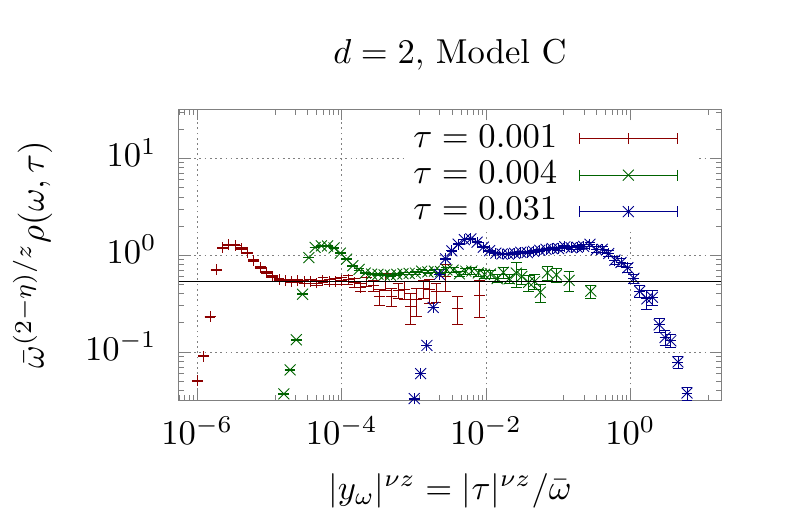}
    \end{minipage}
    \begin{minipage}[t]{.5\linewidth}
        \includegraphics{\fdir/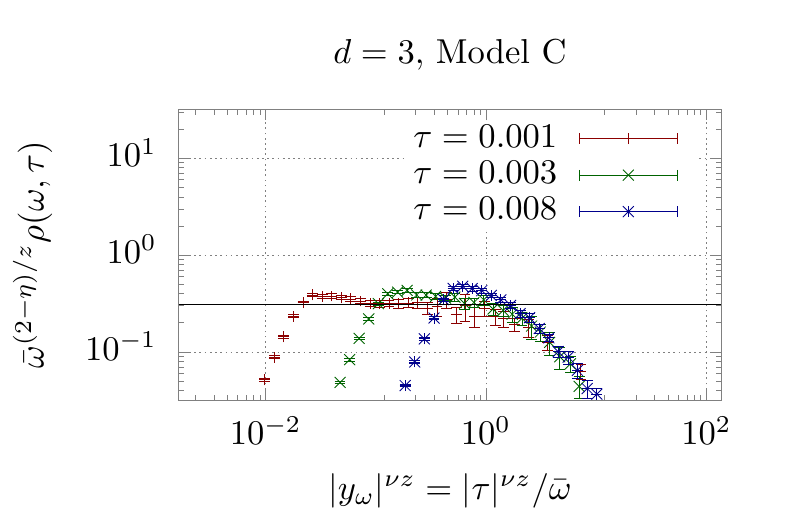}
    \end{minipage}
    \begin{minipage}[t]{.5\linewidth}
        \includegraphics{\fdir/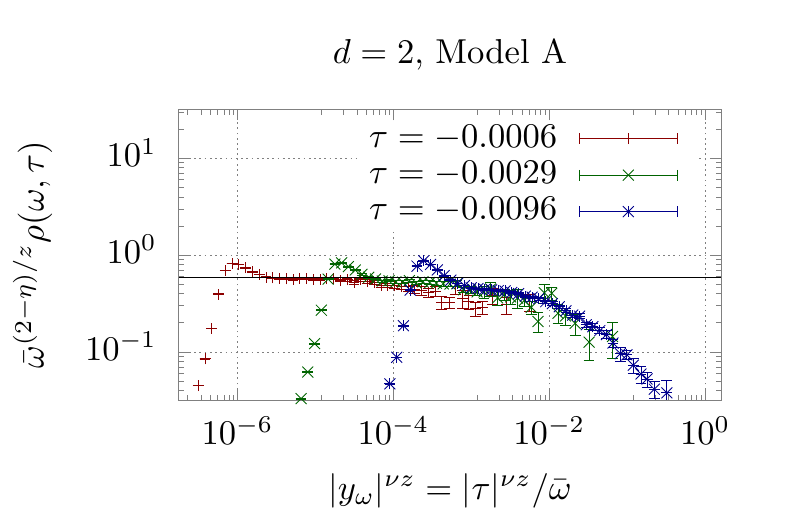}
    \end{minipage}
    \begin{minipage}[t]{.5\linewidth}
        \includegraphics{\fdir/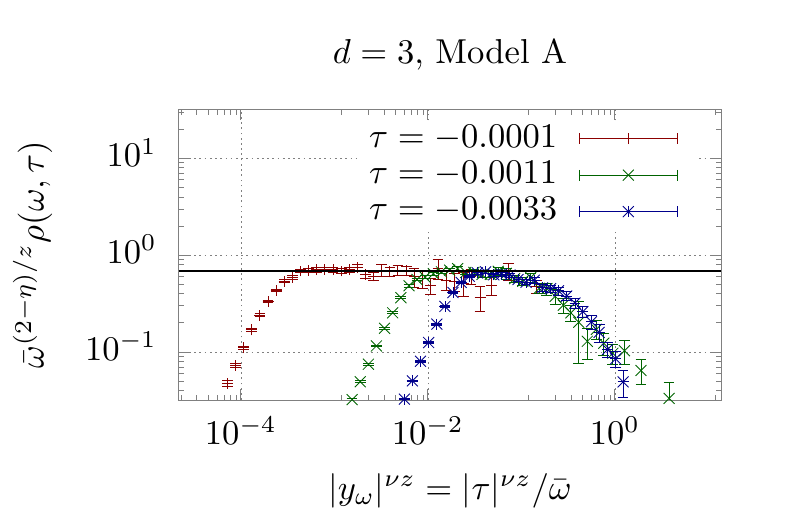}
    \end{minipage}
    \begin{minipage}[t]{.5\linewidth}
        \includegraphics{\fdir/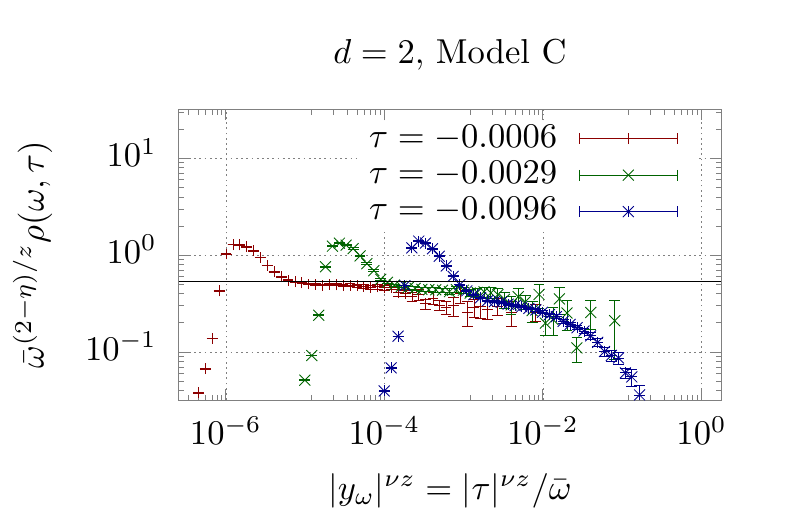}
    \end{minipage}
    \begin{minipage}[t]{.5\linewidth}
        \includegraphics{\fdir/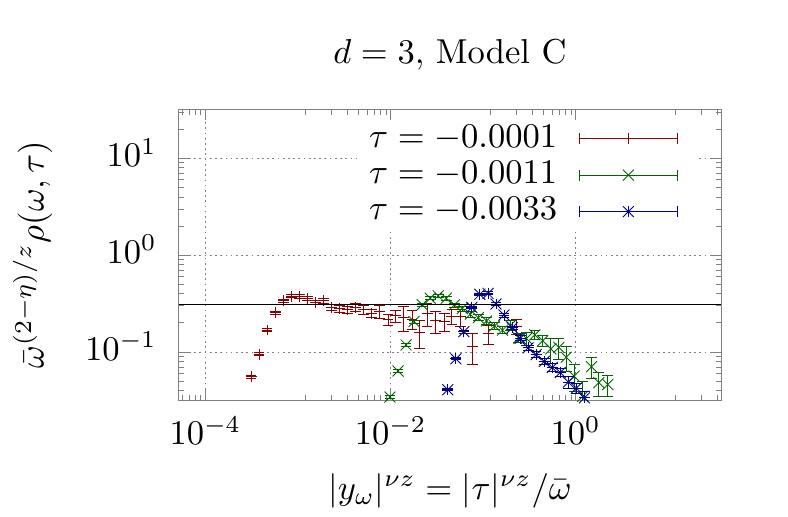}
    \end{minipage}
    \caption{Scaling function $f_{\omega}(0, y_{\omega})$ of the spectral function at vanishing spatial momentum for the four different critical scenarios.
        Solid black lines represent the amplitude $f_{\omega}$ resulting from the fit of $f_{p}(x_{p},0)$ at large $x_{p}$ in \cref{fig:p_scaling_functions}, which also describes the behaviour of $f_{\omega}(0, y_{\omega})$ when the second argument is small, indicating that the limits $p \to 0$ and $\tau \to \pm 0$ commute.
        Note that based on the available lattice sizes, finite-size effects (both in spatial and temporal direction) start to play a role at very small $\tau$.
        The data for $d=3$ spatial dimensions was generated on lattice volumes of $192^3$.}
    \label{fig:f_omega_t}
\end{figure}

\begin{figure}
    \graphicspath{{\fdir}}
    \begin{minipage}[t]{.5\linewidth}
        \includegraphics{\fdir/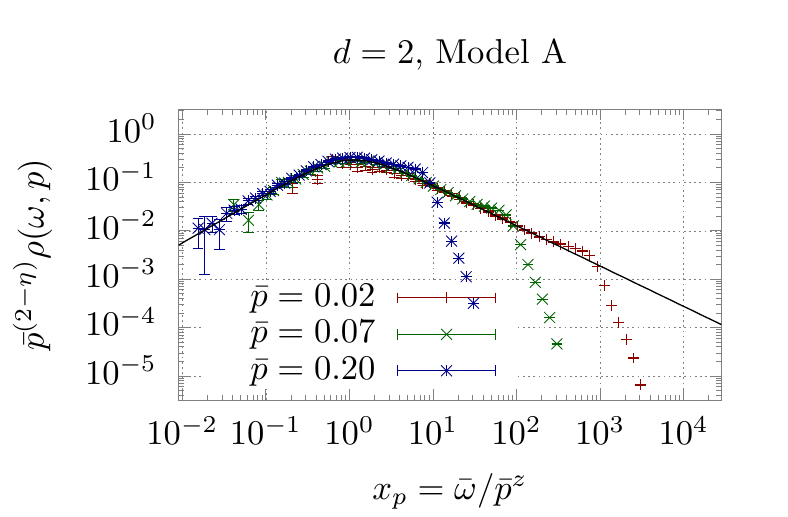}
    \end{minipage}
    \begin{minipage}[t]{.5\linewidth}
        \includegraphics{\fdir/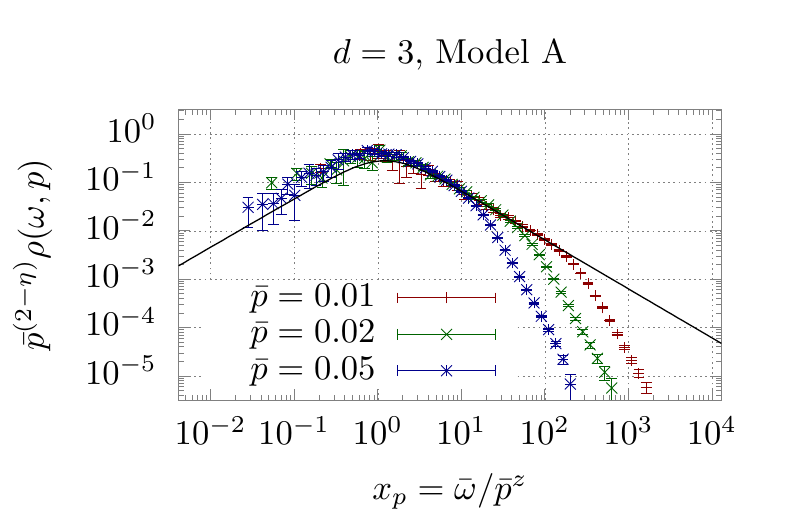}
    \end{minipage}
    \begin{minipage}[t]{.5\linewidth}
        \includegraphics{\fdir/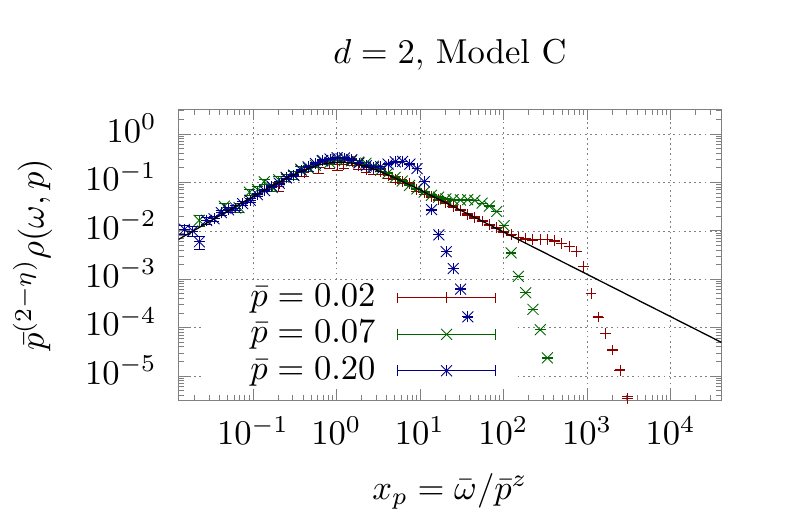}
    \end{minipage}
    \begin{minipage}[t]{.5\linewidth}
        \includegraphics{\fdir/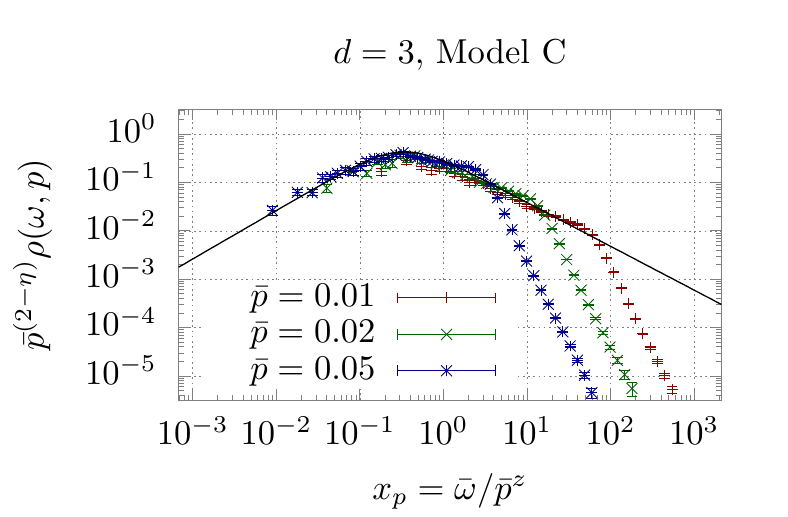}
    \end{minipage}
    \caption{Scaling function $f_p(x_p, 0)$ of the spectral function for the four different critical scenarios. 
        Due to availability of data, we use $\tau = 0.0009(2)$ ($d=2$) resp.~$\tau = -0.00008(5)$ (d=3) as a proxy for the critical temperature.
        Solid lines represent the fit to the ansatz in \cref{eq:rho_scaling_ansatz_p}, we fit $f_{\omega}$ and $a_p$ as free parameters, and the exponent of the power law for large $x_p$ is given by $(2-\eta)/z_{\text{comb.}}$.
        Note that finite size effects are not relevant here due to finite spatial momentum.
        To achieve the comparatively very low $p$ in $d=3$, a single set of data at $N=512$ very close to $T_c$ was generated.}
    \label{fig:p_scaling_functions}
\end{figure}

\begin{figure}
    \graphicspath{{\fdir}}
    \begin{minipage}[t]{.5\linewidth}
        \includegraphics{\fdir/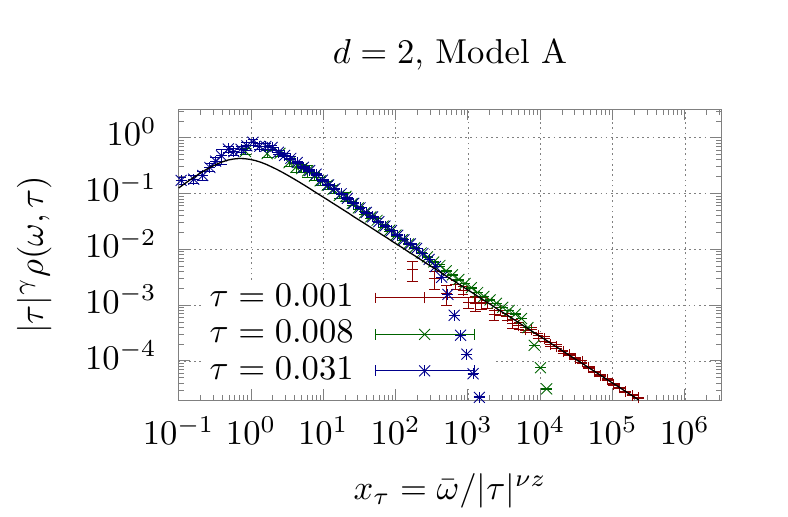}
    \end{minipage}
    \begin{minipage}[t]{.5\linewidth}
        \includegraphics{\fdir/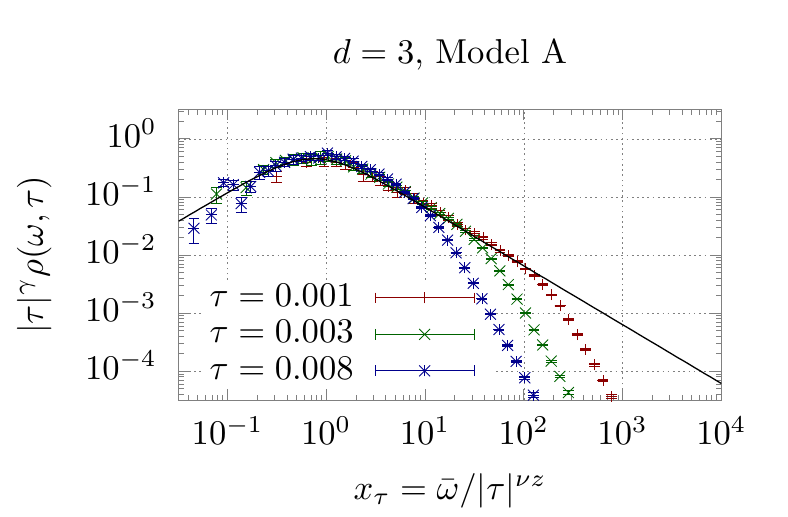}
    \end{minipage}
    \begin{minipage}[t]{.5\linewidth}
        \includegraphics{\fdir/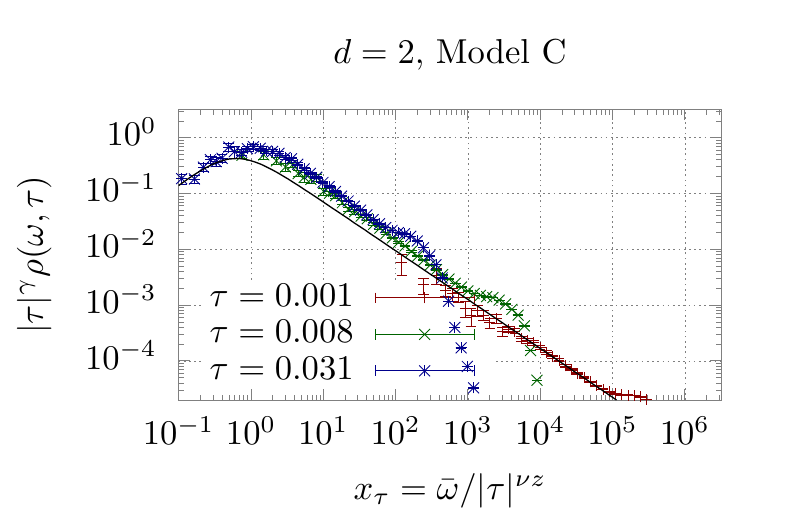}
    \end{minipage}
    \begin{minipage}[t]{.5\linewidth}
        \includegraphics{\fdir/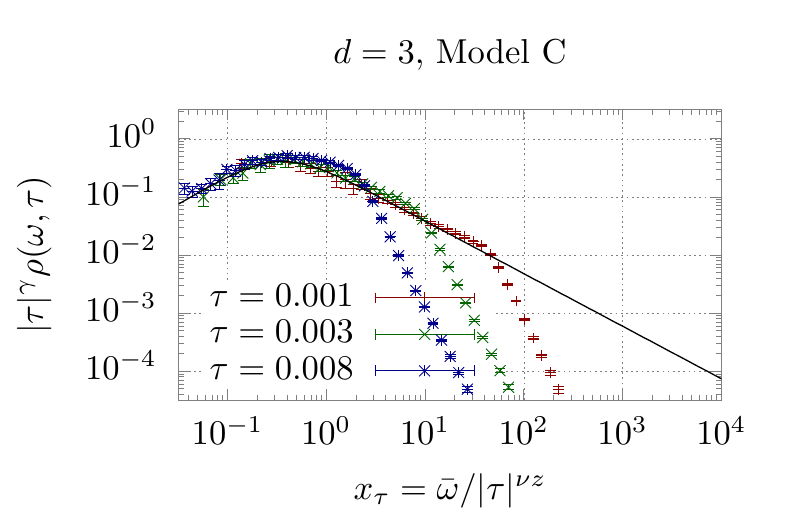}
    \end{minipage}
    \begin{minipage}[t]{.5\linewidth}
        \includegraphics{\fdir/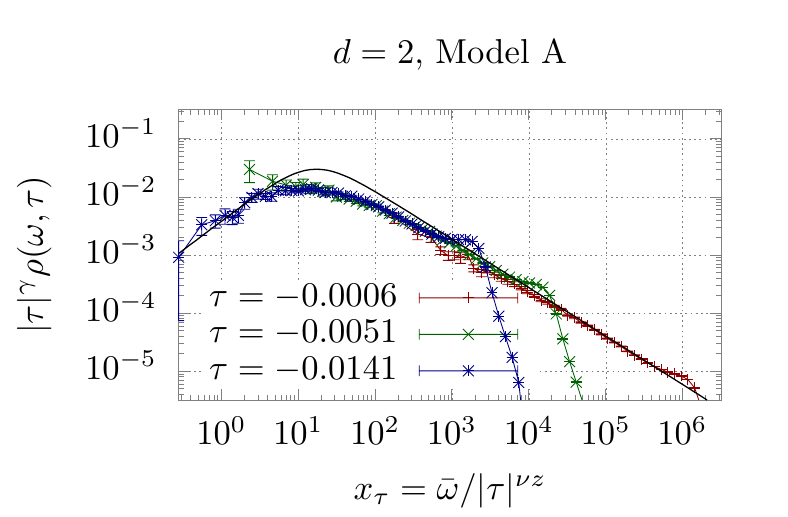}
    \end{minipage}
    \begin{minipage}[t]{.5\linewidth}
        \includegraphics{\fdir/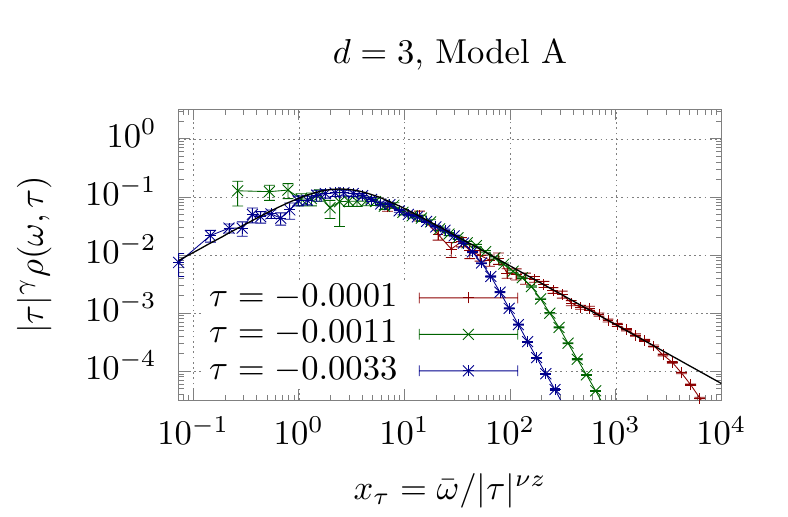}
    \end{minipage}
    \begin{minipage}[t]{.5\linewidth}
        \includegraphics{\fdir/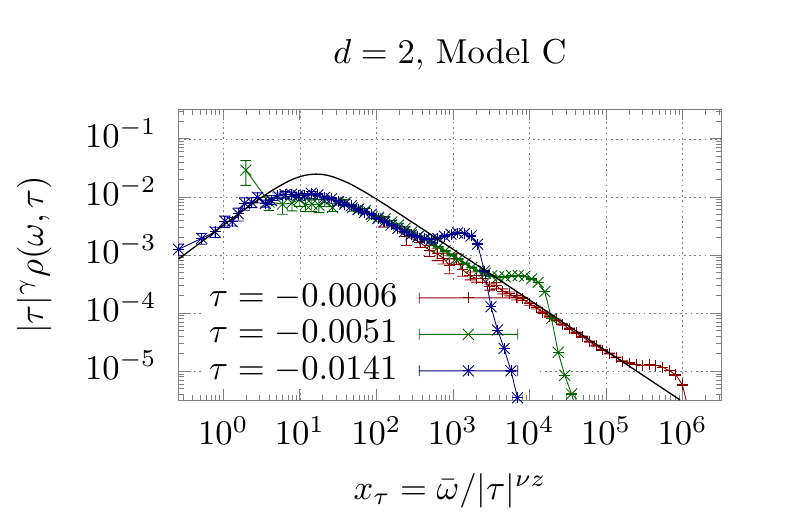}
    \end{minipage}
    \begin{minipage}[t]{.5\linewidth}
        \includegraphics{\fdir/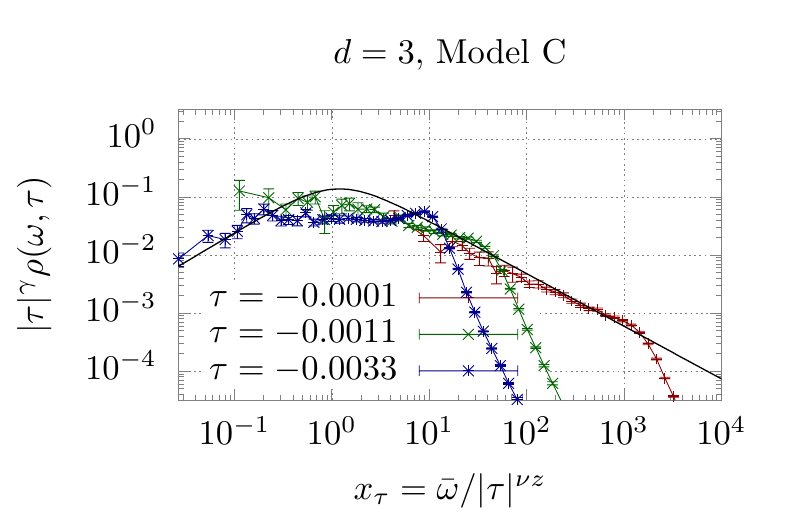}
    \end{minipage}
    \caption{Scaling functions $f_{\tau}^{\pm}(\bar \omega / |\bar \tau|^{\nu z}, 0)$ of critical spectral functions at vanishing momenta above (top four) and below (bottom four) $T_c$ for the four different scenarios. 
        Solid lines represent the fit to the ansatz in \cref{eq:rho_scaling_ansatz_t}, where we obtain $a_{\tau}^{\pm}$ by fitting \cref{eq:rho_scaling_ansatz_t} to the data at small $x$, keeping $f_{\omega}$ as obtained from the ansatz in \cref{eq:rho_scaling_ansatz_p} to the data in \cref{fig:p_scaling_functions}, and $z=z_{\text{comb.}}$.
        Note that based on the available lattice sizes, finite-size effects (both in spatial and temporal direction) start to play a role at very small $\tau$.
        The data for $d=3$ spatial dimensions was generated on lattice volumes of $192^3$.}
    \label{fig:t_scaling_functions}
\end{figure}

\begin{figure}
    \graphicspath{{\fdir}}
    \begin{minipage}[t]{.5\linewidth}
        \includegraphics{\fdir/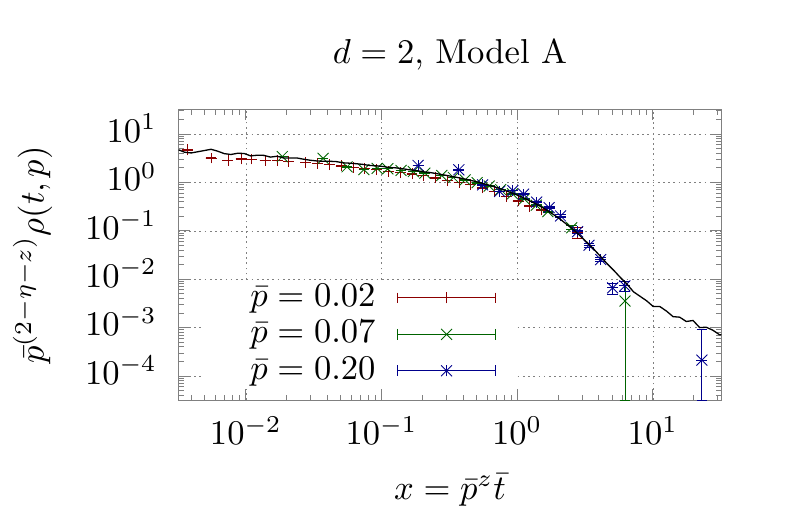}
    \end{minipage}
    \begin{minipage}[t]{.5\linewidth}
        \includegraphics{\fdir/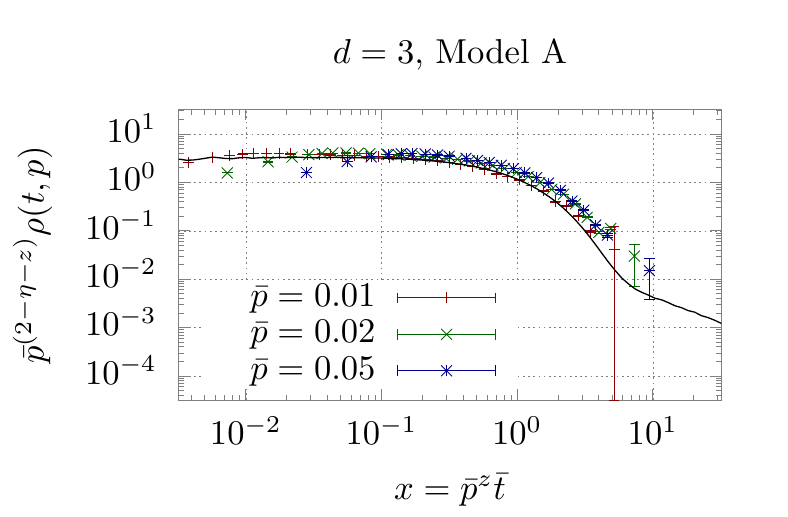}
    \end{minipage}
    \begin{minipage}[t]{.5\linewidth}
        \includegraphics{\fdir/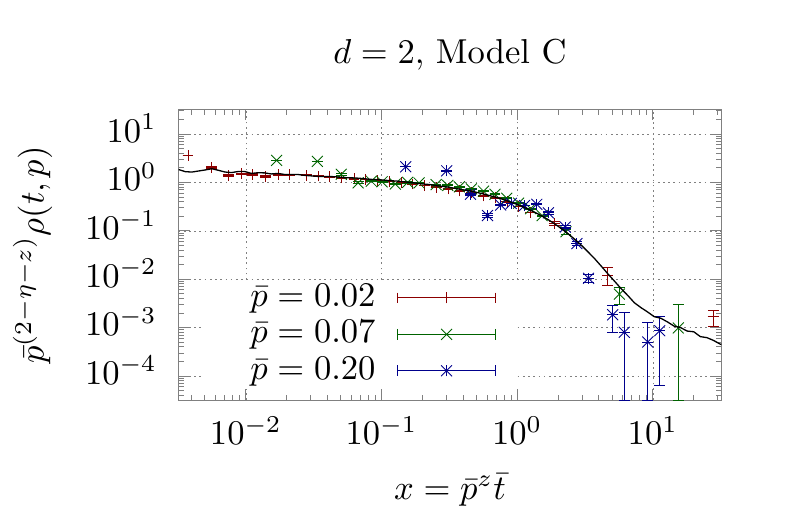}
    \end{minipage}
    \begin{minipage}[t]{.5\linewidth}
        \includegraphics{\fdir/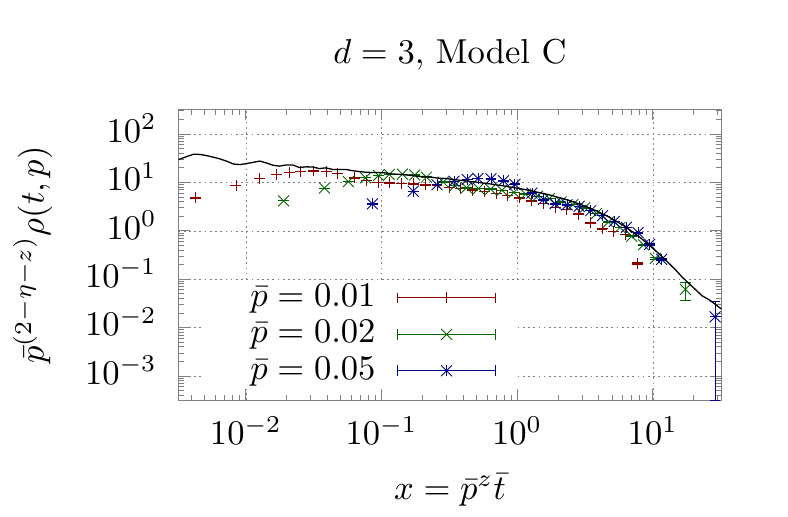}
    \end{minipage}
    \caption{Scaling functions $\tilde{f}_p\left( \bar p^z \bar t, 0 \right)$ of the critical spectral functions in the time domain for the four different critical scenarios; cf.~\cref{eq:t_rho_scale_p}.
    Solid lines represent the numerical Fourier transform of the fit to the data in frequency space.}
    \label{fig:p_scaling_functions_t}

    \begin{minipage}[t]{.5\linewidth}
        \includegraphics{\fdir/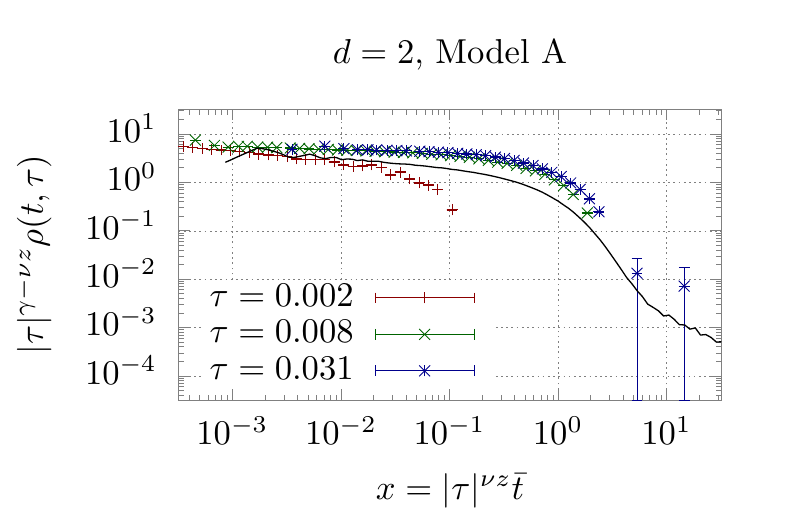}
    \end{minipage}
    \begin{minipage}[t]{.5\linewidth}
        \includegraphics{\fdir/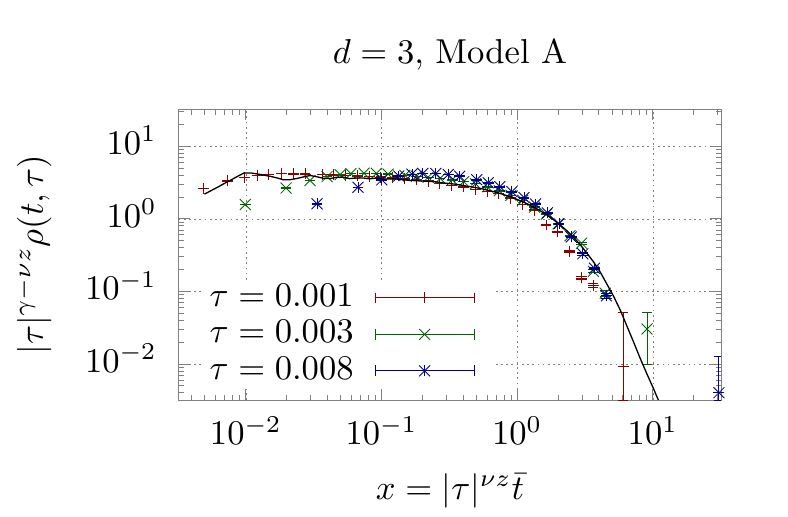}
    \end{minipage}
    \begin{minipage}[t]{.5\linewidth}
        \includegraphics{\fdir/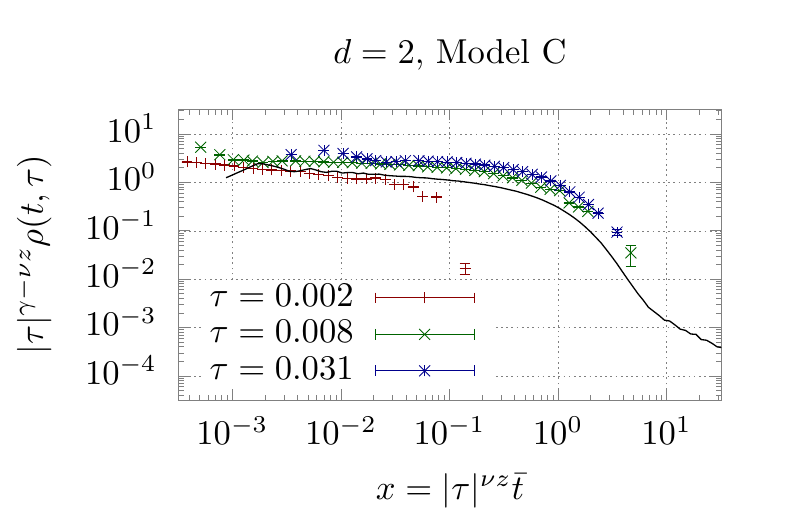}
    \end{minipage}
    \begin{minipage}[t]{.5\linewidth}
        \includegraphics{\fdir/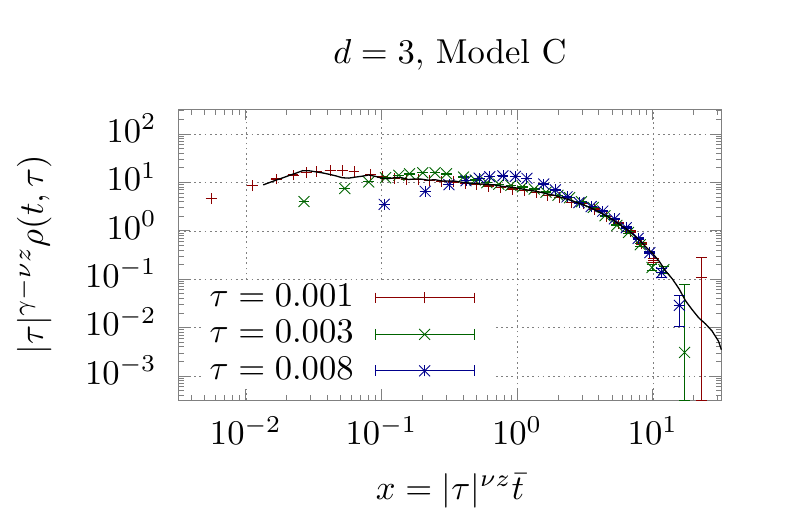}
    \end{minipage}
    \caption{Scaling functions $\tilde f_{\tau}^{\pm}(\bar t / \bar |\tau|^{-\nu z}, 0)$ of the spectral functions at vanishing spatial momentum in the time domain for the four different critical scenarios; cf.~\cref{eq:t_rho_scale_t}.
        Solid lines represent the numerical Fourier transform of the fit to the data in frequency space.
        Finite size effects again shift the data at small $\tau$ away from the scaling function.
    The data for $d=3$ spatial dimensions was generated on lattice volumes of $192^3$.}
    \label{fig:t_scaling_functions_t}
\end{figure}

\subsection{Dynamic scaling functions}

Since the spectral function is derived directly from the two-point correlation function, one expects the critical behavior to be governed by the following scaling form \cite{berges_dynamic_2010}
\begin{equation}
    \rho \left(\omega, p, \tau\right) =  s^{2-\eta} \rho \left( s^z \omega, s  p, s^{\inv \nu} \tau \right),
    \label{eq:rho_scaling}
\end{equation}
in the limit $\omega,p,\tau \to 0$, and we omit any residual dependencies on the finite volume.
If not stated otherwise, we will restrict ourselves to positive frequencies $(\omega > 0)$ to compactify notation, noting that the behaviour for negative frequencies $(\omega < 0)$ is trivially obtained from the symmetry of the spectral function $\rho(-\omega, p, \tau) = - \rho(\omega, p, \tau)$.

The scaling law in \cref{eq:rho_scaling} allows us to define three alternative scaling functions $f_\omega$, $f_p$ and $f_\tau^{\pm}$ according to
\begin{align}
    \rho\left(\omega, p, \tau\right) &= {\bar \omega}^{-(2-\eta)/z}   \, f_\omega \Big(\bar p^z/\bar \omega, \tau/\bar \omega^{1/\nu z} \Big) \label{eq:rho_scale_omega},\\
    \rho\left(\omega, p, \tau\right) &= {\bar p}^{-(2-\eta)}\, f_p \Big( \bar \omega/\bar p^z, \tau/\bar p^{1/\nu} \Big) \label{eq:rho_scale_p},\\
    \rho\left(\omega, p, \tau\right) &= |\tau |^{-\gamma} \, f^\pm_\tau \Big( \bar \omega/|\tau |^{\nu z}, \bar p^{1/\nu}/|\tau |\Big) \label{eq:rho_scale_tau},
\end{align}
where $\gamma $ is the static susceptibility exponent with $\gamma = \nu (2-\eta)$ from static scaling relations.
Indeed, the scaling behaviour in \cref{eq:rho_scale_omega,eq:rho_scale_p,eq:rho_scale_tau} is clearly visible in our classical-statistical simulations, as can be seen from \cref{fig:f_omega_p,fig:f_omega_t,fig:p_scaling_functions,fig:t_scaling_functions,fig:p_scaling_functions_t,fig:t_scaling_functions_t}, where we present results for the scaled spectral functions in the vicinity of the critical point.
We note that in order to perform the axis re-scaling in \cref{fig:f_omega_p,fig:f_omega_t,fig:p_scaling_functions,fig:t_scaling_functions,fig:p_scaling_functions_t,fig:t_scaling_functions_t}, one also needs the value of the dynamic critical exponent $z$, and if not stated otherwise, we employ the values in last row of \cref{tab:z_results}, labeled \emph{combined}, with exception of $d=3$ Model C, where we used the analytic value of $z \approx 2.17$.

Since the scaling functions in \cref{eq:rho_scale_omega,eq:rho_scale_p,eq:rho_scale_tau} are all derived from the scaling behavior of the spectral function, the scaling functions $f_\omega$,$f_p$ and $f_\tau^{\pm}$ are not independent.
Denoting the natural arguments of the respective scaling functions as 
\begin{eqnarray}
x_{\omega}=\bar{p}^{z}/\bar{\omega}\;,~ y_{\omega}=\tau/\bar{\omega}^{1/\nu z}\;, \qquad 
x_{p}=\bar{\omega}/\bar{p}^{z}\;,~ y_{p}=\tau/\bar{p}^{1/\nu}\;, \qquad
x_{\tau}=\bar{\omega}/|\tau|^{\nu z}\;,~ y_{\tau}=p^{1/\nu}/|\tau|\;,
\end{eqnarray}
one finds the following relations between the scaling functions
\begin{align}
    f_p(x_p,y_p) &=  x_p^{-(2-\eta)/z} \, f_\omega\big(1/x_p,y_p/x_p^{1/\nu z}\big)\, ,\\
    f_p(x_p,y_p) &=   |y_p|^{-\gamma} \, f_\tau^\pm\big(x_p/|y_p|^{\nu z},1/|y_p|\big)\, , \\
    f_\tau^\pm(x_\tau,y_\tau) &= x_\tau^{-(2-\eta)/z} f_\omega\big(y_\tau^{\nu z}/x_\tau,\pm 1/x_\tau^{1/\nu z}\big)\, , 
\end{align}
where the super-script in $f^\pm_\tau $ is used to distinguish $\sgn(\tau)=\sgn(y_{p})=\pm 1$ above/below $T_c$.
Similarly, the Fourier transform of \cref{eq:rho_scaling} yields in the time domain
\begin{equation}
    \rho \left( t, p, \tau \right) = s^{(2-\eta-z)} \rho\left( s^{-z} t, s p, s^{1/\nu} \tau \right),
    \label{eq:rho_scaling_t}
\end{equation}
from which one can derive the time-domain scaling functions analogous to \cref{eq:rho_scale_omega,eq:rho_scale_p,eq:rho_scale_tau}.

Based on the results of \cite{berges_dynamic_2010}, we expect the scaling function $f_\omega(x_\omega,y_\omega)$ to be regular in the origin, with a constant value  $f_\omega \equiv f_\omega(0,0)$, such that the critical zero-momentum spectral function obeys the infrared  power law
\begin{equation}
    \rho(\omega, 0, 0) = f_\omega \, {\bar \omega}^{-(2-\eta)/z} . \label{IR_power_omega} \\
\end{equation}
This expectation is verified in \cref{fig:f_omega_p,fig:f_omega_t}, where we present simulation results for $f_{\omega}(x_{\omega},0)=\omega^{(2-\eta)/z} \rho(\omega,p,0)$ as a function of $x_\omega$ at $\tau \,\propto \, y_\omega = 0$  (\cref{fig:f_omega_p}) and respectively $f_{\omega}(0,y_{\omega})=\omega^{(2-\eta)/z} \rho(\omega,0,\tau)$ as a function of $y_\omega$ at $p\, \propto\,  x_\omega =0$ (\cref{fig:f_omega_t}).
Generally, one observes an excellent scaling collapse of the data for different momenta(reduced temperatures), with deviations only at very small values of $x_\omega$($y_\omega $) where the residual contributions from the quasi-particle peak at finite momentum(reduced temperature) effectively acts as a cut-off.
Specifically, for small arguments $x_\omega$ or $y_\omega$ the two data sets in \cref{fig:f_omega_p,fig:f_omega_t} appear to converge towards the same value $f_{\omega}$ indicated by a solid black line, demonstrating that $f_{\omega}$ is indeed finite and that the limits $p\to 0$ and $\tau \to 0$ commute at small but finite frequency $\bar{\omega}\neq 0$.

Due to the above relations between the scaling functions $f_{\omega},f_{p}$ and $f_{\tau}^{\pm}$, we can further exploit the regularity of $f_{\omega}$ near the origin $(x_{\omega},y_{\omega})=(0,0)$ to deduce the behavior of the scaling functions $f_{p}$ and $f_{\tau}^{\pm}$ for large values  of the arguments $x_{p}=1/x_{\omega}$  and $x_{\tau}=1/|y_{\omega}|^{\nu z}$ as
\begin{align} 
    \bar{p}^{2-\eta} \rho(\omega,p,0) =~f_p(x_{p},0) & \stackrel{x_{p} \to \infty}{\to} f_\omega \, x_{p}^{-(2-\eta)/z} \, ,\nonumber \\
    |\tau|^{\gamma} \rho(\omega,0,\tau) = f_\tau^\pm(x_{\tau},0) &\stackrel{x_{\tau} \to \infty}{\to}  f_\omega \,  x_{\tau}^{-(2-\eta)/z} \, . \label{SFs_largex}
\end{align}
This fixes the high-frequency ($\omega \gg 1/\xi_{t}(\tau,p) $) behavior of the critical scaling function $f_p(x_{p},0)$ at very low momenta, and that of the zero-momentum scaling function $f^\pm_\tau (x_{\tau},0)$ very close to criticality at the same time. The characteristic time $\xi_t(\tau,p)$ for this asymtpotic behavior, from the definition of our scaling variables, behaves as $\xi_t(0,p) \sim 1/p^z$ and $\xi_t(\tau,0) \sim 1/|\tau|^{\nu z}$ at small $p$ and $\tau$, respectively.

Conversely, the low-frequency $(\omega \ll 1/\xi_{t}(\tau,p))$ behavior of the spectral function is determined by the behavior of $f_\omega(x_{\omega},0)$ and $f_\omega(0,y_{\omega})$ at asymptotically large values of $x_{\omega}$ and $y_{\omega}$, and can be determined from the behavior of $f_p(x_{p},0)$ and $f_\tau^\pm(x_{\tau},0)$ at small values of the arguments $x_{p}=1/x_{\omega}$  and $x_{\tau}=1/|y_{\omega}|^{\nu z}$.
Numerically, we find from \cref{fig:p_scaling_functions,fig:t_scaling_functions} that the leading behavior at small $x_{p}$ and $x_{\tau}$ is well described by 
\begin{align}
    f_p(x_{p},0) &= a_p \, x_{p} \, + \, \cdots \, , \nonumber \\
    f_\tau^\pm(x_{\tau},0) &= a_\tau^\pm  \, x_{\tau} \, + \, \cdots \, , \label{SFs_smallx} 
\end{align}
which corresponds to
\begin{align}
    f_\omega(x_{\omega},0) &\to a_p \, x_{\omega}^{-(2-\eta)/z-1} \,, \;\; x_{\omega} \to \infty\,,  \nonumber \\
    f_\omega(0,y_{\omega}) &\to a_\tau^\pm \, |y_{\omega}|^{-\gamma - \nu z} \,, \;\; y_{\omega} \to \pm\infty\, .\label{eq:fw_large_xy} 
\end{align}
Specifically, for the critical spectral function ($\tau=0$) at finite spatial momentum, \cref{eq:fw_large_xy} implies an infrared behavior, valid asymptotically for  $\bar\omega \ll \bar p^z$,
\begin{equation}
    \rho(\omega,p,0) = a_p \, \bar p^{-(2-\eta)} \, \bar\omega/\bar p^z \, + \, \cdots \,.\label{IR_omega_p}
\end{equation}
Comparing \cref{IR_omega_p} with \cref{IR_power_omega}, we see that the limits  $p\to 0$ and $\omega\to 0$ do not commute, as the critical spectral function is non-analytic in the origin.
In particular, we have $\lim_{\omega \to 0} \rho(\omega,0,0) = \infty $, while $\lim_{p\to 0 } \rho(0,p,0) = 0$.
Physically, this has the intuitive  interpretation that any finite momentum $p$ introduces an effective IR cutoff for the correlation length $\xi \sim 1/p$, which in turn is associated with a finite correlation time $\xi_t(0,p) \sim \xi^z\sim 1/p^z$, that defines the characteristic frequency $\bar\omega \sim \bar p^z $ where the power law changes from (\ref{IR_power_omega}) to (\ref{IR_omega_p}).

Similarly, for the zero-momentum spectral function off criticality we conclude that the infrared behavior for $\bar\omega \ll |\tau|^{\nu z} $ is modified as 
\begin{equation}
    \rho(\omega ,0,\tau) = a_\tau^\pm \, |\tau|^{-\gamma} \, \bar\omega/|\tau|^{\nu z} \, + \, \cdots , \label{IR_omega_tau}
\end{equation}
so that for all $\tau \not=0$ we also have $\lim_{\omega\to 0} \rho(\omega,0,\tau ) = 0$, with the characteristic frequency where the power law changes from (\ref{IR_power_omega}) to  (\ref{IR_omega_tau}) given by $\omega \sim |\tau|^{\nu z}$.
Once again, this simply reflects the finiteness of the correlation time at non-vanishing $\tau$ where we now have $\xi_t(\tau,0) \sim \xi^z \sim 1/|\tau|^{\nu z}$.

So far we have analyzed the limiting behavior of the scaling functions for very large and very small arguments.
Now, in order to interpolate between these two limits, we observe that the inverse of the scaling functions $f_p(x,0)$ and $f_\tau^{\pm}(x,0)$ is globally very well described by a sum of the reciprocal power laws (\ref{SFs_largex}) and (\ref{SFs_smallx}) for large and small $x$.
We therefore use the following parametrizations of the scaling functions to fit our data,
\begin{align}
    f_p(x_p,0)    &= \frac{1}{(a_p \, x_p)^{-1} + f_\omega^{-1}\, x_p^{(2-\eta)/z}} \, \;\;
    \mbox{for} \;\; x_p = \bar\omega/ \bar p^z \, ,
    \label{eq:rho_scaling_ansatz_p}\\
    f_\tau^\pm(x_\tau,0)  &= \frac{1}{(a_\tau^\pm\, x_\tau) ^{-1} + f_\omega^{-1}\, x_\tau^{(2-\eta)/z}} \, \;\;
    \mbox{for} \;\; x_\tau =   \bar\omega/ |\tau|^{\nu z} \, .
    \label{eq:rho_scaling_ansatz_t}
\end{align}
Our results are compactly summarized in \Cref{fig:t_scaling_functions,fig:p_scaling_functions}, where we show these parametrizations for the scaling functions $f_p(\bar\omega/\bar p^{z} , 0) $ and  $f_\tau^\pm(\bar\omega/|\tau|^{\nu z}   ,0) $ fitted to our data for $\bar p^{(2-\eta)} \rho(\omega, p, 0)$ and $|\tau|^\gamma \rho(\omega, 0, \tau)$  respectively.
The regions where the points overlap are used to determine the universal scaling functions.
The numerical results for the parameters $a_p,\, a_{\tau}^{\pm},\, f_\omega$ obtained from the fits are listed in \cref{tab:scaling_function_amplitudes}.
Corresponding plots for the scaling laws in the time domain are shown in \cref{fig:t_scaling_functions_t,fig:p_scaling_functions_t} where we plot the Fourier transformed scaling functions $\tilde f_p(x,0) $ and $\tilde f_\tau^+(x,0)$, from the analogous definitions 
\begin{align}
    \rho(t,p,\tau) &=  \bar p^{-(2-\eta-z)}\, \tilde f_p(\bar p^z \bar t, \tau/\bar p^{1/\nu})\,, \label{eq:t_rho_scale_p}\\
    \rho(t,p,\tau) &= |\tau|^{-(2-\eta - z)} \, \tilde f_\tau^\pm( |\tau|^{\nu z} \bar t, \bar p^{1/\nu}/|\tau |)\, .\label{eq:t_rho_scale_t}
\end{align}
By looking at \Cref{fig:t_scaling_functions,fig:p_scaling_functions}, one finds that scaled data sets for the spectral functions at different momenta/reduced temperatures overlap with each other to rather good accuracy, and that, apart from $f_{\tau}^{\pm}(x_\tau, 0)$ in $d=2$ spatial dimensions, the ans\"atze in \cref{eq:rho_scaling_ansatz_p,eq:rho_scaling_ansatz_t} match the overlapping data points exceptionally well.
Moreover, the characteristic frequencies  $\bar \omega_c \sim \bar p^z$ and  $\bar \omega_c \sim \tau^{\nu z}$ mentioned above can be read off directly from the coinciding maxima in the respective scaling variables $x_{p}$ and $x_{\tau}$.
The above fits (\ref{eq:rho_scaling_ansatz_p}) and (\ref{eq:rho_scaling_ansatz_t}) in turn imply that the frequency scaling function $f_\omega(x_{\omega},y_{\omega})$, either at criticality ($y_{\omega} = \tau/\bar\omega^{1/\nu z} = 0 $) or at zero momentum ($x_{\omega}= \bar p^z/\bar\omega  =0 $), is equally well described by
\begin{align}
    f_\omega(x_\omega,0) &= \frac{f_\omega }{1 + (f_\omega/a_p)  \, x_\omega^{(2-\eta)/z+1}} \, ,\\
    f_\omega(0,y_\omega) &= \frac{f_\omega }{1 + (f_\omega/ a_\tau^\pm) \, |y_\omega|^{\nu z+ \gamma}} \, ,
\end{align}
with $a_\tau^\pm $ for $\sgn(y_\omega) = \pm 1 $ above and below $T_c$, which provides a complete description of the scaling function $f_{\omega}(x_\omega,y_\omega)$ along the two coordinate axes.

\begin{table}
    \centering
    \caption{Extracted fit parameters from fitting \cref{eq:rho_scaling_ansatz_p,eq:rho_scaling_ansatz_t} to the scaling functions.
        The first line corresponds to the results of the fits of \cref{eq:rho_fit_func} with $z$ as a fit parameter, with an estimate of the systematic error.
        The other lines represent the values obtained by fitting the data in \cref{fig:p_scaling_functions,fig:t_scaling_functions} with a fixed value for the dynamic critical exponent, corresponding to the last row of \cref{tab:z_results} labeled \emph{combined}, see below.
        For the latter ones we only give the statistical fit error.
    }
    \label{tab:scaling_function_amplitudes}
    \begin{tabular}{r || l l | l l}
        & \multicolumn{2}{c}{2D} & \multicolumn{2}{c}{3D}\\
        & Model A & Model C & Model A & Model C \\
        \hline\hline
        \rule{0pt}{2.4ex}\rule[-1.2ex]{0pt}{0pt}
        $f_\omega$ & 0.73(9) & $0.50_{-0.14}^{+0.23}$ & 0.56(35) & $1.14_{-0.56}^{+0.25}$\\
        $f_\omega(z_{\text{comb.}})$ & 0.592(4) & 0.536(2) & 0.69(1) & 0.544(7)\\
        $a_p$ & 0.56(1) & 0.51(1) & 0.46(37) & 1.13(10)\\
        $a_t^+$ & 1.30(9) & 1.43(6) & 1.20(10) & 1.13(16)\\
        $a_t^-$ & 0.0038(2) & 0.0032(1) & 0.111(8) & 0.14(1)\\
    \end{tabular}
\end{table}

By comparing our results for the scaling functions in \Cref{fig:t_scaling_functions,fig:p_scaling_functions}, we find that in general, the scaling regions are larger in $d=2$ than in $d=3$, and similarly larger for Model A than for Model C.
Specifically, for $d=3$, it seems that in some cases the standard volumes ($N=128$) are still not large enough to show critical effects of reasonable strength.
Hence, in order to achieve some sufficient overlap for extracting the universal scaling function at finite spatial momentum $p$, we have generated a single additional data set at $T\approx T_c$ with $N=512$ in $d=3$, which allows us to investigate very small spatial momenta, with $p < 0.05$ in lattice units.
By inspecting e.g.~the upper right panel in \cref{fig:p_scaling_functions}, it is then clear that for very small $p$, the overlap region does indeed extend to the right of the maximum, where the slope in the logarithmic plot is determined by the critical power law.

One important advantage of working with finite spatial momenta $p>0$ is that finite volume effects are essentially irrelevant, as the relevant infrared cut-off is set by the momentum $p$ rather than the system size.
Conversely, at $p=0$, finite volume effects inevitably appear close to criticality, i.e.~for very small $\tau$, as can be seen e.g.~in \cref{fig:t_scaling_functions,fig:t_scaling_functions_t}.

By closer inspection of the results in \Cref{fig:t_scaling_functions,fig:p_scaling_functions}, e.g.~when comparing models A (upper row) to models C (lower row), one further notices that the spectral functions $\rho(\omega,p,\tau)$ start to deviate from the scaling function when reaching the remnants of the quasi-particle peak, which appear as additional ``shoulders'' at the high-frequency end.
Clearly, this effect is more pronounced in Model C, where high-frequency fluctuations do not receive the additional damping due to the coupling to the heat bath.
By looking at the results for $\tau<0$ in \cref{fig:t_scaling_functions} we also note that the universal scaling of the spectral functions, when approaching criticality from the ordered phase, does not appear to emerge from the remnants of the quasi-particle peak, even in close vicinity of the critical point.
Instead, as previously alluded to in the context of the discussion of \cref{fig:sf-overview}, it rather seems that the universal critical behavior of the spectral function emerges from the soft collective low-frequency excitation in the ordered phase.

\subsection{Extraction of the dynamic critical exponent}\label{ssec:z_extraction}
Having established the dynamic scaling behavior of the spectral function, we now turn to the extraction of the dynamic critical exponent $z$.
Generally speaking, there are multiple possibilities to extract the dynamic critical exponent $z$ from the data, and we will explore three different methods in the following, which we will refer to as the {\em scaling} method, the critical {\em IR power law} method, and the divergence of the {\em correlation time} $\xi_t$ method.

Evidently, to obtain the plots of the scaled spectral function in \cref{fig:f_omega_p,fig:f_omega_t,fig:p_scaling_functions,fig:t_scaling_functions,fig:p_scaling_functions_t,fig:t_scaling_functions_t}, one needs the correct value of $z$ that maximizes the overlap of the data points.
Vice versa, we can exploit this scaling behavior to extract the dynamical critical exponent $z$ by minimizing the deviations from perfect scaling, which we quantify in terms of the $L^2$-norms of pairwise distances of rescaled functions over some frequency interval $[\omega_l,\omega_h]$.
Specifically, for the $p$-rescaled critical spectral functions as depicted in \Cref{fig:p_scaling_functions}, based on \cref{eq:rho_scale_p}, this amounts to minimizing the quantity
\begin{equation}
    \Delta^2(z) = \sum_{p_i}\sum_{p_i < p_j} \int\limits_{\omega_l}^{\omega_h} \intd \omega \frac{\left| p_i^{2-\eta}\rho(p_i^{z} \omega, p_i, 0) - p_j^{2-\eta}\rho(p_j^{z} \omega, p_j, 0) \right|^2}{ \left( p_i^{2-\eta}\Delta\rho(p_i^{z} \omega,p_i, 0) \right)^2 + \left( p_j^{2-\eta}\Delta\rho(p_j^{z} \omega, p_j, 0) \right)^2 }\;,
    \label{eq:z_fdiff}
\end{equation}
where $\Delta \rho$ denotes the statistical error of the measured spectral functions, which is used to weight the deviations.
Similarly, an analogous functional is minimized to optimize the scaling of the $\tau$-rescaled spectral functions at vanishing momentum in \Cref{fig:t_scaling_functions}.

Even though this procedure is in principle very robust, we were not able to completely eliminate the dependence on the upper limit $\omega_h$ of the frequency interval.
Since this dependence is particularly strong for the $\tau$-rescaled spectral functions, we have disregarded them in the final estimate.
By the principle of least sensitivity, i.e.~by looking for a plateau in the results for different $\omega_h$, we can then estimate reasonable values for $\omega_h$ in case of the $p$-rescaled spectral functions, which yield a set of plausible values for the dynamic critical exponents that are shown in \cref{tab:z_results} in the row labeled {\em scaling}.
However, since the systematic uncertainties associated with this procedure are still somewhat uncontrolled, we have also explored two alternative methods to calculate the dynamic critical exponent $z$.

Our second method of extracting the dynamic critical exponent $z$ from our data exploits the large $x_{p}$ and $x_{\tau}$ behavior of the scaling functions $f_p(x_{p}, 0)$ and $f_{\tau}^{\pm}(x_{\tau}, 0)$.
Based on the scaling form of the spectral function in \cref{eq:rho_scale_omega}, we extract the infrared power law (\ref{IR_power_omega}) from either the critical spectral function at $\tau = 0$ and some sufficiently small momentum $p$, or that at zero momentum sufficiently close to criticality.
For example, \cref{eq:rho_scaling_ansatz_p} entails that the frequency dependence of the critical spectral function ($\tau = 0$), at a given fixed value of $p$, is of the form
\begin{equation}
    \rho(\omega) = \frac{1}{(a\omega)^{-1} + b \omega^\sigma } \, .\label{eq:rho_fit_func}
\end{equation}
Likewise, cf.~\cref{eq:rho_scaling_ansatz_t}, the same form should also describe the frequency dependence of the zero-momentum spectral function at fixed small $\tau\not=0$, with $\sigma = (2-\eta)/z$ in each case.
We note that the power-law amplitude $b = {f_t^+}^\sigma/f_{\omega}$ is fixed by the constant $f_\omega$ governing the regular behavior of $f_{\omega}(x_{\omega}, y_\omega )$ near the origin at $x_\omega=y_\omega=0$, whereas the parameter $a$ is related to the amplitudes governing the leading small $x_p$ resp.~$x_{\tau}$ behaviour of $f_p(x_p,0)$ resp.~$f_{\tau}(x_{\tau},0)$ via the relations
\begin{align}
    a_p &= a \left( \bar p_{\text{fit}} \right)^{z+2-\eta}, \label{eq:ap_from_a} \\
    a_{\tau}^{\pm} &= a \left|\tau_{\text{fit}}\right|^{\nu z+\gamma}, \label{eq:at_from_a}
\end{align}
where $\bar p_{\text{fit}},\, \tau_{\text{fit}}$ designate the spatial momentum resp.~reduced temperature where the fit was performed.

By fitting this ansatz with the amplitudes $a,\,b$ and the exponent $\sigma$ as free parameters to the critical spectral functions ($\tau=0$) at fixed momentum we obtain reasonably stable results for the dynamic critical exponent $z$, which are listed in the second row of \cref{tab:z_results} with the label {\em IR power law}.
In practice, we simultaneously fit the spectral functions for the smallest two ($d=3$) or three ($d=2$) spatial momentum indices to improve statistics.
Unfortunately, in $d=3$ even for our largest possible lattices we can not reach small enough spatial momenta $p$ to obtain a reasonably well constrained signal for $z$, especially in Model C.

In order to estimate the uncertainty in $z$, we vary the upper limit of the frequency interval where we fit \cref{eq:rho_fit_func} to the data in a sensible range.
We eliminate one third of the results with the largest $\chi^2/$d.o.f..
Of the remaining results, we take the highest and lowest values for $z$ as bounds on the confidence interval, and calculate the average of $z$ weighted with the statistical uncertainty.
If the weighted average of $z$ is not near the center of the confidence interval, we separately note the uncertainties in both directions.

If one naively repeats this process for the zero-momentum spectral functions at small $\tau$, one arrives at implausible values of $z$, which drift towards the result at finite spatial momentum upon approaching the critical point at $\tau=0$.
This is due to the fact that, as can be seen in \cref{fig:t_scaling_functions}, the universal scaling function $f_{\tau}^{\pm}$ deviates from the asymptotic power law for intermediate $x$, but converges to it for large $x$.
We take the convergence of $f^{\pm}_{\tau}$ for large $x$ to the same power law that describes $f_p(x, 0)$ for $x \gtrsim 1$ as an indicator that this power law describes the true asymptotic behaviour of the universal scaling functions.
Nevertheless, since $f_p$ converges much earlier, we believe that the results for $z$ from fits to the spectral function at $\tau = 0$ and small momentum $p$ are much more reliable.

\begin{figure}[t]
    \graphicspath{{\fdir}}
    \begin{minipage}[t]{.5\linewidth} 
        \includegraphics{\fdir/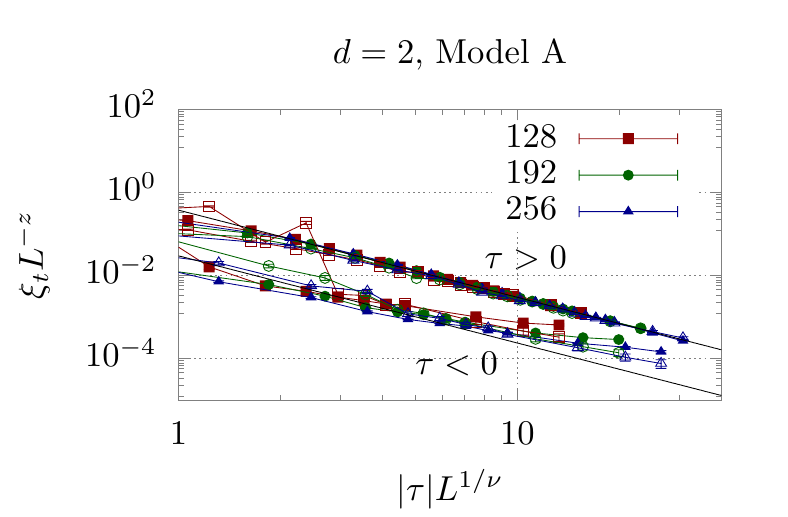}
    \end{minipage}
    \begin{minipage}[t]{.5\linewidth} 
        \includegraphics{\fdir/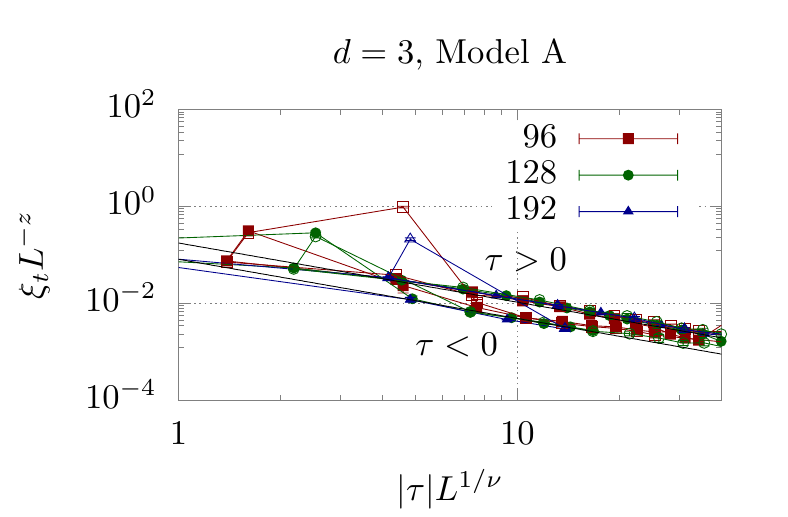}
    \end{minipage}
    \begin{minipage}[t]{.5\linewidth} 
        \includegraphics{\fdir/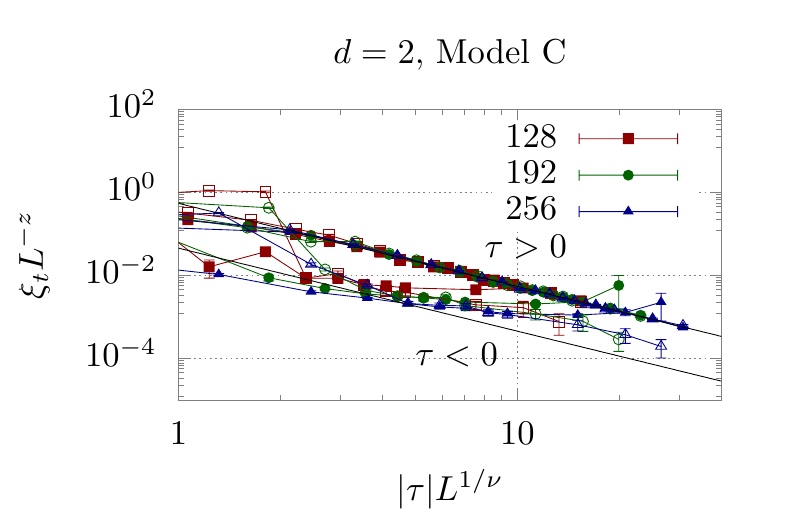}
    \end{minipage}
    \begin{minipage}[t]{.5\linewidth} 
        \includegraphics{\fdir/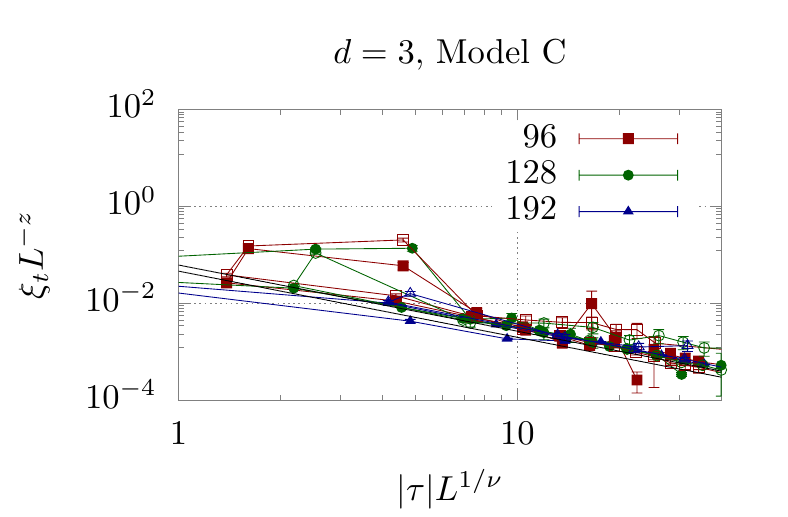}
    \end{minipage}
    \caption{
        Behaviour of the auto-correlation time as a function of the finite-size scaling variable $|\tau|L^{1/\nu}$.
        Different symbols correspond to extractions of the correlation time using fits (filled symbols) and integration (open symbols) of the spectral function; upper curves in each panel correspond to $\xi_t(\tau > 0)$, while lower curves correspond to $\xi_t(\tau<0)$.
        Solid lines in each panel correspond to a power law fit according to \cref{eq:fss_xit_power}, from which we extract the dynamic critical exponent $z$ along with the non-universal amplitudes $f_{t}^{\pm}$.
    }
    \label{fig:xit_tau}
\end{figure}

\begin{figure}[t]
    \graphicspath{{\fdir}}
    \begin{minipage}[t]{.5\linewidth} 
        \includegraphics{\fdir/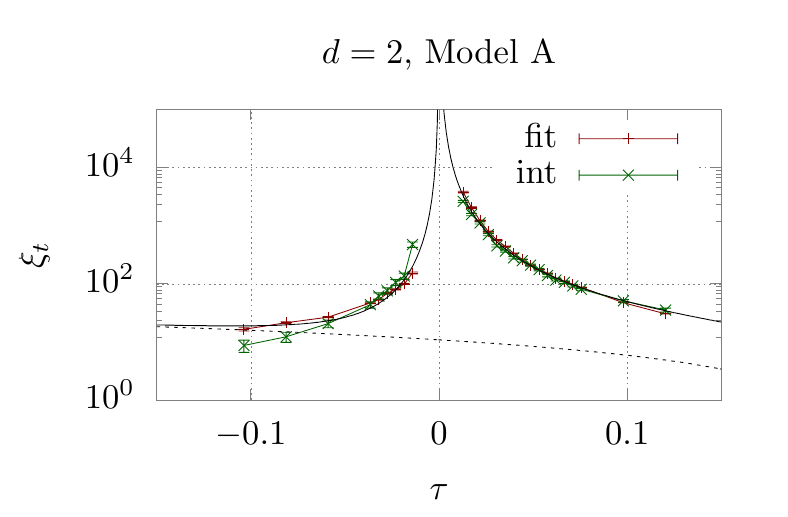}
    \end{minipage}
    \begin{minipage}[t]{.5\linewidth} 
        \includegraphics{\fdir/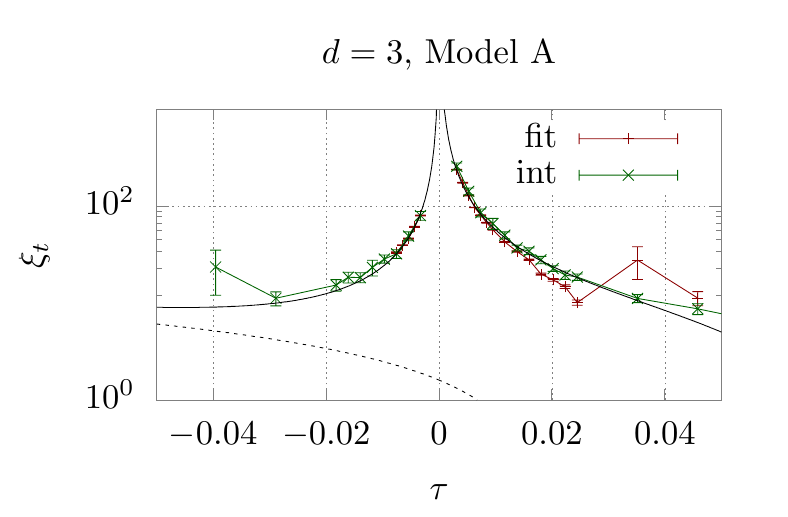}
    \end{minipage}
    \begin{minipage}[t]{.5\linewidth} 
        \includegraphics{\fdir/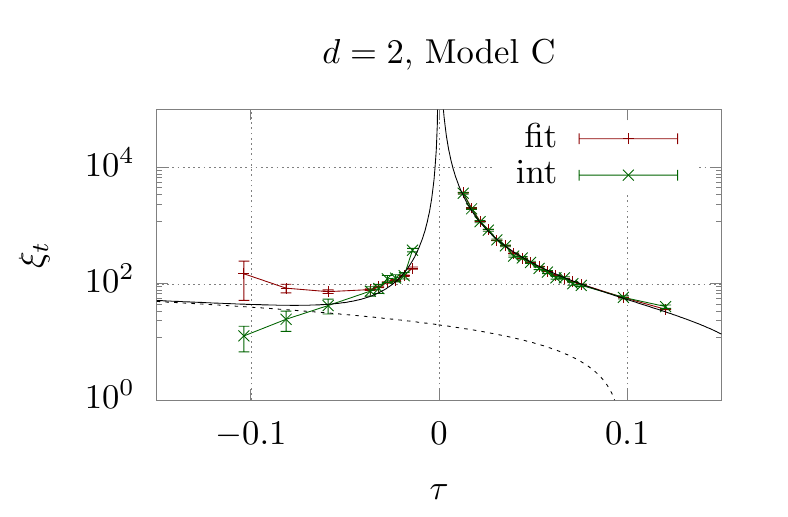}
    \end{minipage}
    \begin{minipage}[t]{.5\linewidth} 
        \includegraphics{\fdir/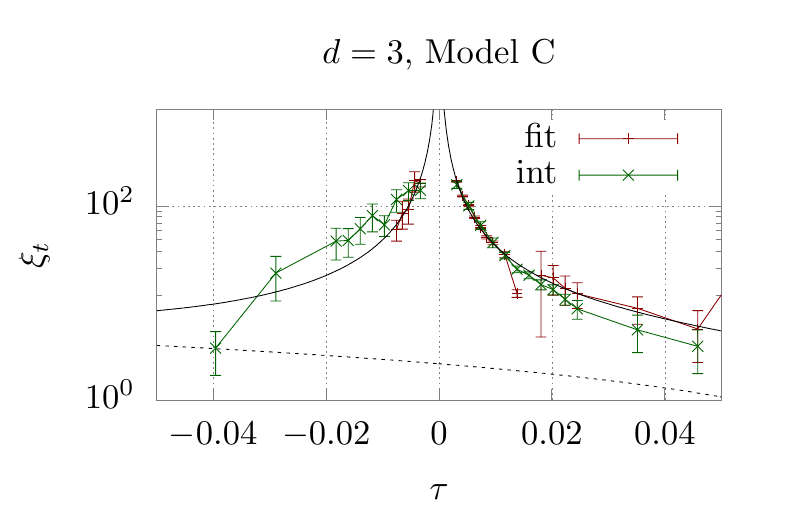}
    \end{minipage}
    \caption{Behaviour of the correlation time $\xi_t(\tau)$ extracted from fits and integration of the spectral function, as a function of reduced temperature $\tau$.
        Note that in order to minimize finite volume effects, we only include results for $|\tau| L^{1/\nu}> 2.5$ for $d=2$,  and $|\tau| L^{1/\nu}> 5$ for $ d=3 $ obtained on $256^2$ and $128^3$ lattices.
        Solid black lines show a fit according to \cref{eq:xit_regular}; the regular contribution is shown separately as a black dashed line.
    }
    \label{fig:xit_lin}
\end{figure}

\subsection{Critical behavior of the auto-correlation time \tops{$\xi_t$}{xi\_t}}
\label{sec:xi_t}

We can also extract the dynamic critical exponent $z$ from the divergence of the auto-correlation time $\xi_t(\tau)$ in the vicinity of the critical point.
By use of \cref{eq:rho_scaling_t} and setting $s=\bar t^{1/z}$, the spectral functions in the time domain satisfy the following scaling relation
\begin{equation}
     \rho \left( t, p, \tau \right) = \bar t^{(2-\eta)/z - 1} \, f_t\Big(\bar p^z \bar t, \tau \,\bar t^{\, 1/\nu z} \Big),
    \label{eq:rho_scale_tt}
\end{equation}
where the presence of the second scaling variable $y = \tau \, \bar t^{\, 1/\nu z}$ explicitly shows that there is a temperature-dependent characteristic time scale $\xi_t \sim |\tau|^{-\nu z}$ associated with the auto-correlation time.
For the sake of simplicity, we focus on the case of vanishing spatial momentum $p=0$,  where the spectral function near criticality at times larger than this characteristic correlation time $(t/\xi_t(\tau) > 1)$, is well described by a product of a power law and an exponential,
\begin{equation}
    \tilde f_\tau^+( \tau^{\nu z} \bar t,0) =  (\tau \, \bar t^{\, 1/\nu z})^{\nu (2-\eta-z)} \,    f_t \left(0, \tau \, \bar t^{\, 1/\nu z} \right) \propto \, \bar t^{\, (2-\eta)/z - 1}\exp\big(-t/\xi_t(\tau)\big)\,,
    \label{eq:rho_scale_t_ansatz}
\end{equation}
as can be seen in \cref{fig:t_scaling_functions_t}.
Near the critical point, the correlation time $\xi_{t}(\tau)$ diverges as the system approaches criticality, and the spectral function then assumes its characteristic power-law form.
Nevertheless, away from criticality the correlations in the system decay exponentially on large timescales $t\gtrsim \xi_t$, and the functional form in Eq.~(\ref{eq:rho_scale_t_ansatz}) can be used to define the auto-correlation time $\xi_t(\tau)$ from fits to numerical data, and infer its non-universal amplitudes $f^{\pm}_t$ and their universal ratio $U_{\xi, t}$ as
\begin{equation}
    \xi_t(\tau) = \xitpm |\tau|^{-\nu z}, \quad U_{\xi, t} \equiv \xitp/f_t^-.
    \label{eq:xi_power_law}
\end{equation}
We further use the results for the auto-correlation time $\xi_t(\tau)$ in the high-temperature phase $(\tau>0)$ to define $\bar t = t/f_t^+$ as our dimensionless time variable when presenting spectral functions in the time domain, such that for $\tau>0$ the scaling variable $\tau^{\nu z} \bar t$ employed in \cref{fig:t_scaling_functions_t} corresponds directly to the ratio $t/\xi_t$.
Besides extracting the auto-correlation time $\xi_t(\tau)$ from a fit to the ansatz in \cref{eq:rho_scale_t_ansatz}, it is also possible to extract $\xi_t(\tau)$ by integrating the spectral functions in the time domain, as explained in detail in Appendix~\ref{sec:integration}.

Before we present our numerical results, some further remarks are in order.
Since we focus on the behavior of modes with vanishing spatial momentum $p=0$, their auto-correlation time $\xi_{t}(\tau)$ diverges at the critical point as $\xi_{t}(\tau) \sim 1/|\tau|^{\nu z}$ an infinite system.
However, in our simulations the divergence of the auto-correlation time is limited by the finite system size $L$, which in the immediate vicinity of the critical point $\tau \approx 0$ becomes the relevant infrared cut-off.
Based on the dynamical finite-size scaling hypothesis, one expects that in this regime the auto-correlation time behaves as
\begin{eqnarray}
    \xi_{t}(\tau,L) = L^{z} g_{\xi}(\tau L^{1/\nu}),
\end{eqnarray}
where $g_{\xi}(x)$ is the finite-size scaling function of the auto-correlation time analogous to the ratio  $R = \xi/L$ used for static finite-size scaling. 
While for all finite values of the finite-size scaling variable $x=\tau L^{1/\nu}$ the divergence of the auto-correlation time is effectively regulated by the finite volume, in order to recover the infinite-volume scaling in \cref{eq:xi_power_law}, one needs asymptotically large values of $\pm x$, 
where this finite-size scaling function satisfies 
\begin{equation}
    g_{\xi}(x\to \pm\infty) \to \xitpm  |x|^{-\nu z}, \label{eq:fss_xit_power}
\end{equation}

 We present our results for the auto-correlation time in \cref{fig:xit_tau,fig:xit_lin}, where we study the dependence of $\xi_{t}(\tau,L)L^{-z}$ as a function of the finite-size scaling variable $\tau L^{1/\nu}$ in \cref{fig:xit_tau} and subsequently estimate the magnitude of singular and regular contributions to the correlation length $\xi_{t}(\tau)$ in \cref{fig:xit_tau}.
Generally, the results of the two different extraction methods (fit and integration) agree very well with each other, although for $\tau<0$ the integrated $\xi_t$ are generally somewhat closer to the power law, and the slope of this power law fit produces slightly smaller results for $z$ in Model A.
Strikingly, one also observes from \cref{fig:xit_tau}  that the data exhibits a clear finite-size scaling across different lattice sizes, which we can exploit to extract the dynamic critical exponent as explained in the following.

In order to obtain the dynamic critical exponent $z$ from the correlation times, we first apply finite-size scaling with a plausible estimate for $z$, to find a region where the data for different lattice sizes shows sufficient overlap.
Based on the results depicted in \cref{fig:xit_tau} it becomes obvious that this hardly works at $\tau<0$, but gives a clear power law at large values of $\tau L^{1/\nu}$, for $\tau>0$.
We then fit the power law in \cref{eq:fss_xit_power} for $\tau>0$ to the un-scaled data in the selected region, to get both the amplitude $\xitp$ and the exponent $\nu z$.
Subsequently, we estimate the amplitude ratio $U_{\xi,t}$ as far as possible by fitting a power law with the exponent obtained earlier to a few data points with $\tau<0$.
Errors are obtained in a similar way as for the power law fit of the IR divergence of the spectral function.
By varying the temperature interval where we fit the power law to the correlations times $\xi_t(\tau)$ and keeping two-thirds of the results with the lowest $\chi^2/\text{d.o.f.}$ as well as eliminating outliers with $\chi^2 > 2\chi^2_{\text{min}}$, we compute the averages weighted by statistical uncertainties and estimate the confidence interval by taking the highest and lowest values for $\xi_t,\, \xitp$ and $U_{\xi_t}$.
We remark that these parameters are strongly correlated, so a large uncertainty in the dynamic critical exponent $z$ leads to large uncertainties in both $\xitp$ and $U_{\xi_t}$.

The results of this procedure for the non-universal amplitude $\xitp$ and the ratio $U_{\xi,t}$ are given in \cref{tab:xit-amp}. Those for the dynamic critical exponent $z$ are shown in the row denoted by ``$\xi_t$ power law'' of \cref{tab:z_results}.
Especially the two large amplitude ratios $U_{\xi,t}$ in $d=2$ seem quite remarkable when compared to the analytically known amplitude ratio of the spatial correlation length $U_{\xi} = 2$~\cite{pelissetto_critical_2002}.
Although the $d=2$ data for $\xi_{t}$ below the critical temperature $(\tau<0)$ does not necessarily justify a power law fit all that well, by looking at \cref{fig:xit_tau} one is led to conclude that we might rather underestimate this ratio.

While in \cref{fig:xit_tau} the data at least in 2+1D perfectly fits a power law above the critical temperature $\tau > 0$, we find that a precise extraction of $z$ remains difficult with the available data.
In order to improve the accuracy, one could generate data closer to $\tau \gtrsim 0$ in large volumes to minimize finite-size effects.
Below the transition temperature in both 2+1D and 3+1D, we find that the data for the correlation time $\xi_t$ deviates strongly  from the expected power law behaviour.
One reason for this is the (much) smaller value of the non-universal amplitudes $f_t^-$, which in combination with relatively large regular contributions leads to a suppression of the critical signal.
We try to capture the regular contributions by fitting a regular function up to linear order in addition to the power law
\begin{equation}
    \xi_t(\tau) = \xitpm |\tau|^{-\nu z} + f_{r,0} + f_{r,1} \cdot \tau
    \label{eq:xit_regular}
\end{equation}
The comparison between the resulting fit and the data is shown in \cref{fig:xit_lin}.
The fit now also describes the data away from $\tau=0$ much better, which is dominated by the regular part, shown in \cref{fig:xit_lin} as a dashed line, especially on the low temperature side.
However, by introducing these additional degrees of freedom in the fit, we lose precision in the estimate of the dynamic critical exponent $z$, both in terms of statistical and systematic uncertainties.

\begin{table}
    \centering
    \caption{Non-universal amplitude $\xitp$ and universal amplitude ratio $U_{\xi,t}$ of the correlation time $\xi_t$, obtained by fits of the data to \cref{eq:xi_power_law}, shown in \cref{fig:xit_tau}.
        Since the extraction of the amplitudes is strongly correlated with the extraction of the exponent, the uncertainties are rather large.
    }
    \label{tab:xit-amp}
    \begin{tabular}{r || l l | l l}
        & \multicolumn{2}{c}{2D} & \multicolumn{2}{c}{3D}\\
        & Model A & Model C & Model A & Model C \\
        \hline\hline 
        \rule{0pt}{2.2ex}\rule[-1.2ex]{0pt}{0pt}$\xitp$ & $0.43_{-0.04}^{+0.10}$ & $0.54_{-0.11}^{+0.07}$ & $0.08_{-0.04}^{+0.11}$ & 0.028(6)\\
        \rule{0pt}{1.2ex}\rule[-1.2ex]{0pt}{0pt}$U_{\xi,t}$ & $8.2_{-0.9}^{+4.2}$ & 9.0(2.7) & $2.7_{-0.3}^{+0.6}$ & $1.2_{-0.2}^{+0.5}$\\
     \end{tabular}
\end{table}

\begin{figure}
    \graphicspath{{\fdir}}
    \begin{minipage}[t]{.5\linewidth}
        \includegraphics{\fdir/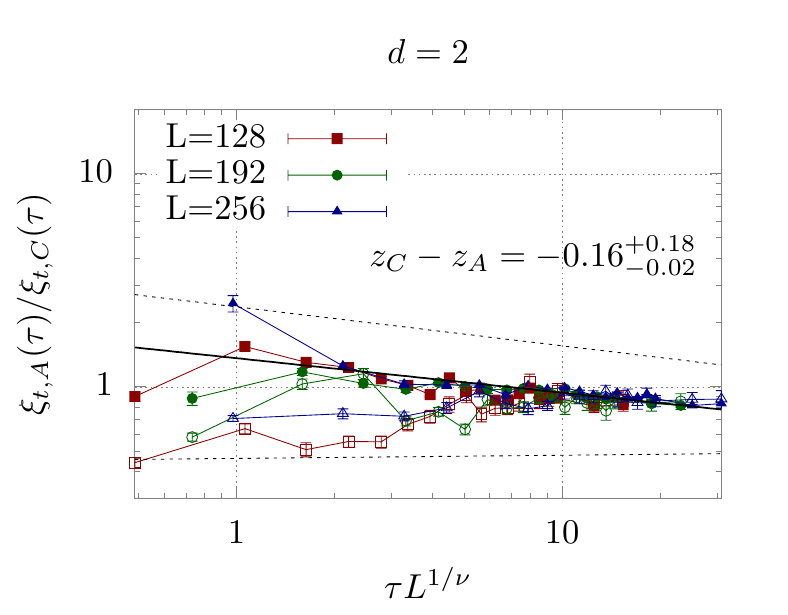}
    \end{minipage}
    \begin{minipage}[t]{.5\linewidth}
        \includegraphics{\fdir/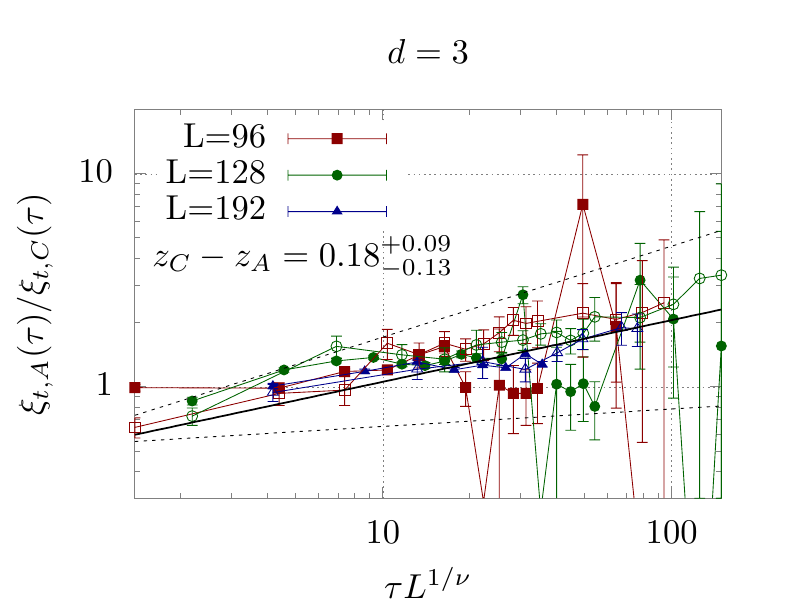}
    \end{minipage}
    \caption{
        Dependence of the ratio of auto-correlation times in Models A and C, on the finite size scaling variable $\tau L^{1/\nu}$.
        Solid lines in the left and right panels show a power law fit of the large $\tau L^{1/\nu}$ behavior, from which we extract the difference $\nu (z_{C}-z_{A})$ of the dynamic critical exponents (cf.~\cref{eq:diff_zazc}).
        Dashed lines indicate the confidence interval of the extraction of $\nu (z_{C}-z_{A})$, which is also presented in the figure.
    }
    \label{fig:xit_ratio_ac}
\end{figure}

Besides providing an alternative means to extract $z$, one additional advantage of the auto-correlation time method is that it allows for a direct comparison of the critical dynamics of different models.
In particular, to estimate the \emph{difference} between the dynamic critical exponents $z$ of Models A and C, one can look at the ratio of the correlation times at the same (reduced) temperature, which satisfy
\begin{equation}
  \frac{\xi_{t,A}(\tau)}{\xi_{t,C}(\tau)} \overset{\tau>0}{=}
    \frac{f^+_{t,A}}{f^+_{t,C}} \cdot \tau^{-\nu\left( z_A - z_C \right)}
    \label{eq:diff_zazc}
\end{equation}
in the infinite volume limit.
Such a direct comparison between Models A and C is presented in \Cref{fig:xit_ratio_ac}, where we show the ratio of the correlation lengths in \cref{eq:diff_zazc} for the symmetric phase ($\tau>0)$ as a function of the finite-size scaling variable $\tau L^{1/\nu}$.
Even though this ratio reveals some tension between the two different extraction methods (exponential fit and integration), the general trends are clearly visible, where in $d=2$ dimensions, $z_C > z_A$, and the power law at large $\tau L^{1/\nu}$  slopes downwards; while in $d=3$ dimensions, the difference changes sign $z_C < z_A$ and the slope of the power law is positive.
By performing a power law fit to the ratio, we can obtain a direct estimate of $z_C-z_A$, which is also indicated in \Cref{fig:xit_ratio_ac}, with the quoted errors obtained in the same way as for the power law fit of the correlation times.

\begin{figure}[b]
    \graphicspath{{\fdir}}
    \begin{minipage}[t]{.5\linewidth}
        \includegraphics{\fdir/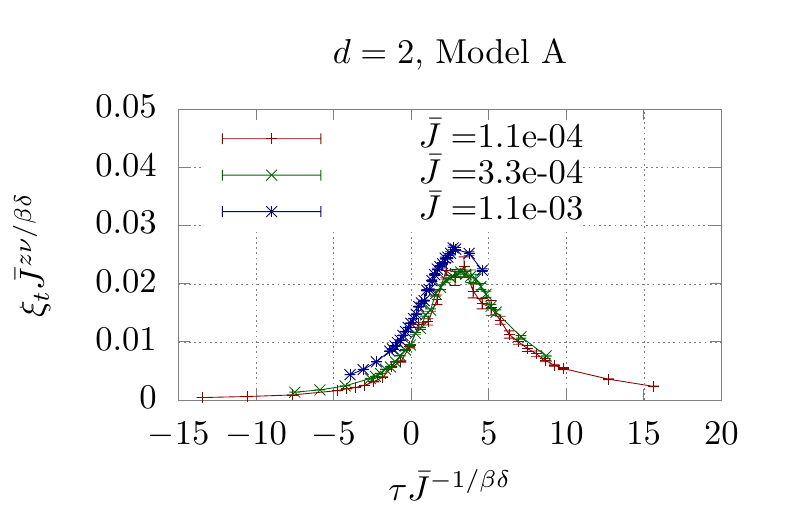}
    \end{minipage}
    \begin{minipage}[t]{.5\linewidth}
        \includegraphics{\fdir/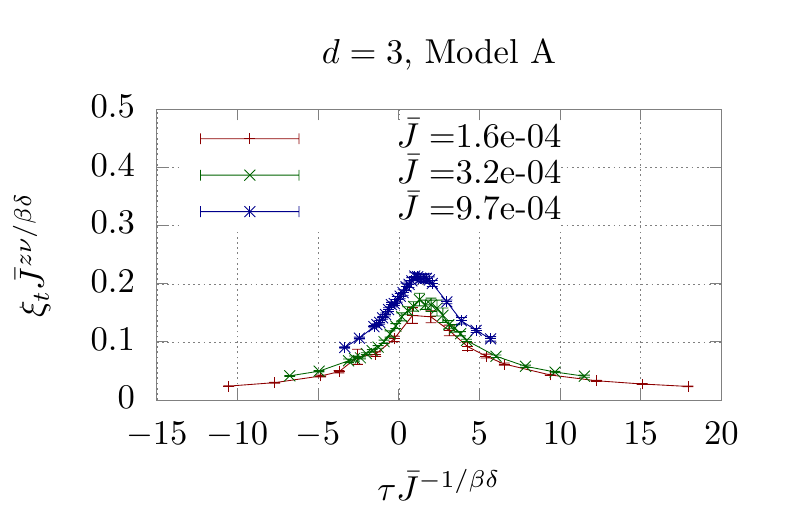}
    \end{minipage}
    \caption{Magnetic scaling of the correlation time $\xi_t(\tau, J)$.
        Overlapping data points correspond to the universal magnetic scaling function of the auto-correlation times $\hat {\xi}_{t,J}(\tau/|\bar J|^{1/\beta\delta})$ (cf.~\cref{eq:magnetic_scaling}).
        Scaling breaks down when $\bar J$ becomes too large, as indicated by the blue data points.
}
    \label{fig:mag_scale}
\end{figure}

So far we have only considered the critical behaviour of the spectral function in absence of explicit symmetry breaking ($J=0$).
When introducing a non-zero explicit symmetry breaking ($J \neq 0$), the magnetic scaling hypothesis states that the singular part of the free energy density can be written as
\begin{eqnarray}
f_{\text{sing}}(\tau,\bar J)=s^{-d} f_{s}(s^{1/\nu} \tau,s^{\beta \delta/\nu} \bar J)
\end{eqnarray}
allowing one to express the singular behavior of the free energy in terms of scaling functions $\hat{f}_{\pm}$, $f_{J}$ (note that $\nu d = 2-\alpha$)
\begin{align}
 f_{\text{sing}}\left( \tau, \bar J \right) &= |\tau|^{2-\alpha}\hat{f}_{\pm}\left( \bar J / |\tau|^{\beta \delta} \right), \\
 f_{\text{sing}}\left( \tau, \bar J \right) &= |\bar J|^{(2-\alpha)/\beta \delta}\hat{f}_{J}\left(\tau/ |\bar J|^{1/\beta \delta} \right),
\end{align}
which depend on a single variable $\tau/|\bar J|^{1/\beta \delta}$ that combines the $\tau$ and $\bar J$ dependence.
Similarly, one expects an analogous magnetic scaling behaviour for unequal-time correlation functions, such that e.g.~the auto-correlation time $\xi_t$ is expected to obey the scaling form
\begin{equation}
    \xi_t(\tau, \bar J) =  s^{z} \xi_{t}(s^{1/\nu} \tau,s^{\beta \delta/\nu} \bar J),
    \label{eq:magnetic_scaling}
\end{equation}
which by analogy allows to define the magnetic scaling function 
\begin{eqnarray}
    \xi_t(\tau, \bar J) = |\bar J|^{-z \nu /\beta\delta} \hat{\xi}_{t,J}(\tau/|\bar J|^{1/\beta\delta}).
\end{eqnarray}
We investigate this behavior in \cref{fig:mag_scale}, where the auto-correlation time $\xi_{t}(\tau,\bar J)$ obtained by integration for different values of $\bar J$ is rescaled to recover the underlying magnetic scaling function $\hat{\xi}_{t,J}$ as a function of $\tau/|\bar J|^{1/\beta\delta}$.
Data points for the two smaller values of $\bar J$ largely overlap confirming the magnetic scaling of the auto-correlation time.
Even though the magnetic scaling starts to break down for the larger $\bar J$, as one departs from the critical region, it is clear that in a certain (model dependent) range of $\bar J$ and $\tau$, one can extrapolate $\xi_t(J,\tau)$ from the overlap region in \cref{fig:mag_scale}.

\begin{table}
    \centering
    \caption{
        Extracted values for the dynamic critical exponent $z$.
        See \cref{sec:discussion} for detailed discussion of the different extraction methods.
    }
    \label{tab:z_results}
    \begin{tabular}{r || l l | l l}
        & \multicolumn{2}{c}{2D} & \multicolumn{2}{c}{3D}\\
        \multicolumn{1}{c||}{$z$} & Model A & Model C & Model A & Model C \\
        \hline\hline
        scaling & 2.03(9) & 2.00(8) & 1.83(17) & 2.13(16)\\
        \rule{0pt}{1.2ex}\rule[-1.2ex]{0pt}{0pt} IR power law & 2.24(8) & $1.98_{-0.21}^{+0.12}$ & $1.88_{-0.27}^{+0.10}$ & $2.47_{-0.25}^{+0.10}$\\
        \rule{0pt}{1.2ex}\rule[-1.2ex]{0pt}{0pt} $\xi_t$ power law & $2.03_{-0.08}^{+0.05}$ & $2.00_{-0.04}^{+0.07}$ & $2.17_{-0.18}^{+0.31}$ & $2.48_{-0.07}^{+0.10}$\\
        \hline
        combined & 2.10(4) & 2.00(5) & 1.92(11) & 2.41(7)\\
     \end{tabular}
\end{table}

\subsection{Discussion of results for the dynamic critical exponent \tops{$z$}{z}}\label{sec:discussion}
We provide a summary of our results for the dynamic critical exponent $z$ in \cref{tab:z_results}, where we give the results from the three different extractions, along with an error-weighted combined average.

For Model A in 2+1D, the overlap method and the power law fit to the correlation times $\xi_t$ give a surprisingly small result for $z$ with a relatively large error.
The power law fit to the infrared divergence of the spectral function yields a result closer to the one of \cite{nightingale_dynamic_1996}, albeit even a bit larger.
Combining the results leads to a $z$ that is closer to, but still smaller than the result of \cite{nightingale_dynamic_1996}. Incidently, our combined result for Model A in 2+1D is fully consistent with the experimentally measured value of $z = 2.09(6)$  from Ref.~\cite{dunlavy_critical_2005}.
For Model C in 2+1D, we find a value that closely matches the analytic result $z=2$, with all methods agreeing within their respective statistical errors.

In 3+1D, the non-critical effects on the spectral function are strong, leading to very large uncertainties in the power law fits.
Since the critical window is smaller than in 2+1D, the uncertainty of overlap method increases as well.
Nevertheless, the combined results in 3+1D for both Model A and Model C are compatible with earlier studies, and with the value pedicted from the scaling relation $z=2+\alpha/\nu$ for Model C.

We note that, in order to obtain more precise results, it would be highly beneficial to consider an improved action and/or larger volumes at temperatures closer to the critical point.
In 2+1D with larger volumes, one could probe smaller spatial momenta at $\tau=0$ and increase the precision of the IR power law method.
When considering 3+1D, our results in \cref{fig:t_scaling_functions} compared to those in \cref{fig:p_scaling_functions} clearly indicate that, with our present setup, we were not able to probe low-enough momentum regimes away from the critical temperature $T\approx T_c$.
In both cases we are limited by small critical amplitudes of the correlation time $\xi_t$, especially below the critical temperature $(\tau < 0)$.
However, this study was not intended to achieve high-precision results for $z$, but to test the framework and lay the ground for upcoming non-equilibrium studies.

\section{Conclusion and outlook}\label{sec:conclusion}
By using the classical field approximation close to a second order phase transition, we performed a first principles calculation of the spectral functions of the relativistic  $Z_2$ model.
By including finite spatial momenta in our analysis, we could provide a comprehensive overview of the behavior of the spectral function in all distinct parts of the phase diagram at finite temperature.
We found that, in the symmetric phase, the spectral function is sufficiently well described by a Breit-Wigner quasi-particle shape, with a relativistic dispersion relation and a weakly momentum-dependent decay rate.
When the  $Z_2$ symmetry is spontaneously broken below $T_c$, we find an additional excitation at low frequencies, with a different spectral shape and dispersion indicative of a soft collective mode, cf.~Appendix \ref{sec:capwav}.
While for sufficiently large spatial momenta, the quasi particle peak behaves continuously across the transition, we found clear indications that it is the second low-frequency mode that transforms into the dominant IR divergence at the critical point.
Since our results have been obtained from first-principles numerical studies in the classical-statistical limit, it may be insightful to compare these results quantitatively to others obtained e.g.~by the use of functional methods in the future.

By examining the static critical behaviour of our model, we have set the scales for the main focus of this study, which has been the quantification of dynamic critical effects.
By analyzing the behavior of the spectral function of the order parameter in the vicinity of the phase transition, we explicitly verified the dynamic scaling hypothesis, and performed a detailed analysis of the scaling properties as a function of frequency, momentum and reduced temperature.
We successfully extracted the corresponding universal scaling functions, which describe the infrared properties of the spectral function of the order parameter, and carefully assessed the implications for the behavior of the spectral function at small frequencies and momenta.
We further developed a complete parametrization of the scaling functions in the special cases where either the spatial momentum or the reduced temperature vanishes.

We also analyzed the divergence of the correlation time $\xi_t$ in the vicinity of the critical point, by performing two different extractions of this quantity, and demonstrated its finite-size and magnetic scaling properties in the vicinity of the phase transition.
By modifying the classical equations of motion to introduce a coupling to a heat bath, we were able to simultaneously study the behavior in the dynamic universality classes of Models A and C.
While away from criticality, the additional coupling to the heat bath in Model A only gives rise to minor changes of the spectral function, we found that it leads to a change in the dynamic critical exponent $z$, as is expected due to the change in conservation laws.

We extracted the dynamic critical exponent $z$ for a total of four different scenarios, corresponding to Hamiltonian (Model C) and Langevin (Model A) dynamics in 2+1D and 3+1D systems.
Clearly, calculating the dynamic exponent $z$ proved to be more of a challenge than expected, as e.g.~deviations of the universal scaling function of the spectral function from their asymptotic behaviour give rise to sizeable systematic uncertainties in the corresponding scaling analysis, and also prohibit a precise extraction of $z$ from fitting the asymptotic power law dependence of the scaling functions.
Similarly, the extraction of the correlation times $\xi_t$ itself is difficult, and small amplitudes hinder a precise determination of the exponent $z$.
Nevertheless, by combining the results of the three different methods to assess the systematic uncertainties, we arrive at plausible values, which are largely compatible with earlier studies and analytic results.
In case of 2D Model A, we find our result closer to the one from experiment \cite{dunlavy_critical_2005} than the Monte Carlo result \cite{nightingale_dynamic_1996}, but not by a significant margin.
For 2D Model C as well as 3D Models A and C, our results do not deviate significantly from scaling relations and earlier results.
In order to achieve a more precise characterization of the critical dynamics based on Monte-Carlo studies, one might consider to explore optimized actions specifically to increase the amplitude $\xitp$ in 3+1 dimensions, for example.

By virtue of magnetic scaling, we are further able to predict the dynamic critical properties of the system in a certain radius around the critical point, which provides the baseline for studying possible signatures of criticality for non-equilibrium systems which approach the critical point in the phase diagram.
Indeed, one particularly appealing feature of the classical-statistical setup is that it can be extended to the study of (non-equilibrium) dynamical phase transitions.
Although the classical-statistical field theory generally exhibits a Rayleigh-Jeans divergence, such that non-universal quantities become dependent on the ultraviolet regularization of the model, the universal critical dynamics is correctly captured and the singular critical contributions to all non-universal parameters such as transport coefficients and auto-correlation times can in principle be calculated to arbitrarily high precision.
Moreover, systematic extensions beyond the classical-statistical limit are possible and currently under development.
The Gaussian state approximation \cite{buividovich_real-time_2018}, for example, takes into account quantum corrections in the spirit of time-dependent mean-field approximations to the dynamics, however at the expense of significantly increased numerical costs.
We believe that such classical-statistical simulations of non-equilibrium phase transitions can provide an important benchmark to different macroscopic descriptions, currently being developed in the context of the QCD critical point search \cite{Mukherjee:2016kyu,Sakaida:2017rtj,Nahrgang:2018afz,An:2019csj,Du:2020bxp,bluhm_dynamics_2020}, and intend to return to this problem in a future publication.

Besides applications to non-equilibrium phase transitions, it should also be possible, with rather limited modifications of the simulation setup, to extend this study to systems with a conserved order parameter (Models B and D) or systems with continuous symmetries such as the relativistic O(4)-model (Model G), to arrive at a more systematic characterization of the dynamic scaling behavior in the vicinity of a critical point.

\section*{Acknowledgements}
We thank G.~D.~Moore, F.~Karsch, O.~Kaczmarek, D.~Mesterh\'azy, and D.~Smith for valuable discussions.
This work was supported by the Deutsche Forschungsgemeinschaft (DFG, German Research Foundation) through the CRC-TR 211 ``Strong-interaction matter under extreme conditions'' – project number 315477589–TRR 211.

\begin{appendices}
\section{Integrating the spectral function}\label{sec:integration}
\begin{figure}
    \graphicspath{{\fdir}}
    \begin{minipage}[t]{.5 \linewidth}
        \includegraphics{\fdir/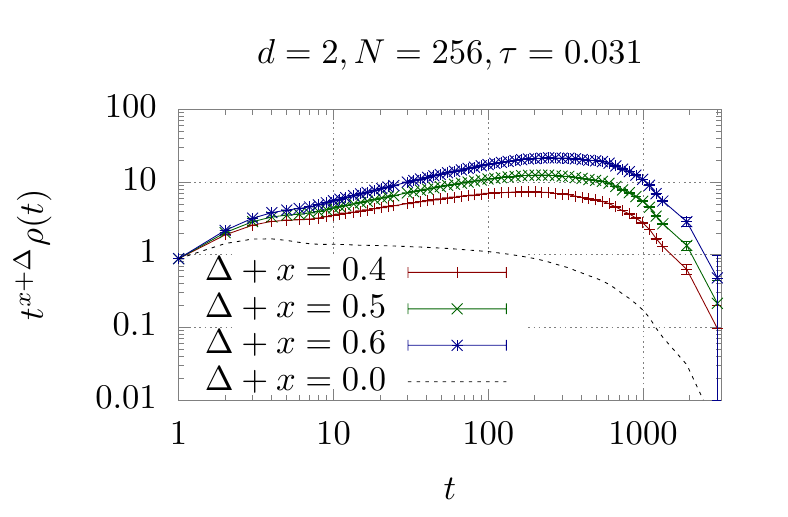}
    \end{minipage}
    \begin{minipage}[t]{.5 \linewidth}
        \includegraphics{\fdir/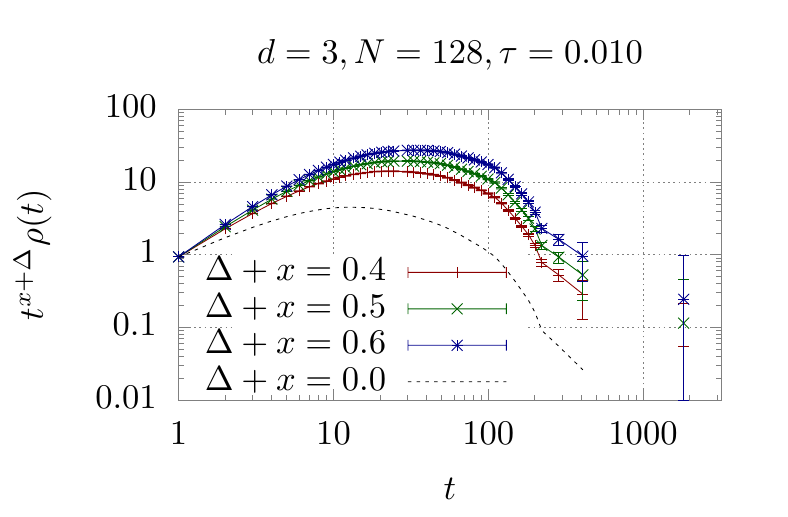}
    \end{minipage}
    \caption{Integrands of \cref{eq:intxit_formula}, in this case Model A spectral functions.
    Clearly, the late-time tail is amplified in comparison to early fluctuations.}
    \label{fig:integratios}
\end{figure}

In the ordered phase, at low temperatures $\tau<0$, the least-squares fit of the ansatz in \eqref{eq:rhofit_t} does not produce satisfying results.
This might be due to strong non-critical oscillations at rather early times.
We try to remove them by integration using the following procedure.

If one integrates the ansatz 
\begin{equation}
    \rho(t) = C t^\zeta \exp(-t/\xi_t),
    \label{eq:rhofit_t}
\end{equation}
one finds by definition of the upper incomplete Gamma function:
\begin{equation}
    \int\limits_{1/\Lambda}^{\infty} t^\zeta \exp \left( \frac{-t}{\xi_t} \right) \intd t = \xi_t^{1+\zeta} \Gamma\left( 1+\zeta, \frac{1}{\xi_t \Lambda} \right).
    \label{eq:integral}
\end{equation}

The result unfortunately is not exactly proportional to the correlation time $\xi_t$, but to a power thereof.
However, if take a ratio of integrals, where we modify the power of the time variable $t$ in one integrand, we get a result with the desired proportionality:
\begin{equation}
    \left(\frac{\int\limits_{1/\Lambda}^{\infty} t^{\zeta + x + \Delta}\exp \left( \frac{-t}{\xi_t} \right) \intd t}{\int\limits_{1/\Lambda}^{\infty} t^{\zeta + x} \exp \left( \frac{-t}{\xi_t} \right) \intd t}\right)^{1/\Delta} \propto \left(\frac{\xi_t^{1+ \zeta + x + \Delta}}{\xi_t^{1+ \zeta + x}} \right)^{1/\Delta} = \xi_t
    \label{eq:intxit_formula}
\end{equation}
By introducing extra exponents $\Delta,x>0$, we allow for emphasizing the long-time exponential over the early, non-critical fluctuations.
This is illustrated in \cref{fig:integratios}: The additional power law shifts the weight of the integrand towards the tail of the exponential, to time scales of the correlation time $\xi_t$.

The proportionality factor then is a power of a ratio of incomplete Gamma functions, which rather strongly depends on the exponent $\zeta$.
Since we cannot calculate this ratio explicitly from the data, we have to limit this method to regimes where we can assume $\zeta$ to be a constant.

\section{Structure of the spectral function at low temperatures}\label{sec:capwav}
In the ordered phase, at $\tau <0$, the spectral functions shows not only a quasi-particle structure, but also a second mode at lower frequencies.
If one plots the spectral function over the squares of spatial momentum and frequency, it becomes apparent that the two structures are separated by the light cone, i.e.~the low-frequency excitation inhabits the space-like region.
Tracking the maximum $\rho(\omega_\mathrm{max},p) = \max \left.\rho(\omega,p)\right|_{\omega^2<p^2}$ of the spectral function in that region, one finds for the dispersion relation roughly a power-law behavior of the form, 
\begin{equation}
    \omega_\mathrm{max}^2 \, \propto \, p^d,
    \label{eq:lowfreq-dispersion}
\end{equation}
in $d$ spatial dimensions. For $d=3$ this agrees with the well-known dispersion relation of thermally driven capillary waves. In $d=2$ spatial dimensions the situation seems less clear. In Ref.~\cite{JChemPhys132} for example, the excitation specrta of two-dimensional fluid droplets have been studied with resulting dispersion relations that depend on the details of the fluid parameters.  

\begin{figure}
    \graphicspath{{\fdir}}
    \begin{minipage}[t]{.50\linewidth}
        \centering{$d=2$}
    \end{minipage}
    \begin{minipage}[t]{.50\linewidth}
        \centering{$d=3$}
    \end{minipage}
    \vspace{-1.2cm}
 
    \begin{minipage}[t]{.5\linewidth} \includegraphics{\fdir/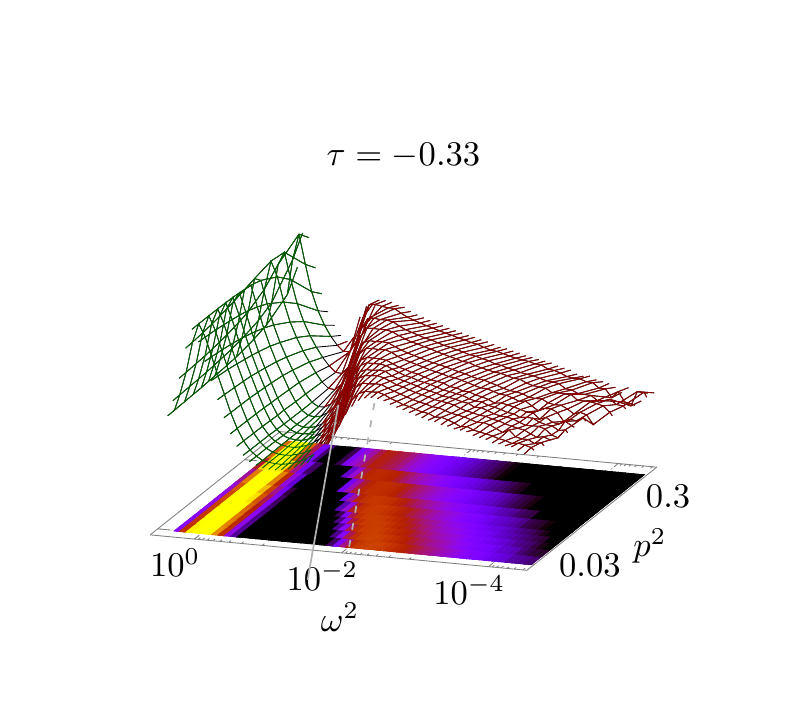} \end{minipage}
    \begin{minipage}[t]{.5\linewidth} \includegraphics{\fdir/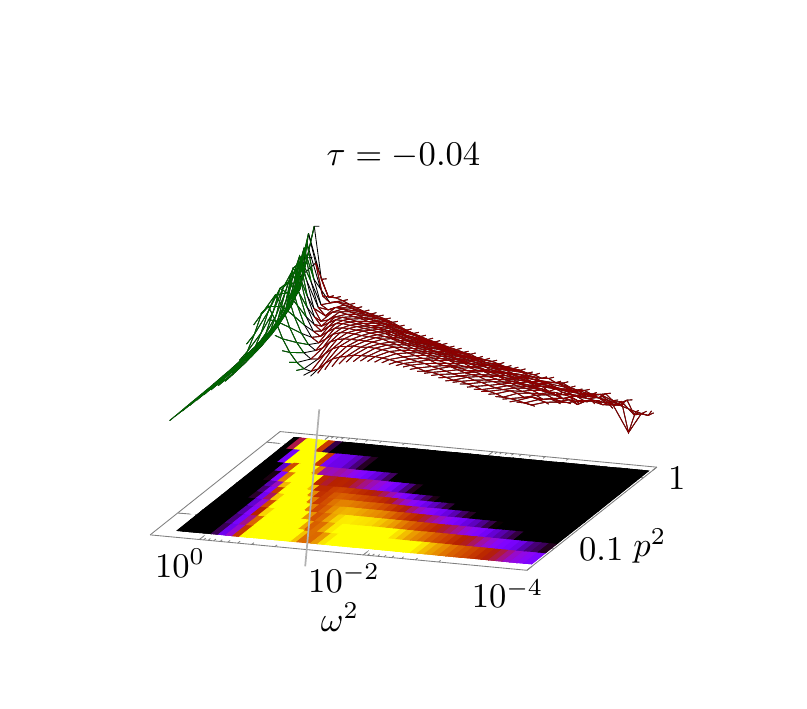} \end{minipage}
    \begin{minipage}[t]{.5\linewidth} \includegraphics{\fdir/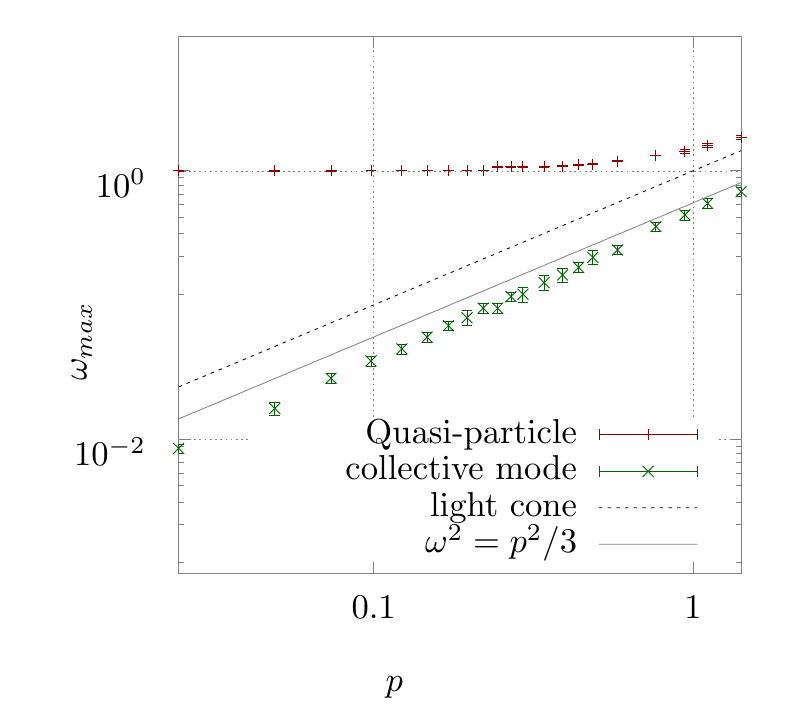} \end{minipage}
    \begin{minipage}[t]{.5\linewidth} \includegraphics{\fdir/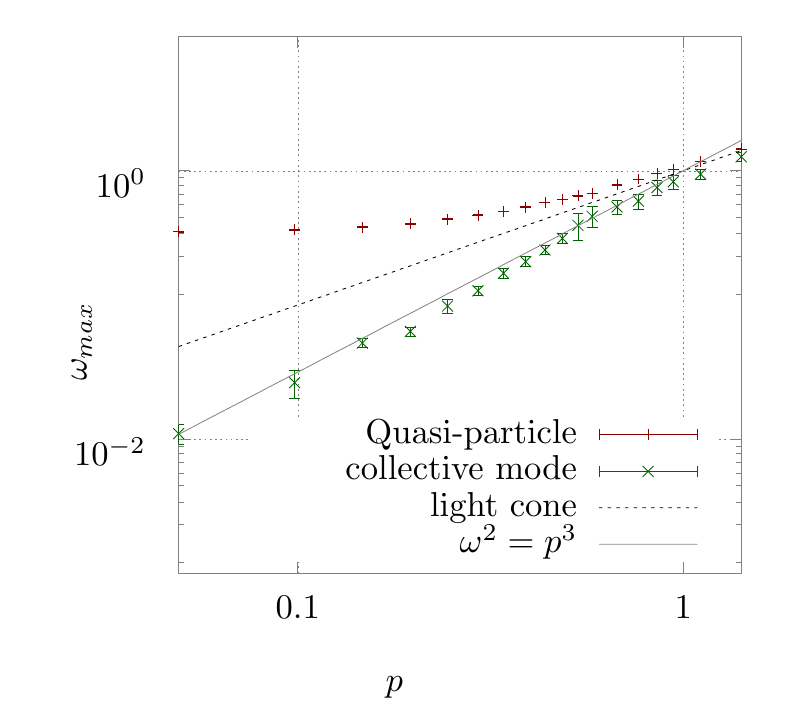} \end{minipage}
    \caption{Spectral functions in the low-temperature phase for Hamiltonian dynamics (Model C, $\gamma=0$) in $d=2$ (left) and $d=3$ (right) spatial dimensions. 
    Time-like parts (where $\omega^2 - p^2 > 0$ are plotted with green lines, space-like parts (where $\omega^2 - p^2 < 0$) with red lines.
     The colormap-projections at the bottom also show solid lines where $\omega^2 - p^2 = 0$, indicating the light-cone.
     In the bottom panels we show the corresponding dispersion relations of the quasi-particle peaks (above) and the soft modes (below the light-cone), by tracing the positions of the respective local  maxima in the spectral function.
     For comparison, we also plot solid lines, for an ideal sound-wave dispersion relation $\omega^2 = p^2/3$ in $d=2$ (left), and for cappillary waves  with $\omega^2 = p^3$ in $d=3$, to guide the eye. At high spatial momenta $p$, the soft mode dissolves in 3+1D.
        }
    \label{fig:sf-structure}
\end{figure}

In \cref{fig:sf-structure} we show exemplary low-temperature spectral functions in $d=2$ and $3$  spatial dimensions, as functions of frequency and momentum squared. The time-like and space-like parts are separated by different colors, and the light-cone is shown as a solid line in the colormap projection in the bottom plane. The light-cone marks the separation between the quasi-particle and the soft mode. In $d=3$ dimensions this soft mode closely follows the power-law dispersion relation of thermally driven capillary waves as seen in the bottom right panel of \cref{fig:sf-structure} where we plot a solid line with $\omega =  p^{3/2}$ for comaprison. Bottom left we show the corresponding quasi-particle mode above and soft mode below the light-cone in $d=2$ dimensions together with an ideal sound-wave dispersion $\omega = p/\sqrt{3}$ to guide the eye. 

Close to the critical temperature, the low-frequency part grows in magnitude and seems to merge with the quasi-particle.
In the symmetric phase, at $\tau > 0$, there is only the quasi-particle peak left.

 \end{appendices}

\bibliographystyle{elsarticle-num}
\bibliography{library}

\end{document}